Fermi National Accelerator Laboratory

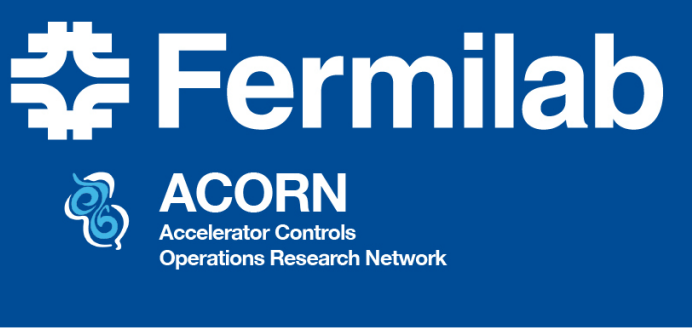

# HUMAN-SYSTEM INTERFACE STYLE GUIDE FOR ACORN CONTROL SYSTEM

Accelerator Controls Operations Research Network

August 2025

Rachael Hill, Casey Kovesdi, Torrey Mortenson, Madelyn Polzin, Zach Spielman, and Dr. Katya Le Blanc

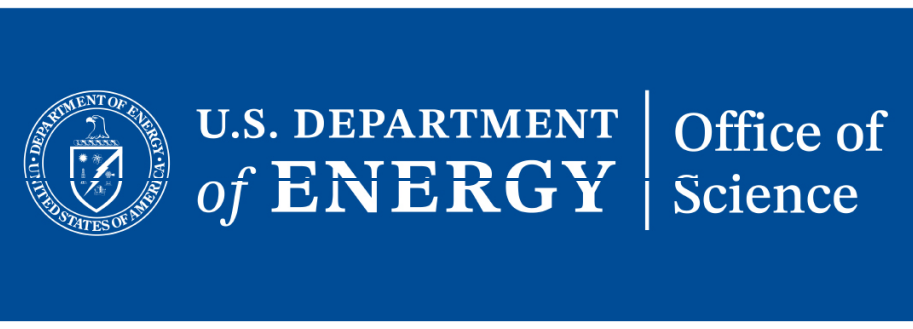

## Revision Log

| Revision | Description | Effective Date |
|---|---|---|
| 0.1 | Initial Version | 08/31/2023 |
| 0.2 | Version 2, Minor improvements | 08/31/2024 |
| 0.3 | Version 3, Major improvements in content and structure | 08/31/2025 |

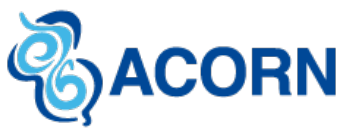

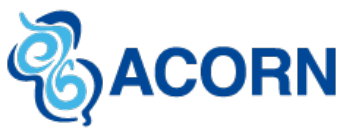

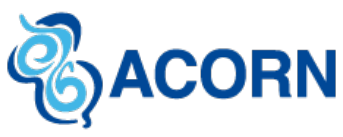

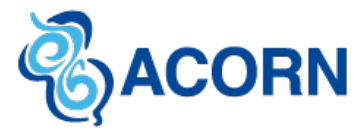

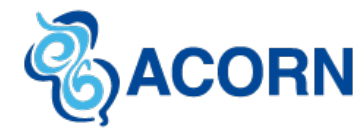

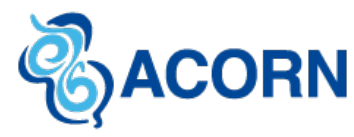

# 1.0 SCOPE OF DOCUMENT

## 1.1 Purpose & Scope

The purpose of this style guide is to provide a clear, consistent framework for the design and development of human system interfaces (HSIs) used throughout the Fermilab accelerator complex. It establishes a shared visual and interaction foundation to ensure that interfaces remain intuitive, effective, and cohesive, regardless of when or by whom they are developed. By adhering to these guidelines, developers can avoid introducing unnecessary deviations that compromise usability or increase system training burden. This consistency is especially critical in long-term, multi-contributor projects where interface continuity and maintainability are paramount.

This document serves as a practical reference for all HSI development activities related to the accelerator control environment. While the guidance provided is comprehensive, it is not exhaustive of every potential design scenario. As such, the style guide is intended to function as a living document, subject to regular review and revision. Updates will be made at least annually to incorporate emerging best practices, operational feedback, and evolving system needs. Areas where detailed guidance is still under development are clearly indicated in gray throughout the document and will be addressed in future revisions according to project priorities.

It is important to note that this guide defines the baseline standards and requirements for all control system interface work, but it is not the sole resource for development-related questions. In cases where unique or complex design challenges arise that are not explicitly covered here, developers are encouraged to consult the UX/human factors team at Fermilab for additional support. This guidance is also intended to be used in conjunction with [ACORN-doc-700: *Design Philosophy for Accelerator Control Rooms*], which outlines the broader design context in which these interfaces operate.

## 1.2 Structure of Document

This document is organized into three integral parts (*see Figure 1*), each building upon the last to provide a comprehensive and scalable foundation for interface design. The purpose of this structure is to ensure alignment with human factors principles, consistency across system displays, and adaptability to future needs as control technologies and operational practices evolve.

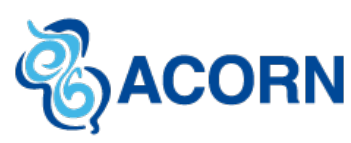

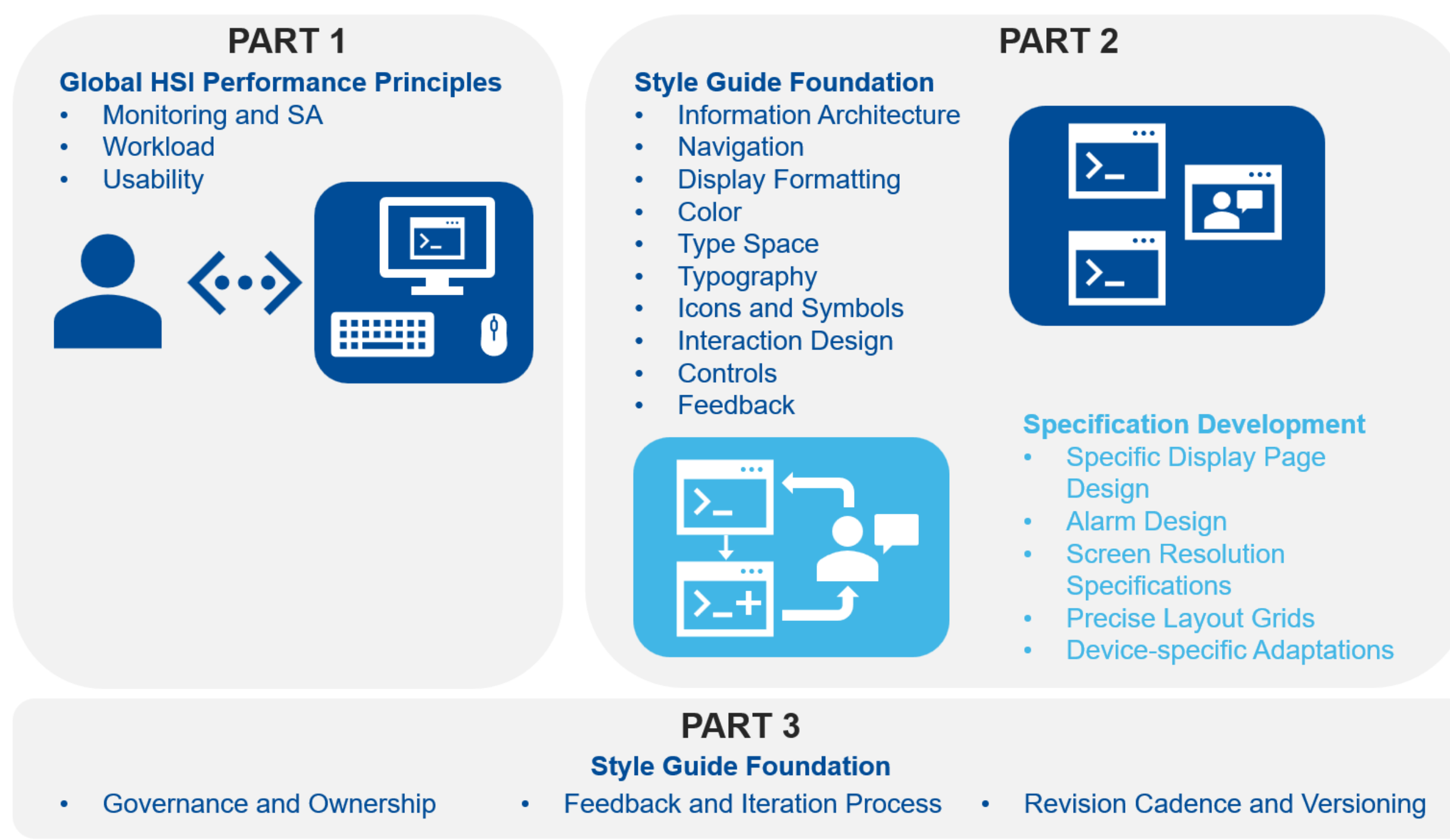

***Figure 1. Illustration of the HSI Style Guide Structure***

Part 1 includes global key principles for HSI design and establishes the overarching human factors philosophy and design intent. This section articulates the cognitive and usability goals that should inform all HSI development activities. It includes foundational guidance derived from sources such as the *High Performance HMI Handbook*, ISO/IEC standards, and the Nielsen Norman Group's usability heuristics. Topics include the importance of human factors functional and design considerations for monitoring, detection, and selection. These principles are written to be system-agnostic and provide a baseline for evaluating any HSI implementation within the accelerator complex.

Part 2 details the style guide for accelerator HSIs and translates those high-level principles into actionable, context-specific design rules. It defines standard elements and visual conventions for accelerator control interfaces, including information architecture, visual hierarchy, navigation structures, color usage, typography, iconography, layout, and system feedback. This section also specifies guidance on interaction, control affordances, and consistent terminology across systems (i.e., labeling). Where applicable, graphical examples and before/after case studies are included to reinforce preferred design practices. This portion of the guide serves as the primary reference for developers, designers, and reviewers during interface creation and modification.

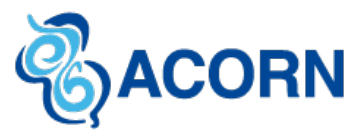

Part 3 describes the next steps and future revisions by outlining the governance, lifecycle, and versioning process of the HSI style guide. It includes recommendations for continuous feedback collection from control room operators, usability testing protocols, and change request workflows. This section encourages periodic reevaluation of standards in response to operational feedback, system upgrades, and evolving best practices. It also provides a navigation map for users of the document to help them understand how to apply, interpret, or challenge specific guidance as the control system environment matures.
Together these three parts create a balanced, flexible structure for guiding the development of intuitive and cohesive interfaces in accelerator control rooms.

## 1.3 Acronyms & Glossary

| Acronym | Term | Definition |
|---|---|---|
| **ACNET** | Accelerator Control Network | Accelerator Control Network is the primary control system utilized at Fermilab. |
| **ACORN** | Accelerator Control Operations Research Network | The project that will modernize the accelerator control system and replace end-of-life power supplies to enable future operations of the Fermilab Accelerator Complex with megawatt particle beams. |
| **BBM** | Beam Budget Monitor | Accelerator machinery that measures beam intensity over time. |
| **Fermilab** | Fermi National Accelerator Laboratory | US Department of Energy particle physics and accelerator laboratory. |
| **HFR** | Human Factors Requirements | Requirements that meet the criteria of human factors functional design. |
| **HSI** | Human-System Interface | The digital interface through which an operator interacts with the accelerator control system. |
| **IA** | Information Architecture | The overall conceptual model used to plan, structure, and assemble system information. |
| **IDA** | Information and Decision-Aiding | Logic-based algorithms of automation that assist operators in decision making. |
| **IxD** | Interaction Design | The practice of designing how users interact with digital systems, focusing on system behavior, user input, and the flow of actions to achieve goals |
| **MCR** | Main Control Room | The central hub where operators monitor, control, and coordinate all accelerator systems to ensure safe and efficient beam operations. |
| | Comfort Display | Comfort displays are human system interfaces that provide the highest level of information on displays that are viewable by all crew members. |
| | Fast Time Plot | A graphical display of parameters (usually 4) of a given machine within a short time frame (0.2 seconds) |
| | Page | A specific view or organization of control system information (e.g., index page) |

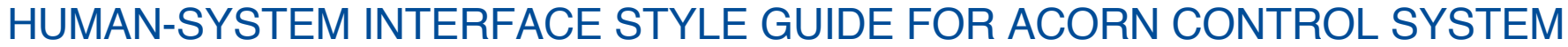
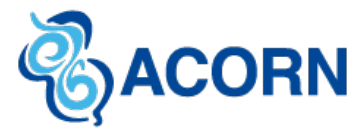

| | | |
|---|---|---|
| | Parameter | A graphical visualization of accelerator data (usually with an x & y axis) |
| | Plot | A specific metric/variable of data across all user applications including plots |
| | User Application | A software tool that provides operators or engineers with an interface to monitor, control, or analyze accelerator systems and data (most often an in-house built program) |

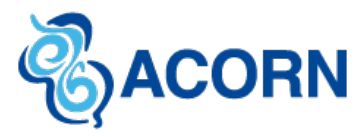

# PART 1: GLOBAL PRINCIPLES FOR HUMAN PERFORMANCE

This section establishes the foundational design principles that guide the development of HSIs across all applications within the accelerator control system. These principles serve as high-level, system-agnostic standards that promote usability, consistency, and operator effectiveness. This section emphasizes the importance of designing interfaces that support situational awareness and minimize cognitive load.

Key topics include designing for primary users first, supporting situational awareness, and embedding usability into every phase of the design.

## 2.0 GLOBAL PRINCIPLES FOR HUMAN PERFORMANCE

### 2.1 Design for Primary System Users First and Additional Users Second

#### 2.1.1 General

In the context of accelerator control systems, design efforts must prioritize the needs of the primary system users (i.e., accelerator operators) above all others. Operators are the individuals who interact with the system continuously, make real-time decisions, and are responsible for system performance, safety, and mission execution. As such, the interface must first and foremost support their workflows, cognitive demands, and situational awareness. This includes optimizing for clarity, responsiveness, efficiency, and error prevention in high-pressure environments.

While additional users such as engineers, physicists, or support staff may also rely on the interface, their needs should be addressed only after the core operator requirements are met. Designing for all users equally from the outset risks diluting usability and introducing unnecessary complexity. Instead, secondary user needs should be supported through role-specific tools, optional display layers, or configurable interfaces that do not interfere with the operator's primary tasks. This approach ensures the interface remains focused, task-aligned, and operationally reliable.

Additional information regarding accelerator roles is included in the following sections.

#### 2.1.2 Accelerator Operators

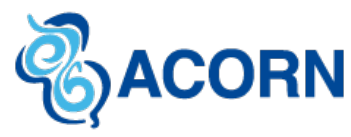

There are multiple types of roles that interact with the accelerator control system including accelerator operators, system experts, and engineering staff. Each of these roles interact with the accelerator control system for specific purposes, and although each role is important, the main control room operators are the primary user of the control system. Operators are the first line of defense for incoming alarms and interact with the control system all day, every day. Two rounds of interviews have been performed thus far. The initial round focused on main control room operators and the second round focused on secondary users: machine and system experts, engineering staff, and physicists.

### 2.1.3 Main Control Room Operators

The main control room operator's primary responsibility is to ensure beam viability by acting as a first responder to any accelerator ailments. Should an unexpected event occur, operators act swiftly to mitigate the incident by performing diagnostics and restorative controls. If the necessary repair is beyond operator capabilities, machine experts are consulted.

In addition to responding to alarms, operators are responsible for tuning the beam and making corrections as deemed necessary to ensure beam losses are minimal and the accelerator complex is running safely. Operators monitor a variety of plots depending on the status of different components in the complex.

The crew chief in the Main Control Room oversees the operators on shift and executes similar tasks to operators, while performing supervisory oversight of the accelerator and facility processes. The crew chief may delegate a task to a specific operator and make decisions for how to proceed in completion of a task. Both operators and the crew chief utilize the same applications in the control system and have very similar workflows, aside from a few additional ones that come with the crew chief role (e.g., timing of beam events).

### 2.1.4 Machine and System Experts

Machine and system experts specialize on specific equipment and are therefore better equipped to solve complex issues relating to their specific devices. The information a machine and system expert may require to diagnose problems with their equipment can differ from the general knowledge held by control room operators. Machine and system experts have similar high-level operational goals (e.g., ensure beam quality) to operators, but they address those goals with access to different levels of information.

### 2.1.5 Engineers

Engineers span different areas of the accelerator complex with focuses including software, hardware, mechanical support, electrical, and instrumentation. Engineers are responsible for developing and maintaining embedded systems (front ends) as well as the data acquisition from the systems. Engineers ensure their systems remain operational through maintaining the installations and monitoring their performance to address any issues.

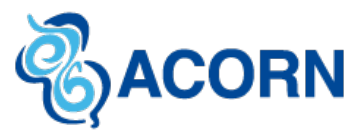

### 2.1.6 Technicians

Technicians provide infrastructural support for the different components (e.g., vacuums) in the accelerator complex. Their focus lies on troubleshooting, maintaining, and monitoring of systems to help ensure continuous and efficient operations. Technician groups will use specific applications for the different systems they oversee. Technicians can occupy both software and hardware roles and often support physical equipment out in the field.

### 2.1.7 Physicists

An accelerator physicist focuses on improving beam performance and exploring global physics topics. The physicist analyzes data through plots to determine the best operational settings.

### 2.1.8 Pre-Requisites

Designing accelerator interfaces with operators as the primary users requires thoughtful preparation and planning. Design teams must start by identifying all user roles, with a clear distinction between primary users (i.e., accelerator operators) and secondary users (e.g., engineers, machine experts, and physicists). Operators should be treated as the primary user group (hereafter referred to as 'operator'). A task analysis is helpful to understand what operators do, when, and how. This includes monitoring systems, responding to alarms, tuning the beam, and collaborating with other staff. Core displays must reflect these tasks and their priority in typical operations. Use cases should be ranked by how often they occur and how critical they are to operations. Design should start with the most common and important operator tasks. Features for other users should only be added once operator's needs are fully supported.

### 2.1.9 Detailed Design Guidance

HFR-PRI-01: Operator Workflows Should Drive Initial Design.
*Additional Information:* The layout, functionality, and data hierarchy of each interface should be based on the workflows and tasks of main control room operators.

HFR-PRI-02: Core Displays Must Be Optimized for Operational Tasks.
*Additional Information:* Interfaces used by both primary and secondary users must prioritize operator-relevant data, controls, and visual structures. Secondary user content must not displace or obscure operational content.

HFR-PRI-03: Secondary User Needs Must Be Role-Isolated or Configurable.

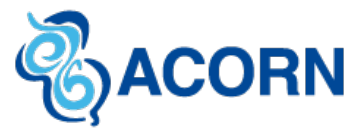

*Additional Information:* Support for engineers, physicists, and system experts must be provided through role-specific tools, alternate views, or user-specific configurations that do not interfere with or disrupt core operator interfaces.

HFR-PRI-04: Shared Systems or Displays Must Default to Operator Mode.
*Additional Information:* In interfaces used by both operators and other users, the default mode of interaction must reflect the operator's perspective, prioritizing alarm management, system health, and recovery actions.

Additional information regarding a task analysis for additional roles as well as key insights from accelerator personnel interview findings can be found in Appendix C.

## 2.2 Support Situation Awareness through Monitoring, Detection, and Selection

### 2.2.1 General

Situational awareness (SA) is a foundational goal of HSI design, particularly in high-stakes environments like accelerator control rooms. Effective situational awareness enables operators to perceive what is happening in the system, understand that information means in context, and anticipate what will happen next (Endsley, 1995). In practical terms, HSI design must support these cognitive processes through the clear presentation of information, meaningful alerts, and intuitive controls.

Monitoring, detection, and selection are central to operator situational awareness and can be strengthened through thoughtful interface design. Monitoring allows users to maintain awareness of key system elements; detection supports the recognition and understanding of abnormalities or meaningful changes; selection enables operators to take appropriate actions quickly and confidently. These functions are not separate steps but interconnected activities that together form a continuous loop of observation, interpretation, and response (see figure 2).

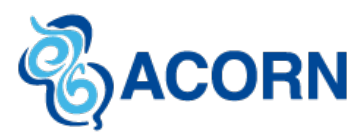

Situational Awareness

Monitoring

Selection

Detection

*Figure 2. Visualization of Situational Awareness in an Accelerator Environment*

### 2.2.2 Pre-Requisites

To design interfaces that support situational awareness, teams must first identify the key system elements that operators must observe and understand. This includes both static and dynamic elements, such as real-time sensor data, machine states, environmental conditions, and interlock statuses. Developers must also understand typical user tasks, failure scenarios, and transitions between operational modes.

Situational awareness can be supported by modeling user workflows and identifying which parts of the system require monitoring (e.g., “What’s the current beamline position?”), detection (e.g., “Why is the cavity overheating?”), and selection (e.g., “Will the beam remain stable?”). Early-stage prototyping and cognitive walkthroughs can help validate whether monitoring, detection, and selection behaviors are effectively supported in the design.

### 2.2.3 Detailed Design Guidance

**HFR-SA-01: Interfaces must present key system elements in a manner that supports rapid perception and comprehension.**
*Additional Information:* Displays should emphasize the most relevant data for each operator role and task, using visual hierarchy, spatial organization, and color to highlight system state and changes over time.

**HFR-SA-02: The interface should support early detection of abnormal or off-nominal conditions.**
*Additional Information:* Alarms, thresholds, and trend deviations must be easily distinguishable from normal background information. Alerting mechanisms must prioritize critical events without overwhelming the user with non-actionable data.

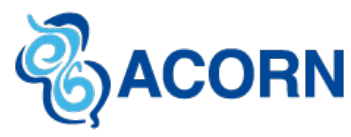

**HFR-SA-03: Interactive controls must support quick, informed selection, and corrective action.**
*Additional Information:* Controls should be clearly associated with the relevant system elements, offer immediate visual feedback, and reflect the current system mode or constraints (e.g., disabled in maintenance mode).

**HFR-SA-04: Displays must support users' ability to anticipate future system states.**
*Additional Information:* Use trend data, predictive indicators, and state transitions to help operators project what will happen next. For example, trends in RF cavity temperature may inform whether action is needed to prevent an interlock trip.

**HFR-SA-05: The interface must maintain context continuity during transitions and task switching.**
*Additional Information:* When users navigate between views or respond to alerts, relevant contextual information (e.g., timestamps, machine mode, system status) should be preserved or re-displayed to maintain situational continuity.

**HFR-SA-06: All aspects of the interface that support SA must be validated through user-centered testing.**
*Additional Information:* Usability testing, simulation walkthroughs, and operator feedback should be used to evaluate how effectively the interface supports perception, comprehension, and projection tasks under both normal and abnormal conditions.

### 2.2.4 Situational Awareness in an Accelerator Environment

In an accelerator control context, situational awareness must be continuously maintained across a highly distributed and time-sensitive system. Operators monitor conditions like beam stability, magnet performance, vacuum integrity, cryogenics, and RF systems, all of which can shift rapidly in response to changing operating modes or environmental factors.

Monitoring displays must therefore be designed to highlight the most critical indicators in real-time, allowing operators to detect subtle deviations before they escalate. Alarm systems must distinguish clearly between routine notifications and events that demand immediate response. Finally, selection mechanisms (e.g., control inputs, mode toggles, or diagnostic views) must be intuitive and aligned with the operator's mental model of system function.

Because the accelerator environment is both complex and dynamic, situational awareness is not achieved through a single display or feature but through a network of visual, interactive, and informational elements that reinforce each other. Supporting this loop of monitoring, detection, and selection is essential to ensuring system safety, operational reliability, and human performance.

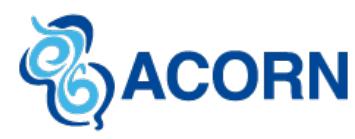

## 2.3 Embed Usability into Every Phase of Accelerator Interface Design

### 2.3.1 General

Many designers and developers are tempted to quickly begin designing interfaces without truly understanding what they are designing and what requirements their designs will meet. Rushing into “the how” without clearly defining “the what” causes designers to improvise. Improvising in design can lead to an endless cycle of iteration where solutions are proposed without clarity, thoughtfulness, and even consensus. This not only creates inconsistency throughout designs, but also disconnection and incoherence. On the contrary, understanding “the what” before even thinking about “the how” provides a clear path for designers towards a cohesive system. “The what” may somewhat change over time which is another reason this document is expected to be revised regularly. However, this document also provides both general guidance of usability principles and specific guidance of interface design specifications. The combination of these guidelines was created through consolidating findings from accelerator personnel interviews, evidence-based human factors design principles, and considerations from other accelerator control system benchmarks. Adhering to these guidelines will lay the foundations for designs that are robust and congruous. This section introduces long-established usability principles. Usability principles are design principles that, when used correctly, enable intuitive and user-friendly interfaces. The stronger the usability of an interface, the better equipped users are to optimally perform. Each of the usability principles introduced should be applied to all concepts of design.

### 2.3.2 Consistency

Consistency in interface design is making elements uniform across a digital system. Consistency creates predictability in an interface which enables user confidence and optimizes overall performance (Krause, 2021). Design consistency is also conforming to a set of standards and persistently applying them throughout all interface designs. The principle of consistency creates the foundation for all design as changing the look and feel of elements from page to page dissociates the entire system. Users should not have to wonder whether different labels, colors, or other elements mean the same thing from page to page throughout a digital system. Failing to maintain design consistency increases cognitive workload and can lead to burnout.

#### 2.3.2.1 Design Consistency Enables Flexible Operations

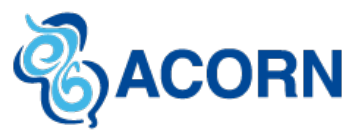

Accelerator environments are complex, and users rely on some level of flexibility to maintain efficient operations. The idea of supporting flexibility might seem contradictory to the principle of consistency, but the opposite is true: properly applied design consistency actually enables flexibility within user operations. Creating consistency in how a system looks and feels directly increases the trust and familiarity the user experiences and thus increases the potential for system mastery (Krause, 2021). The more proficient a user is within a system, the greater their capability is to use the system in a way that best fits their needs (i.e., flexibly).

### 2.3.3 Detailed Design Guidance

**HFR-CONS-01: Interface elements (e.g., labels, colors, icons, layout zones) must be applied consistently across all pages and subsystems to support user expectations and task fluency.**

**HFR-CONS-02: Design systems and visual standards (e.g., widget libraries or pattern sets) must be maintained and referenced in all interface updates to support long-term cohesion.**

### 2.3.4 Familiarity

Jakob's law states that the time users spend interacting with technology outside of work influences how they expect the technology they interact with at work to function (Nielsen, 2000). Therefore, designing the interaction platform for the Fermi accelerators should incorporate familiar elements found across other interactions with websites, technology, or system platforms. Doing so can improve both the inherent and apparent usability of a system. The inherent being the objective usability of a design and the apparent being the perceived usability of a design. Employing familiar concepts, icons, layouts, imagery, and tools can be especially helpful for novice users working with the system. The way the system behaves and the user's expectation of system behavior will align better and earlier in the training process. This is also referred to as designing to match the user's mental model.

### 2.3.5 Detailed Design Guidance

**HFR-FAM-01: Common interface patterns (e.g., breadcrumbs, dropdowns) should be used where applicable to match user expectations based on prior experience.**

**HFR-FAM-02: Terminology, labeling, and symbology yust follow known conventions unless specific domain exceptions are required, in which case the UX/HF team should be involved**.

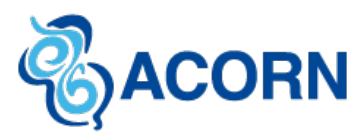

### 2.3.6 Simplicity

Designing simple interfaces can be thought of as the practice of maximizing meaning and information while minimizing the amount of “ink” on the page (Valdez, et. al. 2015). An effective simple design means the functionality required to perform a task is included, and how to perform that task is clear. Furthermore, a simple display is one that is not easily susceptible to misinterpretation. Simplicity does not refer to removing features or capabilities for the sake of simplicity. The number of tasks and amount of information required to run an accelerator may call for full screens and interfaces that make many capabilities always present and available. Natural mappings to interactions can improve the simplicity of an interface as well. Grouping related parameters and designing proximity-compatibility relationships between controls and parameters can improve simplicity

Also, when designing for simplicity, consider the mix of expert and novice users. The interface should focus on supporting novice user tasks directly on the interface in a clear and obvious way. Tasks suited or geared for expert users may require an additional step to access or slightly more nuanced interaction. As users grow from novice to expert users these features will become more explored as the transition occurs offering a natural development from novice to expert. Designing for gradual engagement of this nature requires specific intention as to how the interface may be designed.

### 2.3.7 Detailed Design Guidance

HFR-SIM-01: Interfaces must prioritize essential information and controls, removing or hiding secondary details unless explicitly requested by the user (e.g., drill-down).

HFR-SIM-02: Controls for novice tasks must be made immediately visible and clearly labeled. Expert-level functionality may be placed in expandable menus or secondary views.

### 2.3.8 Abstract & Aggregate Data

Abstracting data to higher-level meaning offers a means to support users more directly by creating immediately meaningful information that requires less cognitive processing than presenting raw data. Abstraction can be applied to labels and telemetry by using natural language or providing context respectively. Both of which reduce interpretation and provide more direct input to the user. Aggregating data is a simple method to take related inputs and combine them into a single indication. Common applications of aggregating data are interlock decision trees. The decision tree can be consolidated to a single indication of ‘met’ or ‘unmet’ clearing space on the screen for other relevant and meaningful elements. If the desired status is not displayed the option to investigate should be present. In any instance of aggregating data, the ability to individually investigate the aggregated elements should be present especially when expert users are involved, this can also be referred to as “drilling down”.

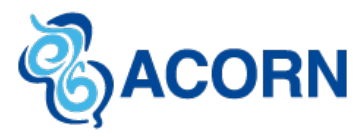

### 2.3.9 Detailed Design Guidance

HFR-AGG-01: Wherever possible, raw data must be aggregated into summary indicators (e.g., beamline interlock condition as "OPEN" or "CLOSED").

HFR-AGG-02: All aggregated or abstracted indicators must include a method for drilling down into underlying data for expert use or troubleshooting.

### 2.3.10 Transparency

The system interface should feel like a direct link to the system. Transparency typically refers to the opacity of an object or the ability to see through an object. When applying transparency to interface design it takes on a slightly different meaning. Transparency is a design concept to overcome the "gulf of execution" (Whitenton, 2018). The gulf of execution refers to the challenge of performing an action to accomplish a task without immediate *a priori* knowledge of the system's internal state. A transparent interface links the user directly to their tasks and the feedback the system provides when performing their tasks. It removes anything that can distract or add extra steps for a user when trying to perform an action. A transparent display is also referred to as having a high correspondence to the system it is designed for. That is the mapping of interface to system domain. If the interface and system have a high correspondence then the transparency of the interface is increased. Correspondence is an interaction between the system and the interface. Therefore, it is equally important that after an action is made the system provides immediate feedback of having received the action.

### 2.3.11 Detailed Design Guidance

HFR-TRAN-01: Interfaces must provide immediate, visible feedback when user actions are received, processed, or rejected by the system.

HFR-TRAN-02: System controls (e.g., buttons, toggles) must reflect their current state and availability (e.g., enabled, disabled, pending) based on operating mode or user role.

### 2.3.12 Visibility

System visibility is the capability of an interface to communicate system status accurately and quickly to the user. A highly visible system overcomes the "gulf of evaluation" or the user ability to understand the system state to decide what action is necessary (Whitenton, 2018). This concept is a parallel to system transparency. Visibility and transparency together are comprehensive in 1) evaluate the system and 2) take proper action. Then the cycle repeats. As highly transparent systems have good correspondence to the system, interfaces with high system visibility have good coherence. Coherence refers to how the graphical elements, layout, and design of the system come together to clearly communicate system status to the user quickly and accurately. One tactic to improve coherence and achieve visibility is by leveraging human perceptual processing.

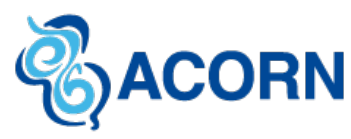

### 2.3.13 Detailed Design Guidance

HFR-VIS-01: Critical system conditions (e.g., alarms) must be visible on all relevant pages without requiring navigation.

HFR-VIS-02: Important information must be presented as distinct (e.g., using layout, color, and formatting) to support rapid recognition, especially during abnormal conditions.

### 2.3.14 Leverage Perceptual Processing

Perceptual processing is a term cognitive psychologists use when referring to how humans sense their environment and then add meaning to it based on current awareness and past experience. When designing an interface this refers to adding inherent meaning to the interface that reduces the need for the user to perform this cognitive processing stage. The symbology, iconography, plots, text, and other design elements should be coherent with the system in operation. For expert users, this strategy creates greater opportunities to see larger patterns or associate deeper meaning to what they are seeing. For novice users, this reduces the need for training as the meaning and feedback provided in the interface acts as a training device.

### 2.3.15 Detailed Design Guidance

HFR-PER-01: Related data and controls must be grouped visually and spatially to support fast scanning and reduce visual search effort.

HFR-PER-02: Pre-attentive features (e.g., high contrast, alignment) may be used to highlight abnormal states but should be sparingly applied to avoid over-alerting or visual noise.

### 2.3.16 Ease of Use (Usability)

Ease of-use is a measure of how users perform their goals with as little burden or intermediary steps between evaluation and execution. It is the coalescence of all other usability principles. to support the operator through monitoring, evaluation, and task execution. Ease-of-use can be a moving target however, especially when users with different levels of experience use the same system. Systems designed for novices typically do not overwhelm the user with available functionality visible on the screen opting to select only the basic functionality that a novice may need. This design may reduce the ease-of-use for an expert or power user. So, trade-offs must be assessed when designing for ease of use.

### 2.3.17 Detailed Design Guidance

HFR-EASE-01: All routine tasks should be executable within a minimal number of steps, ideally no more than three interactions for common workflows (Valdez, et. al. 2015).

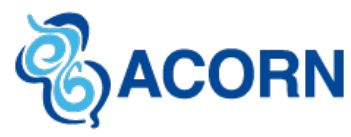

HFR-EASE-02: The interface must be designed to accommodate both novice and expert users. Where appropriate, allow for progressive disclosure of advanced functionality as users become more proficient.

## 2.4 Accessibility in Control System Design

### 2.4.1 General

An important ethical and professional responsibility of digital designers is to ensure that interfaces are inclusive, usable, and non-discriminatory. Accessibility in digital systems has increasingly become a central focus within the design community, emphasizing the need to support users with diverse physical, sensory, and cognitive abilities. Historically, many digital systems were developed without clear accessibility standards, resulting in interfaces that unintentionally excluded some users. However, the growing maturity of frameworks such as the World Wide Web Consortium’s Web Content Accessibility Guidelines (WCAG) has brought some needed clarity and structure to this area. These guidelines outline concrete, actionable principles for designing systems that are perceivable, operable, understandable, and robust across a broad range of users and contexts.

While designing for accessibility is now recognized as best practice, it can still present challenges when adapting legacy systems or specialized, high-stakes environments such as accelerator control systems. Many control interfaces were developed long before accessibility was a formal consideration, and retrofitting inclusivity into such complex infrastructures often requires thoughtful planning and iterative design. Building new systems with accessibility in mind from the outset is generally more straightforward, but modernization efforts must still strike a balance between operational precision and human-centered design.

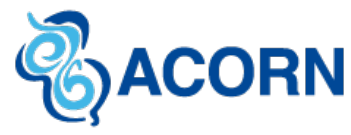

As part of the ACORN initiative, the control system style guide includes a carefully evaluated color palette in which contrast ratios meet or exceed WCAG 2.1 AA standards for text and interface elements. The selected colors have been chosen with common forms of color blindness in mind as well. These design choices not only enhance accessibility but also improve overall visual clarity for every operator, regardless of abilities.

Beyond adherence to visual standards, accessibility is best achieved through continuous engagement with diverse users. Developers and designers are encouraged to actively seek opportunities to make interactions, layouts, and data visualizations more accessible throughout the design process. Involving a variety of users that span different experience levels, cognitive styles, and physical abilities in user testing helps identify unintentional barriers early and guides inclusive design decisions. This iterative, user-centered approach ensures that accessibility is not treated as an afterthought but as a foundational quality of the interface.

In the context of accelerator operations, accessibility is not only an ethical imperative but also a practical advantage. A more accessible control interface broadens participation by enabling capable individuals with different abilities to contribute effectively, while also improving overall usability for all operators. As development of the design progresses, opportunities to develop specific accessibility requirements should be thoroughly considered and executed. Additionally, accessibility reviews should also be embedded in software functionality and design critiques to ensure inclusivity remains a continuous, measurable design goal rather than a one-time requirement.

## PART 2: STYLE GUIDE FOR ACCELERATOR HUMAN-SYSTEM INTERFACES

This section presents the current design standards and conventions for HSIs used throughout the accelerator control environment. This section translates the global design principles from Part 1 into specific guidance for interface development and includes elements such as color usage, typography, iconography, layout structures, controls, and interaction. These standards are intended to promote consistency, usability, and operational efficiency across all systems, regardless of the platform or development team involved.

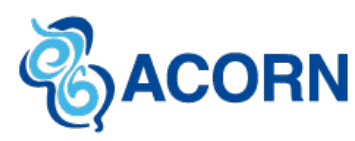

While much of the guidance in this section is well-defined and ready for application, it is important to note that Part 2 is an evolving resource. Several details such as screen resolution specifications, precise layout grids, or device-specific adaptations are not yet finalized. These areas will be developed iteratively, through methods such as prototyping, user testing, and feedback during the early phases of implementation (for more information on this, see Part 3). Sections that are still under consideration or in preliminary stages are marked accordingly and will be updated in future revisions as project needs evolve.

This section is intended as both a reference and a foundation, enabling immediate application of established standards while remaining flexible enough to incorporate future insights.

# 3.0 STYLE GUIDE FOR ACCELERATOR HUMAN-SYSTEM INTERFACES

## 3.1 Information Architecture

### 3.1.1 General

Fundamentally, information architecture (IA) is the identification and definition of a website or application's content and functional components (Rosenfeld, et al., 2015). It also extends to the organization, structure, and naming conventions used within the application. An effective information architecture can have positive impacts across other human factors and UX design principles and, therefore, is an important component of effective design. As an analogy, IA is the steel within a concrete form or a skeleton within a creature. We do not see it directly, but its form is apparent still.

A robust IA will enable a consistent and impactful user experience because the IA creates the structure for all the content that a user will engage with. The purpose of including IA in design efforts is largely to ensure that the application being developed will be understandable and usable to the range of users and experiences that they will have. For example, during routine operations, an operator may need to rapidly navigate from a high-level overview of beamline performance to a specific RF cavity's status. If the interface is not organized around a clear, task-driven hierarchy, this process can become slow, confusing, or error-prone, especially under time pressure. Conversely, when a system expert is troubleshooting a vacuum subsystem, they may require detailed telemetry and historical logs that are irrelevant or somewhat overwhelming for operators during normal shifts. A strong IA supports both user groups by structuring the application in layers and offering intuitive top-level navigation for operational tasks, while providing deeper paths or filters for expert use.

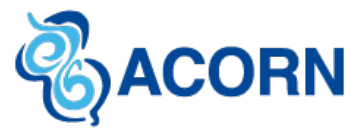

IA serves as a scaffolding or architectural component of a complex application to ensure that the developed application contains what is needed in the proper order, placement, and arrangement. Sometimes a developer may create an application specifically for a very niche use or just for a small team, when the entire user base occupies a “super user” role then IA may not be as relevant. However, if the application is used by more than this limited user group then IA is a critical component. The application’s adherence to the system IA increases ease-of-use for all users despite the niche purpose.

IA is an incredibly important feature of any digital design and can be developed even before the first line of code is written. This section will cover some of the relevant information you need to employ IA effectively in your projects. It is also important to acknowledge that the more complex of a system you are designing, the more complex the IA will become necessarily. This works in the opposite direction as well, so the simpler the design will have a simpler IA that may require less effort to develop. For more in-depth information on IA principles and methods see (Brown, 2010; Covert, 2014).

### 3.1.2 Pre-Requisites

Before beginning a digital design some initial determinations regarding content, inputs, outputs, and use will inevitably be made. This is the most critical time to employ an IA, if you are past that point then the next most important time is right now. There are specific tasks that a human factors or user experience professional can assist with that will help either develop the IA from scratch or validate the initial assumptions regarding the content organization and hierarchy. The data collected from these research tasks serve as the initial inputs into the IA design and are not optional.

### 3.1.3 Detailed Design Guidance

**HFR-IA-01*:* Understand the role of operator context in developing organizational principles of any interface.**

*Additional Information:* Accelerator interfaces should support the breadth of tasks an operator may take at a given time. Interfaces should possess contextual organizational structures that are either top-down from the user archetypes of the system or bottom-up that allow users to develop their own organizational scheme.

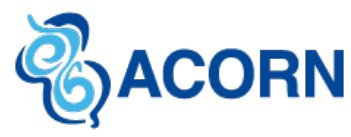

**HFR-IA-02: Operational interfaces should include a “priority” section which enables quick orientation to emergent, novel, or alarm-based information.**

*Additional Information:* Interfaces should always include an area which allows for immediate situation awareness of emergent, novel, or alarm information to consistently allow for orientation to a safety situation. This includes global parameters of interest, as well as role-based interface components, e.g., a machine expert may need machine health parameters in an easily accessible way.

### 3.1.4 Information Architecture in an Accelerator Environment

An important consideration is the context of accelerator operations. At all times lower-level operations, measurements, and evaluations take place within the broader context of the health and performance of the accelerator systems. These types of integrated operational contexts may include some design assumptions that are a result of operational experience, regulatory compliance, or safety. Therefore, developers should evaluate their assumptions regarding interface usage with real operational users to ensure that the interface (regardless of function) fits within and does not impede the overall operational performance of the accelerator. In developing the IA, ensure that this context is reflected where necessary.

Note, as the design becomes more defined through prototyping and iterative development, additional requirements concerning information architecture should be added here once specific decisions have been finalized.

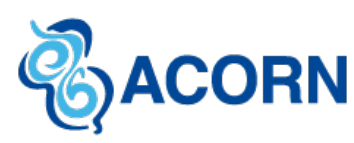

## 3.2 Display Hierarchy

### 3.2.1 General

An important consideration that relates to information architecture, particularly in the context of a control room environment is a display hierarchy. To accommodate the complexity and variety of information included in an accelerator complex as well as present that information in ways that are intrinsically meaningful to operators, a well-planned hierarchy is needed. Similar to how information architecture arranges and orders information, or content, a display hierarchy arranges and orders operator displays. A display hierarchy supports the different levels of awareness that operators are expected to have, such as top-level awareness like mutual or overview system awareness to subsequent-level individual awareness related to an operator's direct tasks.

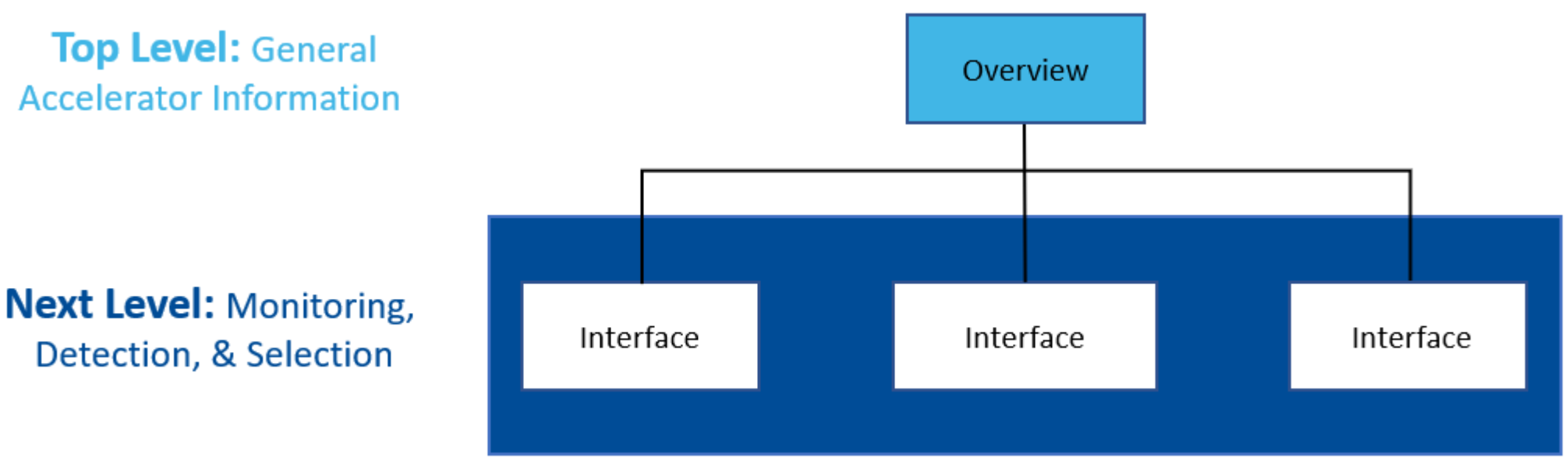

*Figure 3. Accelerator tree structure*

### 3.2.2 Pre-Requisites

Specific prerequisites to designing contexts where a display hierarchy is needed are similar to that of information architecture, you will need to have a thorough understanding of the relevant displays and the specific types of awareness they are intended to support. If developing a novel display architecture for a new system implementation that does not have significant existing operating experience then you will need user testing results to better understand the relevant mental models that the users will be bringing into the environment. This work will also enable development teams to capture critical points of user feedback that can inform the overall design.

### 3.2.3 Detailed Design Guidance

**HFR-DH-01: Understand the specific forms of awareness the displays are intended to support—supported by user research results—and ensure that all forms of awareness designs are reviewed in user testing.**

*Additional Information:* In a particle accelerator control system environment, two levels support intuitive and hierarchically organized displays including, comfort displays (high level, overview, i.e., Level 1) and operator interfaces (lower-level information and operator control; Level 2). This is illustrated in Figure 3.

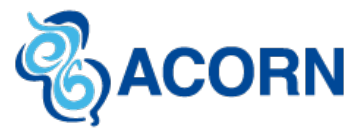

**HFR-DH-02: Ensure that the top level, or overview ("comfort"), information provides general system and accelerator information that supports the broad situation and mutual awareness of equipment and system health.**

*Additional Information:* Operators use "comfort displays" to provide overview information. Parameters included on these screens should be limited to critical system, equipment, or beam information that is necessary for the safe and efficient operation of the facility.

**HFR-DH-03: The design of the top level, or overview ("comfort"), displays should support "at-a-glance" information orientation by operators, supported by user testing results.**

*Additional Information:* Operators should be able to quickly and accurately ascertain the health and current state of the entire accelerator system and critical components. Operators should rely on these displays for initial notifications of any degraded, abnormal, or emergency conditions.

**HFR-DH-04: The next level, or operator task focused, displays should directly support the operator in performing the tasks and functions assigned to them.**

*Additional information:* This level of accelerator interface should support a variety of operator tasks. Each interface should be organized contextually according to user responsibilities and the appropriate level of information and functionality to accommodate said responsibilities

Note, as the design becomes more defined through prototyping and iterative development, additional requirements concerning display hierarchy should be added here once specific decisions are finalized.

### 3.2.4 Display Hierarchy in an Accelerator Environment

The current "comfort display" and operator workstation paradigm demonstrates an existing concept of display hierarchy in the accelerator facility control room. Therefore, it will be critical to evaluate any modifications or updates to these frameworks with robust user research tasking as soon as possible in the development lifecycle. It will be necessary to capture how displays relate to each other to minimize confusion or cognitive fatigue that can accompany large scale design upgrades. Additionally, frequent user testing during design iterations is critical to not only build familiarity and adoption openness, but also to validate the effectiveness of the proposed design.

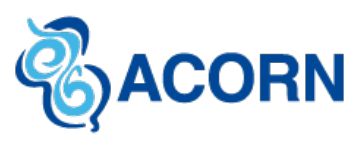

### 3.2.5 References

Foundational references and standards will be compiled and added to the style guide in the next revision to ensure that the team can provide the best references for this section.

## 3.3 Navigation

### 3.3.1 General

The purpose of navigation in a digital system is to guide users to information or functions that they are seeking. Navigation is one of the most important aspects of a digital system, particularly any control system, as it scaffolds the entire user experience. While there are simpler or more streamlined single page style applications, most digital systems contain enough depth to require some form of navigation for users. Navigation design is the discipline of creating, analyzing, and implementing ways for users to navigate through a system. Navigation is how a user can get from Point A to Point B in the most intuitive, expeditious, and least frustrating way possible.

Navigation can take many different forms, and there are many ways to achieve similar usability results. It is important that designers are deliberate in their navigation designs and validate decisions with user research. IA informs the framework for the navigation structure. Navigation design serves to guide users through the taxonomy and hierarchies. Generally, designers should seek a strong alignment between navigation schemes and the relevant IA, however as with all design tasks, user research must be present to validate these assumptions.

Additionally, many navigation icons and display elements have become commonplace, e.g., hamburger menus or breadcrumbs, therefore familiarity will play a role in designing navigational features as users will be expecting navigation to largely work like these known paradigms. If an existing and common navigation structure works for your purposes it is more beneficial to use that than attempt a wholly novel navigation approach to best leverage design inheritance and users' mental models.

### 3.3.2 Pre-Requisites

Like the section on information architecture there is an initial requirement to understand the specific information and interaction structures that are present in your designs. It can be helpful for the design and development team to work from the IA structure and make initial designs and recommendations based on the structure, number or alternatives, and user needs. Initial designs will need to have initial user research both to inform the designs and to validate initial design choices to drive iterations. Especially if navigation is an important consideration in the context, such as in a control room, where finding the needed information or function can be a critical task there needs to be robust usability testing throughout the design lifecycle to ensure that users can perform the necessary tasks within an acceptable time scale and without getting lost in other areas of the system. In terms of specific UX or HF methods to use, the same methods used in developing the IA are useful for navigation design efforts, so refer to the IA section for more information.

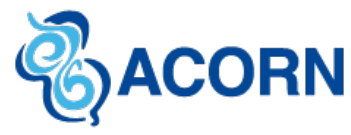

### 3.3.3 Detailed Design Guidance

**HFR-NAV-01: Top or system level navigation should be persistent and visible on all pages.**

*Additional Information:* A key experience to avoid is that users cannot back out of a section of an application or go back to "home." This creates a claustrophobic and frustrating experience for users. Designing a persistent toolbar or header that remains as the users navigate an application or page below is the most common solution for this obstacle.

**HFR-NAV-02: The systems must provide cues that inform the user of their location in the system and in the overall system informational and navigational hierarchy.**

*Additional Information:* Users should always know where they are in a digital system. There are multiple ways to indicate current location to a user. A common approach to this is to design a traceable path (e.g., tabs, pagination, and breadcrumbs).

**HFR-NAV-03: Navigational elements should have a common and persistent design throughout the application.**

*Additional Information:* Navigational elements should be unique to their function and the design should persist. For example, the "home" button or functionality should not change based on different areas or types of applications loaded. Further the "home" functionality should be distinct from other navigational elements.

**HFR-NAV-04: Designs should group similar navigational elements together.**

*Additional Information:* Grouping like elements together is a simple way to communicate commonality to users and helps users learn and use the interface at a high level of success.

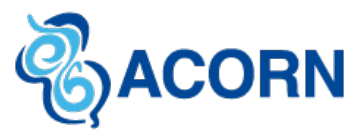

**HFR-NAV-05: Whenever possible, navigation elements should align with user mental models and common digital design practice. Deviation from these designs or styles should only be done when informed by user research.**

*Additional Information:* Due to the massive impact navigation has on a user's overall experience of a system, it is critical that designers leverage existing mental models whenever possible. Novel designs can be interesting to try and ideate, but unless users seem to show similar or greater affinity for the novel designs than with traditional designs the team should stick to traditional methods.

**HFR-NAV-06: Accessing system areas that are within one level of a user's current position should only require one interaction to navigate to.**

*Additional Information:* As the navigational hierarchies are developed, functionality for navigating within the information or functional categories should be developed such that for a user to traverse through the system, the levels of traversal count should match the number of interactions required for such a traversal.

Note, this section should be updated with finalized navigation requirements, including specific menus and widgets, as the design becomes more defined through prototyping and iterative refinement.

### 3.3.4 Navigation in an Accelerator Environment

The accelerator control system navigation should closely follow the IA structure. The closer the navigation scheme mirrors a well-designed control system IA; the more intuitive a user's interaction will be. Global navigation should allow users to access their role-based interface. Local navigation allows the user to focus on specific system information relevant to their task. The accelerator control system navigation should support flexibility for search-based navigation. A navigation menu should reflect the structure of the navigation scheme. The navigation included in these displays should be organized in a hierarchy (i.e., top-down manner) to optimize usability and accessibility. The proposed IA includes two main types of navigation organization: global navigation and local navigation. Global navigation is the top tier (i.e., general) information. It includes the highest level of organization in the navigation hierarchy. Global navigation is how users access their role-based interface. Local navigation is at a lower level (i.e., more specific) information that is contextual to the user's current task. This type of navigation includes child categories to the global navigation.

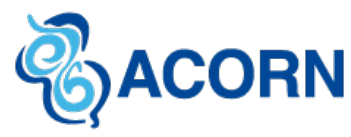

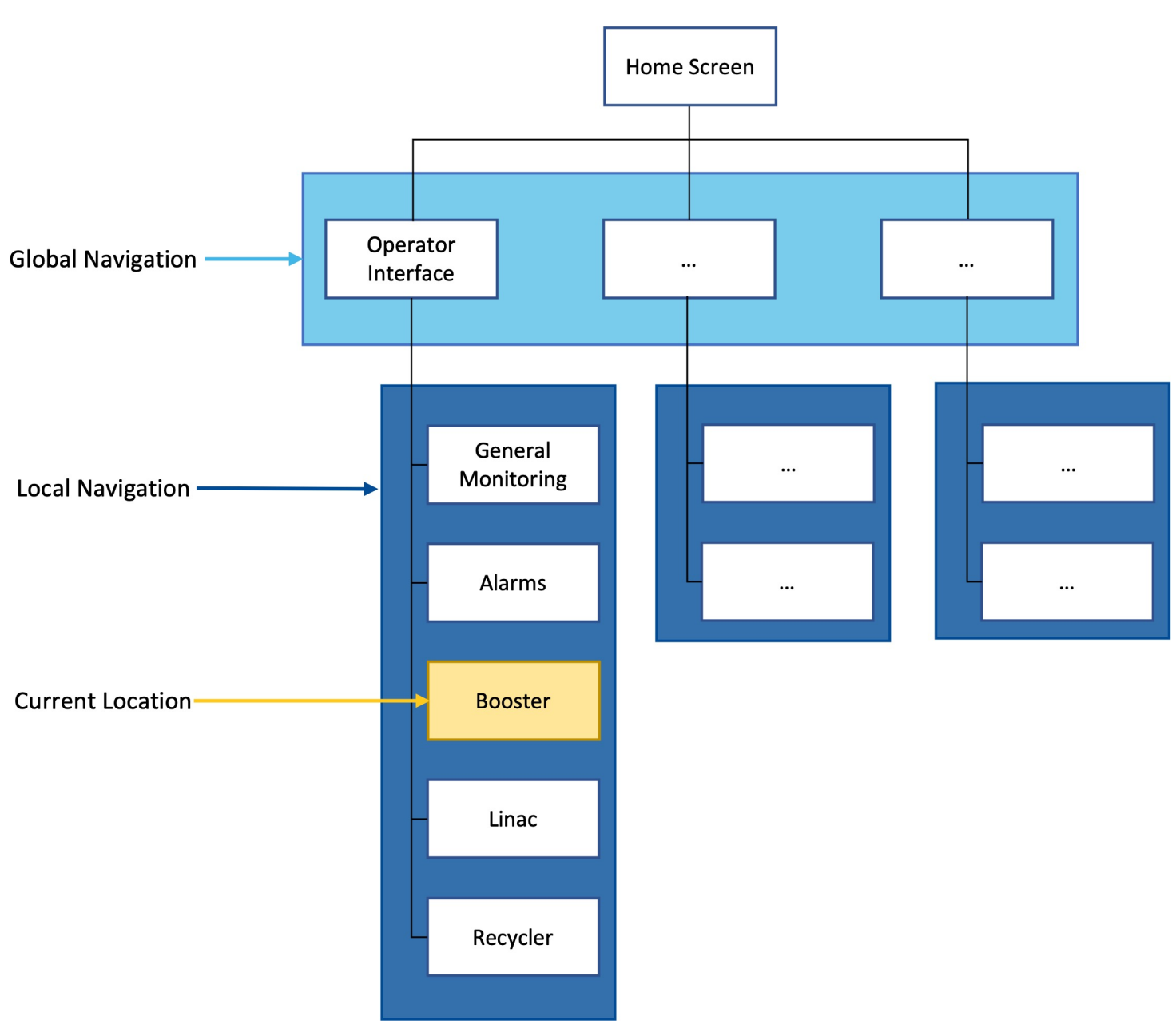

*Figure 4. Example of accelerator navigation hierarchy*

### 3.3.5 References

Design Philosophy for Accelerator Control Rooms (ACORN-doc-700)

## 3.4 Display Formatting & Layout

### 3.4.1 General

Display formatting and layout are fundamental components of effective HSI design. They directly affect how users perceive information, understand system status, and execute control tasks. In accelerator environments where operators must interpret complex data, respond rapidly to anomalies, and manage multiple systems simultaneously, the layout of visual elements should reduce cognitive load, support task flow, and enhance situational awareness.

This section provides guidance for standardizing display structure, with specific requirements for headers, navigation, breadcrumbs, and primary content areas. Adhering to these formatting standards ensures consistency across applications and makes interfaces easier to learn, navigate, and use effectively.

### 3.4.2 Pre-Requisites

Before implementing formatting and layout guidance, user roles and tasks should be identified to understand the workflow of those who use the display (e.g., operators, engineers, system experts) and what tasks they are performing. Display design should reflect these workflows. The hierarchy and content should also be defined by determining the structure of the information being shown, including priority, grouping, and relationships between data types. This supports logical and meaningful layout decisions. Additionally, navigation structure must be established and the layout should align with the system's broader navigation framework, ensuring seamless transitions between related pages or subsystems. Finally, screen and hardware constraints must be known by understanding the dimensions, resolution, and display environment (e.g., control room monitor, mobile field device) where the interface will be used. The layout should be optimized for readability and usability in those contexts.

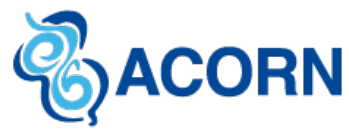

### 3.4.3 Detailed Design Guidance

**HFR-DFL-01: All display pages should contain a header with a unique title at the top of the page.**
*Additional Information:* Each display shall be labeled with a concise and descriptive title. This supports user orientation and quick recognition of system context. The title must be unique within the application and placed at the top of the screen, either left-aligned or centered. This helps avoid confusion with other similarly named applications.

**HFR-DFL-02: All display pages should provide a navigation menu at the top left within the header.**
*Additional Information:* Navigation should be accessible and consistent, located within the top bar. Menu elements must be grouped logically, labeled clearly, and visually distinct from non-navigational controls.

**HFR-DFL-03: All display pages should provide a selectable breadcrumb.**
*Additional Information:* Breadcrumbs support wayfinding and give users a traceable path through the system. Even in flat structures, breadcrumbs enhance orientation and task flow.

HFR-DFL-04: All display pages should provide a primary canvas area that is consistently sized to support the user's primary task.
*Additional Information:* The canvas must be proportioned appropriately to fit the task and content type. Avoid overcrowding, wasted white space, or inconsistent layouts that reduce visibility or usability.

**HFR-DFL-05: The means of saving a display configuration should be explicitly visible to the user.**
*Additional Information:* Whether saving is done automatically or manually, the method should be clearly communicated to the user. The control should be consistently located and provide visible confirmation when saving occurs.

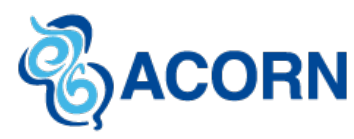

Note, as the design evolves through prototyping and iterative development, additional requirements related to display formatting and layout should be added here once specific decisions have been validated and finalized.

### 3.4.4 Display Formatting and Layout in an Accelerator Environment

In accelerator control systems, layout and formatting decisions must support fast decision-making and continuous situational awareness. For example, operators in the main control room rely on consistent page structure to manage transitions between global system overviews and subsystem-specific diagnostics. A predictable layout with clearly labeled headers, anchored navigation, and primary content zones helps users stay oriented as they shift focus during operations.

Additionally, accelerator interfaces often serve multiple user roles. Engineers and system experts may require access to deeper levels of detail without interfering with the operator's real-time workflow. Maintaining a stable, role-appropriate layout across all pages ensures that system control remains intuitive for primary users while remaining flexible for support staff.

Display formatting also plays a key role in information prioritization. For instance, alarm indicators and interlock statuses should remain in fixed, prominent positions across all screens. Plot areas and numeric data should be sized and spaced for readability in shared environments, such as multi-monitor control rooms.

## 3.5 Color

### 3.5.1 General

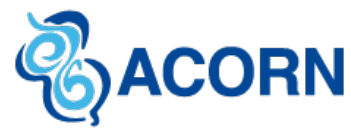

Colors serve multiple functions such as to indicatemeaning, draw a user's attention, and visually separate parts of an interface. In environments as complex as an accelerator control system, it might seem logical to use a different color for every different field, object, or piece of equipment or data to distinguish their meaning. However, there is a limit to the total number of unique colors that a human can readily recall when associating specific meaning to them. Further, color has a strong propensity for attracting the user's attention, particularly highly saturated and salient colors. Therefore, colors should be carefully selected and prioritized for important information, such as events that require immediate attention.

Having too many colors on a display can have the unintended impact of 'disguising' important information as basic information. (e.g., if 'everything' appears important then nothing really is important). Color is a powerful tool to add saliency to a display element. However if every display element has equal color properties, then saliency is lost. For example, alarms can go undetected on an interface that liberally uses highly saturated colors. The amount of colors appropriate for a digital display according to International Society of Automation Standards (ISA 5.5) state that the number of colors should be limited to the minimum required for the display's objective. If a display objective is to detect alarms and determine a mismatch between an input value and output value then color should be reserved for those two objectives, all other background information should be uniformly formatted. In other words, no irrelevant color should be used on a display. Intentional use of color reduces user attentional burden. Avoiding overuse or irrelevant use of color also strengthens the color-coding associations of an interface thereby improving its usability. Specific guidelines concerning color use for accelerator interfaces are included below.

### 3.5.2 Pre-Requisites

There are no explicit pre-requisites for the application of this guidance related to color.

To note, including color in interface design does not guarantee improved performance and user comprehension. Randomly adding or including colors that haven't been pre-approved and screened for accessibility (i.e., included in the palette) can inhibit performance and comprehension by causing visual clutter and confusion. Additionally, poor use of color across the entirety of an interface will diminish the effectiveness of colors conveying high priority information. This directly affects a user's ability to interpret the interface and respond to time sensitive information.

If developers suspect that existing colors aren't satisfactory and introducing a new color is justified, human factors/usability experts can review requests to introduce a new color into the existing palette. If the request is approved, the new color and descriptions will be incorporated into the style guide for future versions.

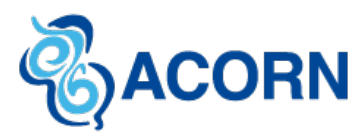

### 3.5.3 Detailed Design Guidance

**HFR-COL-01: Color should be used appropriately to indicate meaning.**

*Additional Information:* Each color included in an accelerator interface should only have one meaning consistently throughout the system. Multiple meanings used for a single-color can cause confusion and performance issues. When multiple colors are used in a single interface, users must remember each of the different meanings in addition to their work tasks. The reliance on a user's memory is even further complicated when a single color is used to convey multiple meanings. To improve the comprehensibility of color-coding associations in accelerator interfaces, each color must be linked to a single meaning or purpose. *Technical Basis:* ACORN-doc-700 (Section 6.3.1); ANSI/HFES 100-2007 (Sections 7.2.5.1 – 7.2.5.10); EEMUA-191:2007 (Section 1.3.8)

**HFR-COL-02: Colors should be consistent with accelerator operator convention and expectations.**

*Additional Information:* Implicit color associations of an accelerator complex should inform overall color usage throughout interface design. An example of this is when Fermilab accelerator structures are painted unique colors and those colors are replicated within the control system interfaces (e.g., the New Muon Lab building interior is painted mint green and the control system interfaces have the same color as a background). A better design for this example would be to include color associated headings (instead of background) to implicitly link accelerator infrastructure to relevant control system interfaces. *Technical Basis:* ACORN-doc-700 (Section 6.3.1); ANSI/HFES 100-2007 (Sections 7.2.5.1 – 7.2.5.10); EEMUA-191:2007 (Section 1.3.8)

**HFR-COL-03: A dull screen color scheme should be adopted to reduce display color saturation to increase the value of salient information.**

*Additional Information:* The "Dull Screen" approach is an interface design concept based on the theory that display elements that indicate normal system behavior should appear "dull" (e.g., gray-scale). When there is change in state indicating abnormal system behavior, salient colors (e.g., red) are used to capture the operator's attention. This strategy helps users rapidly detect events that require their detailed attention (Braseth et al., 2004). This concept can also help reduce the amount of saturated colors included in an interface which improves a user's ability to differentiate between levels of information and focus on what is most important. *Technical Basis:* ACORN-doc-700 (Section 6.3.1); ANSI/HFES 100-2007 (Sections 7.2.5.1 – 7.2.5.10); NUREG-0700 (Section 1.3.8); Braseth et al., 2004

**HFR-COL-04: Saturated colors should be reserved to indicate special meaning.**

*Additional Information:* Each color included in an interface competes with other colors and other display elements (e.g., text) for the user's attention. In line with the dull screen concept, highly saturated colors should be reserved for special or critical elements of a display. Special or

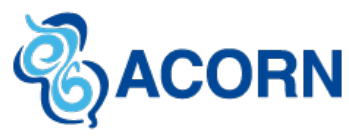

critical elements of a display are those that must effectively draw a user's attention. In an accelerator control system, this can include multiple elements such as live data that must be continuously monitored or alarm events that require immediate attention. However, when too many elements on a display use color to convey special meaning, all information included in the display is disguised as equally important. In other words, if everything appears special, nothing appears special. Reserving saturated colors to indicate special meaning for certain display elements improves overall user performance and strengthens the comprehensibility of interface color-coding associations. *Technical Basis:* ACORN-doc-700 (Section 6.3.1); ANSI/HFES 100-2007 (Sections 7.2.5.1 – 7.2.5.10); NUREG-0700 (Section 1.3.8)

**HFR-COL-O5: Highest priority information (e.g., text or other display elements) must be tested for color blind safety.**
*Additional Information:* Color blind safety is a concept that encourages certain types of color use to accommodate color blindness.

According to the National Eye Institute, color blindness affects about eight percent of males (approximately 10.5 million) and less than one percent of females (2015). There are two major types of color blindness: those who have difficulty between red and green, and those who have difficulty distinguishing between blue and yellow. A challenge in designing to accommodate color blindness is trying to accommodate the unknown. Not only do color-blindness types vary, but the level of color discernment ability varies between individuals as well. Although estimates of afflicted individuals are known, it is difficult to ascertain which color-blindness type potential users might have as well as what their level of color discernment is. Research models have been established to predict or calculate how colors are perceived by color-blind people, but they are not completely accurate. In other words, it isn't possible to predict future end user color blindness type and variability with 100% accuracy. Fortunately, there are some baseline recommendations that regardless of color blindness type and variability, will help accommodate color blindness.

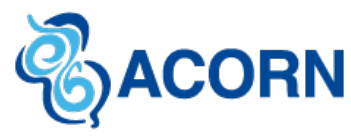

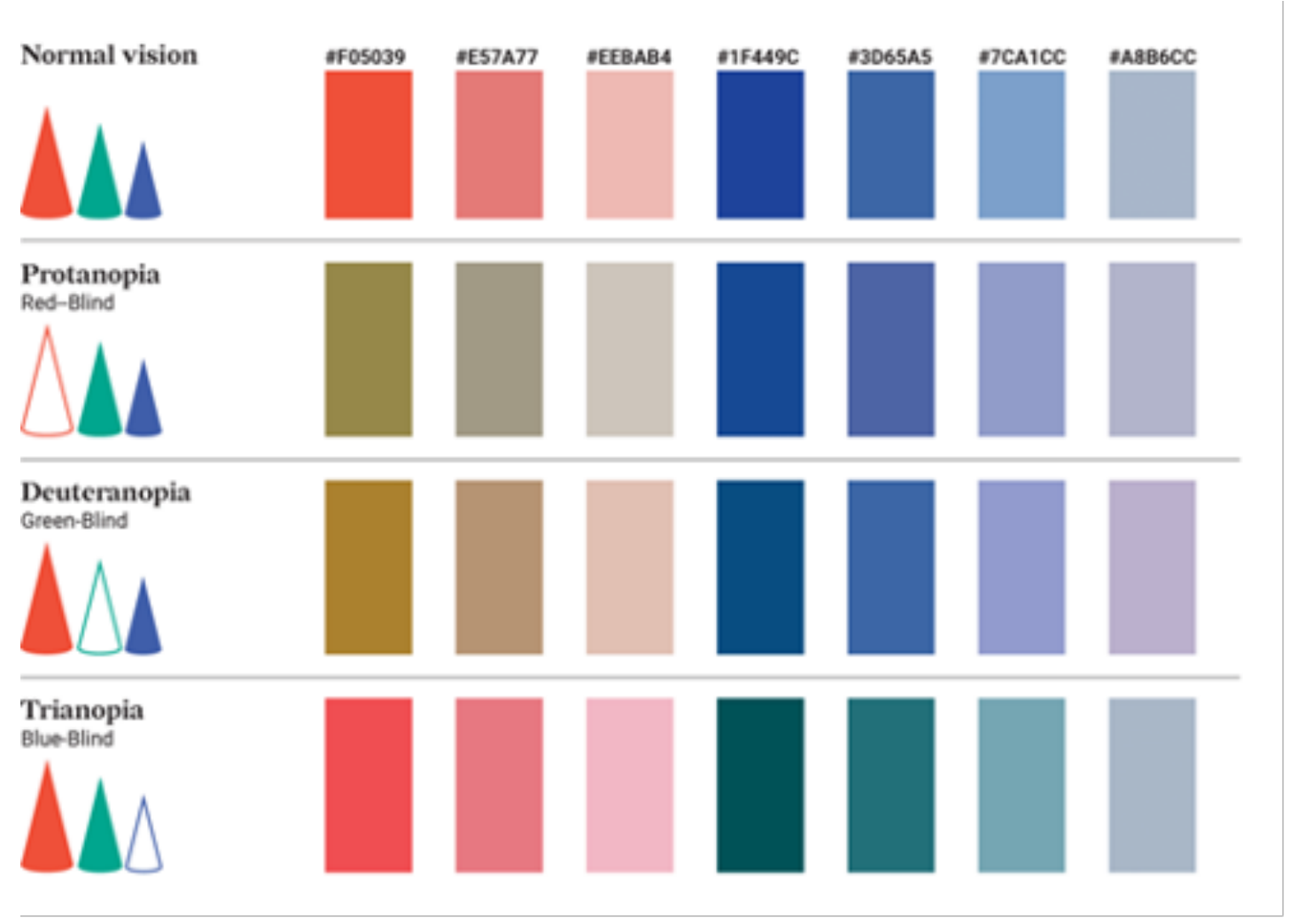

*Figure 5. Color blindness types and color combinations to avoid.*

The more important the interface content is, the more essential it is to make it color blind safe. Color blind safety is a concept that encourages certain types of color use to accommodate color blindness. The colors most detectable by anyone with color blindness are black and white (e.g., black text/elements on a white background and vice versa). This is because these colors have the highest contrast ratio compared to all other colors and are easily discernable from each other. If additional colors must be used, interface content areas should be monochromatic with the interface element color and background color at the opposite ends of the color saturation poles. Refer to the color palettes presented in Table 1 and Table 2. *Technical Basis:* ACORN-doc-700 (Section 6.3.1); ANSI/HFES 100-2007 (Sections 7.2.5.1 – 7.2.5.10); NUREG-0700 (Section 1.3.8)

**HFR-COL-06: The system should apply the project defined color palette consistently across control system display pages.**
*Additional Information:* A color-coding scheme reveals what category a display element fits in as each primary color conveys meaning to the end-user. A properly selected color-coding scheme enables quick identification of display information such as knowing a relevant category quickly without having to search to understand its contents first. *Technical Basis:* The selection of colors presented in Table 1 through Table 2 are based on reviewing the existing color palette in terms of its suitability to be legible when applied to a template that incorporates the color guidelines previously discussed (e.g., HFR-COL-01 – HFR-COL-05). The alpha channel is assumed to be set at ‘1’ in all colors provided (no transparency); this ensures maximum opaqueness. Where existing conventions were no longer suitable, alternative color selections were determined. The colors selected, as documented in this revision, are still subject to refinement through subsequent usability testing. To note, additional colors (i.e., such as for plots) may be selected using the RColorBrewer® data visualization colorblind-friendly color palettes *(Appendix E, Figure 20*). *References:* ACORN-doc-700 (Section 6.3.1); ANSI/HFES 100-2007 (Sections 7.2.5.1 – 7.2.5.10); NUREG-0700 (Section 1.3.8)

Note, a common color palette is under development for graphs (e.g., plots) that meets accelerator convention (HFR-COL-01 & HFR-COL-02), legibility needs (HFR-COL-03 & HFR-COL-04), and color blind safety (HFR-COL-05).

| Color | RBG Value & Hex Number | Primary Function | Description | Additional Information |
|---|---|---|---|---|
| Light Theme Colors | | | | |
| Red | R:178, G:34, B:34 Hex: #B22222 | Alarm, Danger Signal | Critical/High-Priority alarms | All types of faults that interfere with safety-critical operations |
| Orange1 | R:248, G:118, B:67 Hex: #F87643 | Alert, Caution Signal | Caution/Lower-Priority alerts | All types of faults that interfere with normal operations |
| Yellow | R:255, G:230, B:0 Hex: #FFE600 | Alert, Acknowledge Signal | Acknowledge alerts | All types of alerts that operators must acknowledge |
| Cyan | R:10, G:250, B:242 Hex: #0AFAF2 | Alert, User Configurable | Auxiliary alerts | All types of alerts that maintenance and engineering may use to perform monitoring and diagnosis of system health |

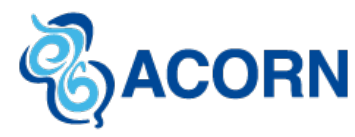

| | | | | |
|---|---|---|---|---|
| Green1 | R:0, G:92, B:0<br>Hex: #005C00 | Live value output | Current live output of a numerical value | |
| Blue1 | R:43, G:102, B:240<br>Hex: #2B66F0 | Setpoint input<br>Dynamic/Clickable element | Setpoint readout value<br>Elements that are dynamic or clickable | |
| Purple | R:165, G:181, B:248<br>Hex: #A5B5F8 | Accessed dynamic element<br>Historical data | Dynamic elements that have been previously clicked by user<br>Elements that convey previous settings<br>Historical data | Previously clicked elements should remain purple until user leaves page |
| Magenta | R:207, G:53, B:209<br>Hex: # CF35D1 | Bad data | Present indication of erroneous data from live value output | |
| White | R:255, G:255, B:255<br>Hex: #FFFFFF | Delineate objects, Text for Critical/High-Priority alarms | Provides perimeter around display elements<br>White space | |
| Very Dark Gray | R:29, G:29, B:29<br>Hex: #1D1D1D | General text (static)<br>Delineate objects | General text default<br>Provides perimeter around display elements | |

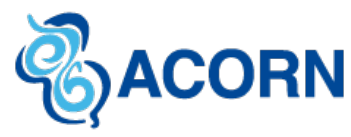

| | | | | |
|---|---|---|---|---|
| | | | | |
| Gray | R:150, G:150, B:150<br>Hex: #969696 | Static elements<br>Process lines<br>Disabled selection fields | Static display elements default<br>Disabled selection fields | Secondary color for process lines and object delineation |
| Light Gray | R:233, G:233, B:233<br>Hex: #E9E9E9 | Background | Provides primary background color, light theme | |

*Table 1. Primary Color-coding associations for accelerator control system (Light Theme)*

| Color | RBG Value & Hex Number | Primary Function | Description | Additional Information |
|---|---|---|---|---|
| Dark Theme Colors | | | | |
| Red | R:255, G:0, B:0<br>Hex: #FF0000 | Alarm, Danger Signal | Critical/High-Priority alarms | All types of faults that interfere with safety-critical operations |
| Orange1 | R:248, G:118, B:67<br>Hex: #F87643 | Alert, Caution Signal | Caution/Lower-Priority alerts | All types of faults that interfere with normal operations |
| Yellow | R:255, G:230, B:0<br>Hex: ##FFE600 | Alert, Acknowledge Signal | Acknowledge alerts | All types of alerts that operators must acknowledge |
| Cyan | R:10, G:250, B:242<br>Hex: #0AFAF2 | Alert, User Configurable | Auxiliary alerts | All types of alerts that maintenance and engineering may use to perform monitoring and diagnosis of system health |
| Green2 | R:82, G:215, B:90<br>Hex: #52D75A | Live value output | Current live output of a numerical value | |
| Blue2 | R:99, G:176, B:2150<br>Hex: #63B0FA | Setpoint input<br>Dynamic/Clickable element | Setpoint readout value<br>Elements that are dynamic or clickable | |
| Purple | R:165, G:181, B:248<br>Hex: #A5B5F8 | Accessed dynamic element<br>Historical data | Dynamic elements that have been previously clicked by user<br>Elements that convey previous settings<br>Historical data | Previously clicked elements should remain purple until user leaves page |
| Magenta | R:207, G:53, B:209<br>Hex: # CF35D1 | Bad data | Present indication of erroneous data from live value output | |

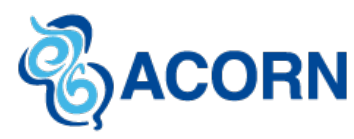

| | | | | |
|---|---|---|---|---|
| | | | | |
| White | R:255, G:255, B:255<br>Hex: #FFFFFF | Delineate objects | Provides perimeter around display elements<br>White space | |
| Light Gray | R:233, G:233, B:233<br>Hex: #E9E9E9 | General text (static)<br>Delineate objects | General text default<br>Provides perimeter around display elements; | |
| Gray | R:150, G:150, B:150<br>Hex: # E9E9E9 | Static elements<br>Process lines<br>Disabled selection fields | Static display elements default<br>Disabled selection fields | Secondary color for process lines and object delineation |
| Very Dark Gray | R:29, G:29, B:29<br>Hex: #1D1D1D | Background | Provides primary background color, dark theme | |

*Table 2. Primary Color-coding associations for accelerator control system (Dark Theme)*

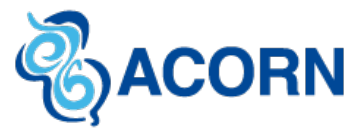

## 3.6 Typography

### 3.6.1 General

Typography is an important consideration for the design of digital systems. There are best practices in font selection, sizing, coloring, and emphasis options which drive some common expectations across the digital design space. The key aspect of typography considerations in digital design are readability and clarity of the text presented to users. A secondary concern can be the aesthetics of the overall design as well, however for the context of this document and designing systems for control room operations of the accelerator facility it is likely that usability will be the primary focus over aesthetic design considerations. With that said there are critical considerations and requirements identified around the text used in digital designs.

### 3.6.2 Pre-Requisites

As with other visual design tasks, the core pre-requisites are user requirements development and user testing to assess readability and clarity of the text. There are specific considerations for informational text in a paragraph format, labels for icons, headings, and overall informational hierarchy. It is important to leverage user expectations whenever possible to ensure a quick adoption of the system. The design guidance that follows are best practices, particularly for higher risk and control interfaces.

### 3.6.3 Detailed Design Guidance

**HFR-TYP-01: Typography should be limited to sans serif font options without user research demonstrating an alternative.**

*Additional Information:* Sans serif fonts are generally better fonts for readability and clarity on digital displays, which are priorities in a higher risk context such as control room operations.

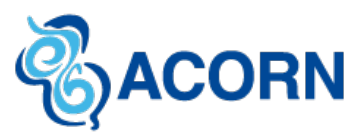

**HFR-TYP-02*:* All alphanumeric text (static and dynamic) should be no less than 9-point font (or 16 minutes of arc) for adequate legibility,** see Table 4.
*Additional Information:* A 9-point font yields of approximately 16 minutes of arc, a unit of angular measurement commonly used to assess visual acuity, when viewing at a seated workstation. Font sizes should be adjusted based on the intended viewing distance (Table 4); for example, text to be read at a distance of 10 feet (e.g., on an OVD) would require 39 point to achieve 16 minutes of arc on a 1920 x 1080 monitor.

| View Distance | Minutes of Arc | Font Height in Inches | Font Height in Pixels | Font Point Size |
|---|---|---|---|---|
| 24 inches | 16 | 0.11 | 11 | 9 |
| 48 inches | 16 | 0.22 | 21 | 16 |
| 62 inches | 16 | 0.29 | 27 | 21 |
| 84 inches | 16 | 0.39 | 37 | 28 |
| 120 inches | 16 | 0.56 | 52 | 39 |
| Based on 1920 x 1080 resolution. | | | | |

***Table 3. Font sizes by viewing distance***

**HFR-TYP-03: All alphanumeric text variations should be consistent throughout all interfaces.**
*Additional Information:* Once a design convention is established, it should be consistently implemented throughout all interfaces and text formatting is no exception. Deviating from the standard text format convention can cause interface confusion and disrupt user workflow. For example, use commonly accepted formatting (e.g., capitalize first letter of titles) throughout all interfaces.

**HFR-TYP-04: Labels for UI elements should use a convention that is consistent and intrinsically meaningful to users.**
*Additional Information:* Labels that are intrinsically meaningful to the user are an example of clear labels. For example, some of the main machines included in the accelerator complex are meaningfully named (e.g., the Booster machine “boosts” particles to increase their velocity). In addition to intrinsic meaning, clear labels also contain plain language that is concise yet descriptive.

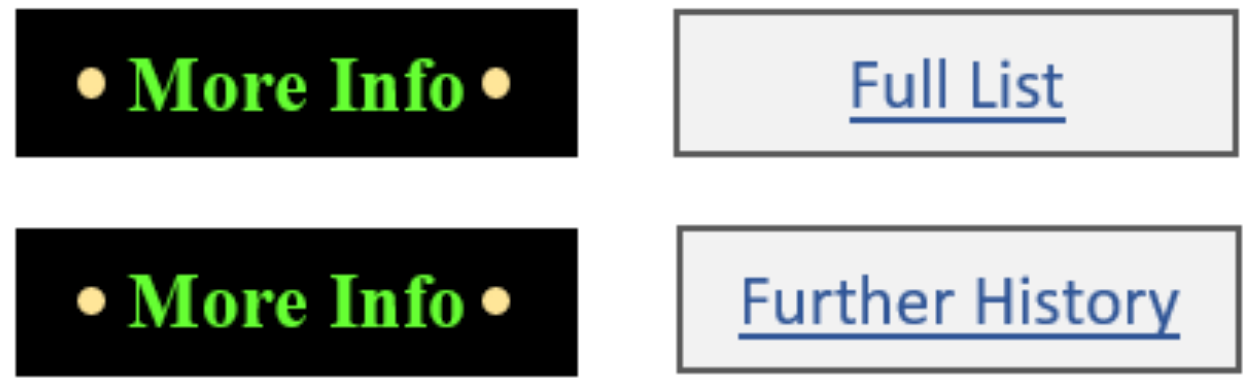

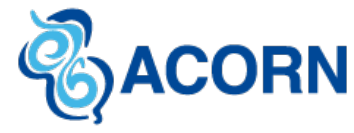

*Figure 6. Unspecific label from ACNET (left) versus descriptive label (right)*

**HFR-TYP-05: All visible labels should be oriented horizontally on display pages.**
*Additional Information:* Labels and information (i.e., words and symbols) should be oriented so that alphanumeric characters are read horizontally from left to right. If labels are static in nature and secondary (e.g., a label for a y-axis plot that is static), then the label may be oriented vertically.

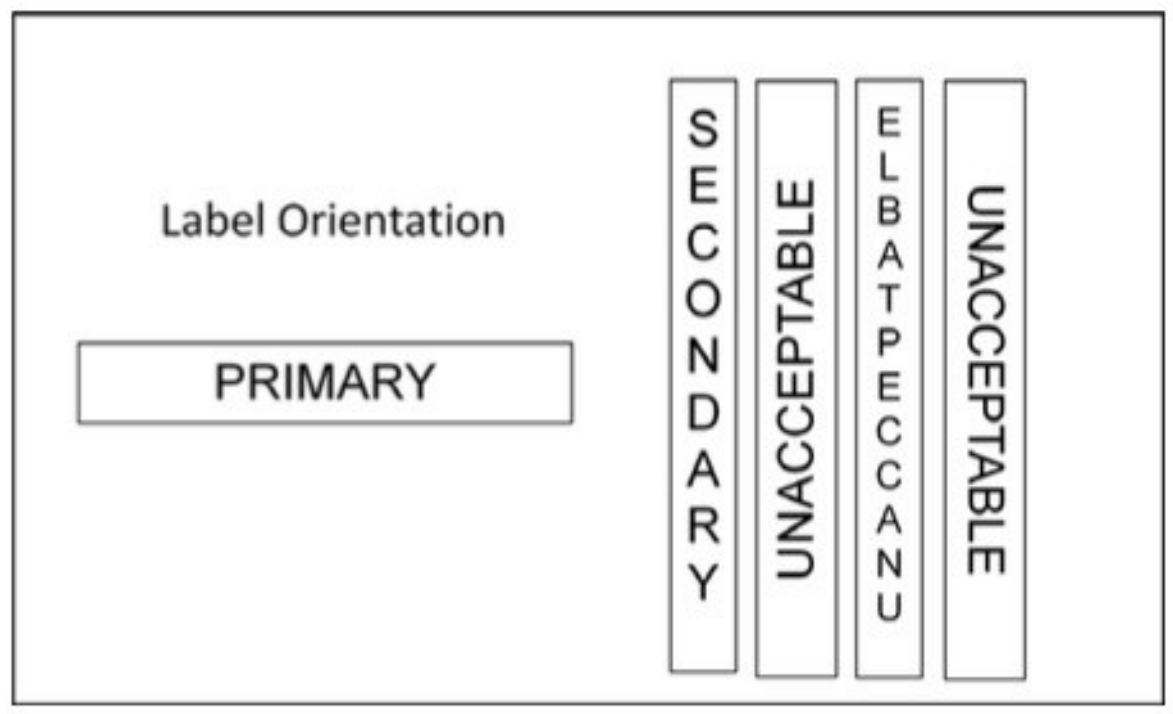

*Figure 7. Orientation of Labels*

**HFR-TYP-06: Labels that contain interaction potential should be visually distinct from information-only labels.**
*Additional Information:* Multiple types of information are included in user interfaces including data that is interactive (i.e., links) and data that is information only. The purpose of visually differentiating between interactive data and information-only data is to provide at-a-glance context which information users can interact with. An example of how to differentiate between interactive and information-only graphics is through visual elements that are intrinsically clickable. In the example included below the text on the left is a clickable link and the text on the right is information only, as shown in the top portion of Figure 7. The link text is not only a different color, but it is also underlined. Blue text that is underlined is a widespread design concept used for digital interfaces such as web pages. Therefore, humans have learned to associate blue, underlined text with links.

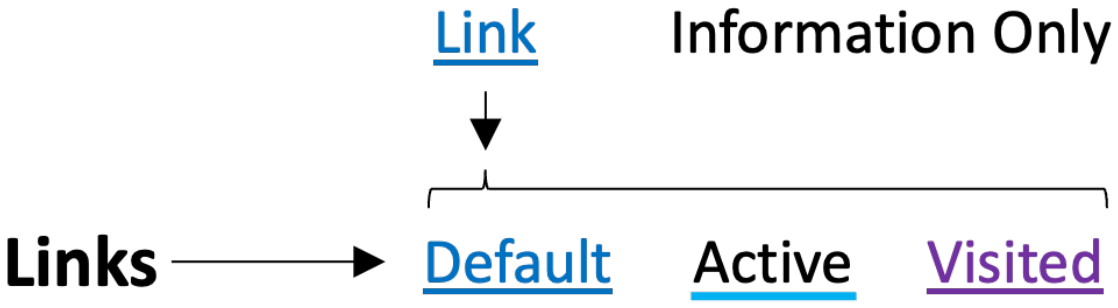

*Figure 8. Indication of interaction markers for links.*

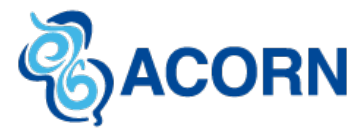

Note, as the design becomes more defined over time, a finalized list of acceptable font type(s) and additional typography specifications should be added here once decisions have been made.

### 3.6.4 Typography in an Accelerator Environment

Typography directly affects how quickly and accurately users can identify subsystem states, recognize warnings, or locate key controls. For example, when scanning multiple beamline plots or evaluating an alarm summary, poorly spaced or inconsistently styled text can slow down response times or cause misinterpretation. In contrast, a clean and consistent typographic system helps users locate relevant data at-a-glance and supports efficient interaction across a wide range of display types (i.e., from workstation displays to wall-mounted overview “comfort” displays).

Given that operators, engineers, and physicists may all interact with the same interface (albeit in different contexts), typography should support both rapid recognition for common tasks and detail precision for expert analysis. Currently there are a limited number of options for the typography in the ACNET system, and where variation is possible it is present as well. Within the context of accelerator control room facility operations there is a critical need for ensuring maximum usability and support for successful human performance whenever possible. Ensuring clear and readable text within the design is one area where the usability can best be realized.

### 3.6.5 References

Design Philosophy for Accelerator Control Rooms (ACORN-doc-700)

## 3.7 Iconography & Symbols

### 3.7.1 General

Icons and symbols are essential tools in the design of HSIs, especially in visually complex environments like accelerator control systems. When used appropriately, they can reduce cognitive load, enhance interface efficiency, and improve overall usability by providing quick visual cues that support operator decision-making. Icons and symbols help communicate meaning with minimal space and can reinforce consistency across different applications and workflows.

Although the distinction between icons and symbols can be subtle, in this context: icons are simplified, often universal, visual representations of actions or functions (e.g., a trash can for delete). Symbols may represent ideas or states and are sometimes more abstract (e.g., a radiation symbol to indicate hazardous conditions).

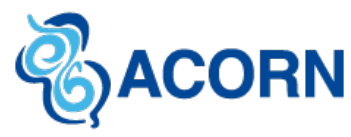

The strength of iconography and symbology lies in their ability to visually encode meaning in a way that supports fast recognition and intuitive use. However, this benefit only holds when icons and symbols are used consistently, sparingly, and with clear intent. Overuse, inconsistency, or unclear meaning can degrade usability and lead to confusion.

### 3.7.2 Pre-Requisites

Before implementing icons or symbols in any interface design, establish a standard icon set. Use a centralized and consistent library of icons, preferably one that leverages familiar or widely recognized designs, unless a unique representation is required. *Note, the UX/HF team must be consulted to validate whether the need for a custom icon is necessary or not.* Additionally, conducting early user research or testing (e.g., preference testing, usability walkthroughs) to confirm that chosen icons are understood by operators and other users. For less familiar icons, especially those not rooted in universal design patterns, it is helpful to include labels or tooltips to reduce ambiguity. Lastly, ensure that icons provide adequate contrast, are not color-dependent alone, and meet accessibility standards (e.g., WCAG 2.1).

### 3.7.3 Detailed Design Guidance

**HFR-ICON-01: Icons and symbols should have clear intentional meaning that is consistent throughout the control system.**

*Additional Information:* An icon or symbol should be recognizable and easy to decipher. Icons are used to simplify actions and functions, and the user should understand the expected behavior of an icon. The meaning and use of an icon should be consistent across the control system. If an icon is used for different actions or functions, it loses meaning and increases cognitive load on the user. Consistent use of an icon allows the user to act fast and quickly understand the expected meaning when they see an icon. Utilizing icons that are commonly known across domains allows a user to transfer that domain knowledge and apply it to the control system.

**HFR-ICON-02: Icons and symbols should be reserved for common functionality.**

*Additional Information:* Icon use should be reserved for common functionality that is repetitive across the control system. If too many icons are used on an interface, it can increase cognitive load on a user by requesting the user to learn and recall a lot of minimized communication. The oversaturation of icons can lead to the development of icons that have intended actions but no clarity to the user. This eliminates the significance of an icon by requiring the user to decipher less clear messaging.

**HFR-ICON-03: If color is used as a signifier, an icon or symbol can increase accessibility.**

*Additional Information:* Color should be paired with an icon or symbol to increase accessibility of consuming the information.

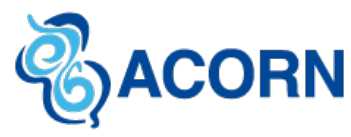

**HFR-ICON-04: Icons and symbols should have contextual content where applicable to communicate intention of use.**
*Additional Information:* Additional context can help communicate icon action or function meaning to the user. Icons that are less known should have additional text to complement the icon. The inclusion of text increases accessibility and conveys meaning to the user when there may be lack of clarity in a less known icon. If an icon is used consistently across the control system and has clear meaning (e.g., a printer for printing), then text can be added via hover for desktop interfaces.

Note, as the design continues to mature, a finalized list of acceptable icons should be added here once iconography decisions are confirmed and standardized across the system.

### 3.7.4 Iconography and Symbols in an Accelerator Environment

Interface iconography must support both rapid use and intuitive understanding for users with varying levels of familiarity (i.e., from full time operators to less frequent users such as physicists). For main control room operators icons can serve as fast-access triggers for common functions like system refresh, alarm acknowledgment, or navigation to subsystems.

For example, an icon depicting a magnet could be used throughout the control system wherever magnet controls or diagnostics are accessed, helping reinforce task location across displays. Similarly, warning symbols tied to radiation, cryogenics, or high-voltage systems must be used consistently to support quick hazard recognition.

Symbols and icons also help minimize interface clutter. For operators monitoring multiple components, small graphical indicators (such as health badges or color-coded system status icons) reduce screen space usage while conveying meaningful summaries. However, these icons should be reinforced with hover-over text or drill-down paths to ensure clarity, particularly for novice users.

Because the control system often serves both generalist operators and system-specific experts, care must be taken to reserve icons for shared, common functions. Avoid developing custom or overly specialized icons that dilute meaning or require memorization. Visual language should always reflect a balance between expressiveness, clarity, and task relevance.

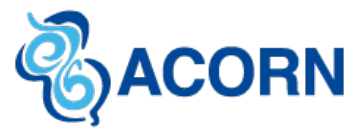

## 3.8 Information Visualization (Dynamic Display)

### 3.8.1 General

Information visualization is part art and science, the art of graphic design and the science of perception and human psychology. Fundamentally, any element of information visualization (e.g., graphs, charts) must accurately and clearly convey the information in the data. In this section, the reference to graphics or graphic elements is explicitly focused on more dynamic, data-driven, diagrams, mimics, drawings, or other representations of systems and components. Particularly in complex environments the use of visual diagrams can enable a much quicker decision making experience as humans process visual information faster than any other form of sensory data. This section will not dive deep into the specific details of human visual perception, though there are resources for those topics if needed, specifically three critical textbooks by Healy (Healy, 2018), Wilke (Wilke, 2019), and Ware (Ware, 2020).

### 3.8.2 Pre-Requisites

Critical to the design of information visualizations is the data structure and understanding how the data is intended to work together to tell its unique story to the user. Further, designers should have a clear understanding of the users' needs related to the data and how the visualization can enhance or support their performance for the task at hand. Having a clear understanding of the task, data, and decision making the developers and designers can develop robust visualizations that support the user's performance.

### 3.8.3 Detailed Design Guidance

**HFR-IV-01: Clearly distinguish contextual information from live plot data.**

*Additional Information:* Presenting operational context within plots (see figure 8) supports rapid recognition of system status. As adding operational context increases the information density of a plot, the design should make distinguishing the operational context from the live data easy to recognize. ACNET often displays new data over the old to help show change over time. However, there is little by way of design to distinguish the entry time of each new reading. Distinguishing the age of the reading in the plot may lead to some emergent features and understanding by users.

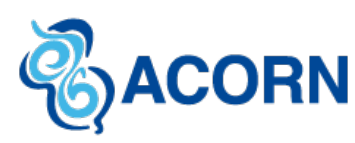

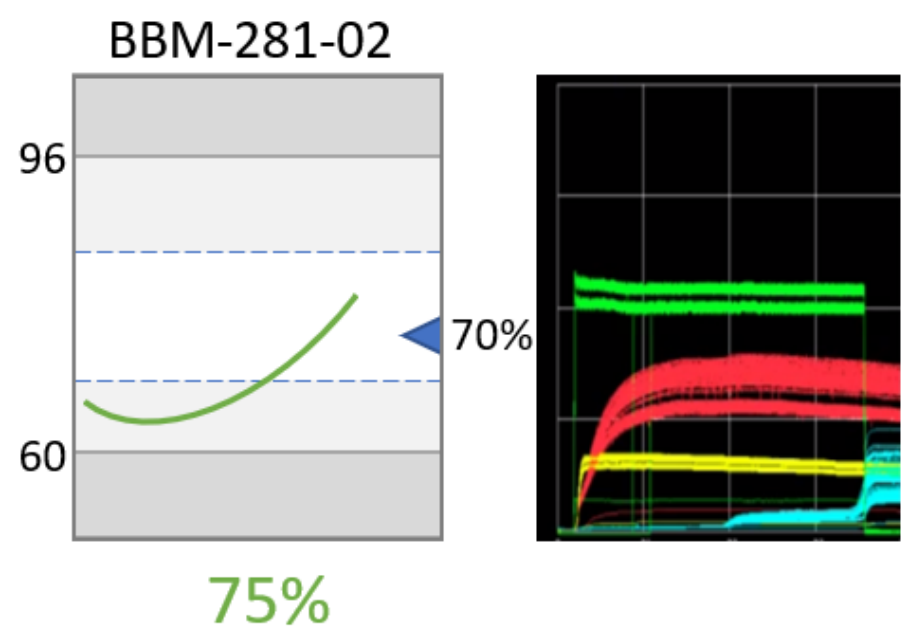

*Figure 9. The left figure shows live data (green, salient) against its operational context (light gray, neutral). The right figure shows data collected over many retrievals, but not indication of the timeline of retrieval*

**HFR-IV-02: Present only necessary data on a plot to improve user time to complete task or understand system status.**
*Additional Information:* Time to complete a task increases with task complexity. The more complex a task the more important it is to determine exactly what information is required. Tendencies to provide extra information can only increase the user need to process and interpret what is important and necessary from the extra information available. Developing a plot to fit a task will improve user's ability to perform the task. The only data displayed on the visualization should be the data required to support the decision making and nothing additional.

**HFR-IV-03: Visualization elements should include labels for its title, axes, parameters, colors, shapes, and engineering units.**

**HFR-IV-04: Visualization elements should support tooltips and hover pop up functionalities expected by users.**
*Additional Information:* A common aspect of digital design is the tooltip, basically a pop up or providing information when a user hovers over an object. In visualization elements this functionality is a powerful way to give additional precision or information to the users in a way that is optional and does not clutter up the display.

**HFR-IV-05: Visualization elements should include a digital readout of the parameter(s) being represented when precise reading is required of the user.**

**HFR-IV-06: Where multiple data points are presented on a single visualization element, each parameter should be coded using color or line type for differentiation.**
*Additional Information:* Consistent with HFR-IV-03, the use of colors, line types, shapes, etc., for differentiation are valuable ways of distinguishing data. However, their use necessitates the inclusion of legends or other descriptive means of identifying the data for the user.

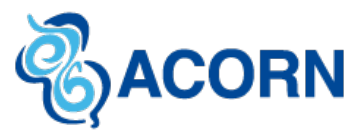

**HFR-IV-07: When multiple parameters are displayed on a single visualization element, the axes of the element should be on the same scale, or if differential scales are necessary, the chart should be broken up or otherwise provide a visual identifier to users that the scales are different and how the scales should relate.**
*Additional Information:* It may be useful or necessary to display different parameters alongside each other which have varying scales, sampling times, or parameter variance. It should be clear to the user how these data points relate and how their decision making should change as a result of the difference in the parameter nature.

**HFR-IV-08: All components, line points, and termination points presented on a mimic display should be labeled.**

**HFR-IV-09: If a visualization element depicts a physical process, such as a diagram of a system, then any physical processes of flow or movement should be depicted by distinctive arrowheads that demonstrate the flow path.**
*Additional Information:* Adapting a physical process diagram into a visualization element with data overlays are a powerful means of supporting operator task performance. However, the design and development teams should be conscientious of the system and process they are designing to ensure that the reality of the process is still captured, such as flow of coolant through a pipe.

Note, as the design progresses through prototyping and iterative refinement, additional requirements concerning information visualization such as finalized interface specifications, widget behavior, and data display formatting should be added here once specific design decisions are confirmed. For example, if a real-time loss monitor is selected for beamline diagnostics, detailed specifications for its plotting intervals, threshold indicators, and color-coding logic should be documented.

### 3.8.4 Information Visualization in an Accelerator Environment

Information visualization is a critical and fundamental component of the accelerator environment. Presenting the data collected and monitored from accelerator experiments is key to success in those efforts. Beyond experimental value, information visualization can play an important role in the operations of a system as complex as a particle accelerator facility. Often data is collected at a frequency or scale that is impossible to process and understand by human operators. By visualizing this information, we allow the operations staff to have a broader understanding of the system and its state more clearly than just representing the data in a raw format.

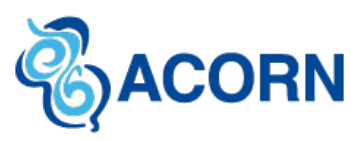

## 3.9 Interaction Design

### 3.9.1 General

Interaction Design (IxD) focuses on the dynamic behavior of systems and how users engage with that behavior over time (Cooper et al., 2014). While visual design addresses what users see and information architecture addresses how content is organized, interaction design is concerned with how systems respond to user input, how states change during task completion, and how feedback guides users through workflows. This "conversation" between the user and the system is what distinguishes IxD while building upon principles from other design domains.

Like other human factors and UX disciplines, IxD benefits from systematic user research and evidence-based design methods. This section will not enumerate all the many principles of IxD but rather the specific user research tasks necessary and some of the more focused items of interest for accelerator operations. For a more in-depth treatment of IxD principles see (Cooper, et al., 2014).

### 3.9.2 Pre-Requisites

Fundamental inputs to these design areas are user requirements documentation, identification of critical interactions for successful task performance, and standard user feedback testing at a minimum. As IxD can get extremely complex quickly it is critical to engage with your human factors or user experience advisor to ensure you are capturing the critical user feedback on interactions that are necessary for your system.

### 3.9.3 Detailed Design Guidance

**HFR-IxD-01: Interactions should be well understood and defined based on the context of the interaction, e.g., what the user is trying to accomplish and how the system will work with the user to achieve that goal.**

*Additional Information:* It can be valuable to consider interactions with a system as a form of communication between the user and system. This can help support a design that enables a coherent functional dialogue between users and their interfaces.

**HFR-IxD-02: The primary interaction modality should be cursor based, and interactions should consider this in their design.**

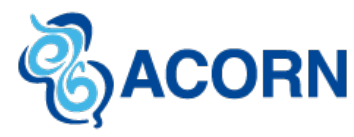

*Additional Information:* Interacting with elements should be directed and executed by either mouse cursor interaction or keystrokes.

**HFR-IxD-03: The system should provide indication of all display elements that include interaction functionality.**
*Additional Information:* There are current instances of "invisible" buttons on the screen that are used during operator tasks but provide zero indication of their presence or function. The new ACORN interface should clearly distinguish the function and presence of all control options available on screen. Further, development team members should take care to ensure the elements deployed into the system align with common, "best" practices for digital design and any deviation from commonplace interaction modalities must be supported by robust user research.

**HFR-IxD-04: Data entry actions should be accompanied by a verification step.**
*Additional Information:* The system should process user's entry only after a subsequent verification or confirmation is executed.

**HFR-IxD-05: Visual feedback should be provided across all user interactions with the system.**
*Additional Information:* Every input by a user should produce a consistent, perceptible visual response output from the computer.

**HFR-IxD-06: Visual feedback should be applied consistently across the control system.**
*Additional Information:* The control system interface should support a variety of feedback interactions to inform users of autonomous system changes and user-initiated changes. Feedback messages should be consistent in style and format across the interface. Similar to color or typography, a common sensory feedback framework ensures a more successful and consistent user experience which increases successful performance likelihood.

**HFR-IxD-07: System latency should be 0.2 seconds or less for real-time responses.**
*Additional Information:* Acceptable response times for key presses and error feedback is 0.2 seconds (Department of Defense, 2019; NRC, 2020; ANSI, 2007; ISO, 1992). If the instrumentation and control architecture does not support such a small response time, then users should be notified to expect a longer wait to prevent doubling up user inputs.

**HFR-IxD-08: Blinking/flashing should be used only for alerting the user to events that require immediate attention.**
*Additional Information:* Flash coding should be employed to call the user's attention to mission-critical events only.

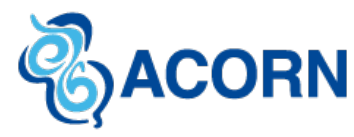

**HFR-IxD-09: No more than two blink/flash rates should be used.**
*Additional Information:* Flash rates should be limited to two types. They should be no more than 5 Hz and no less than 0.8 Hz; the difference should be greater than 2 Hz. The flash that is of higher importance should be of greater frequency.

Note, as the design becomes more defined through prototyping and iterative development, additional requirements concerning interaction design such as the behavior and visibility of interactive components should be added here once specific decisions are finalized.

### 3.9.4 Interaction Design in an Accelerator Environment

As particle accelerator environments are a unique and specific environment there may be unique or specific interactions that are found by the design and development teams. This section provides some general information on designing interactions but only through testing and evaluation will their value to an accelerator control room bear out. Careful consideration should be paid to the specific tasks that operators will be asked to execute, perhaps via task analysis methods, to understand what constitutes those actions and what feedback would be expected or necessary for success. There can be many varieties of interactions that deliver the same information. By developing a deep understanding of the specific requirements of the context will help guide design teams to interactions that are most effective, and those assumptions are validated through user evaluation.

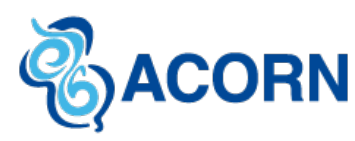

## 3.10 Controls

### 3.10.1 General

In the context of this style guide, a control is defined as an interaction that commands, directs, or regulates the behavior of physical devices on the system. The guidance in this section applies to aspects of the system where operator input directs action on the accelerator system. When discussing these topics, it is important to note that a control is a *type* of interaction, interaction is the parent concept. However, in an operational environment a control is so fundamental that it merits its own section and guidance.

Controls can be input fields where operators change values or set targets for the system to maintain (see Figure 9), or they can be software interfaces that mimic the look and feel of physical control devices (see Figure 10). The guidance applies to both types. However, guidance in other sections may have impacts on these topics as well. Prior, in the Interaction Design section user input requirements are covered, so this section will focus more specifically on control actions where the user is issuing a command to the system.

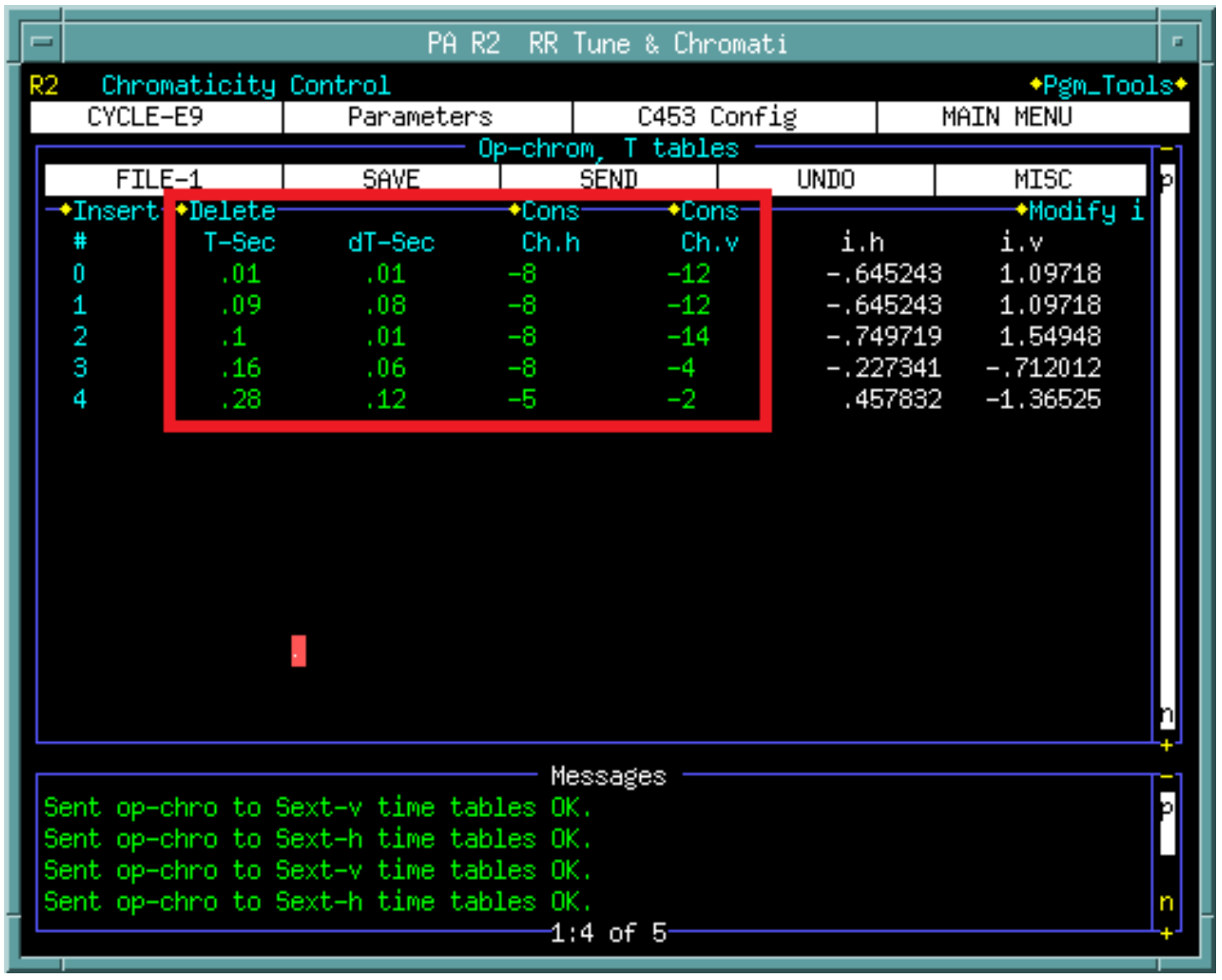

*Figure 10. Example of control input fields marked with the red box*

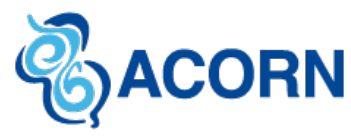

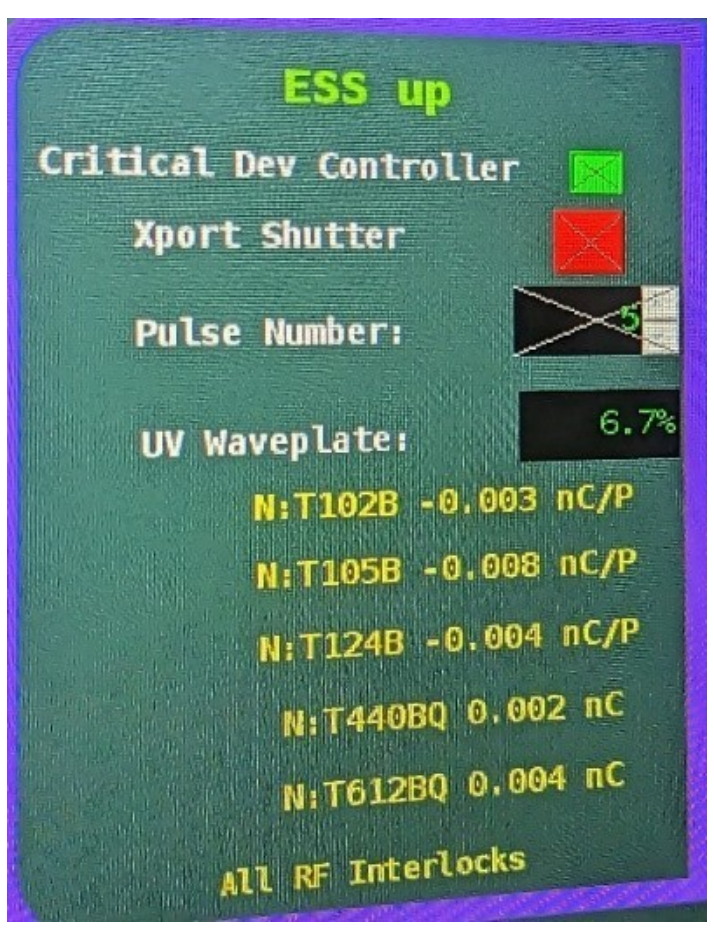

*Figure 11. Example of control faceplate*

When controls have multiple options or modes (e.g., automatic, or manual) make it clear what mode the controls are currently in and how that affects the functionality of the control (i.e., what inputs are available in a particular mode). For example, in the left side of Figure 11, Mode A is selected and there are only two control parameters that an operator can manipulate. The empty input field and the cursor indicating that no input has been selected. In the right-side example, mode B is selected, which has two input options. The second option is not shown when mode A is selected. The visual design of controls should be consistent with the overall design. Especially with control faceplates it is tempting to mimic physical devices by adding 3D effects or texture to the design via a design philosophy known as skeuomorphism. This can add unnecessary visual clutter to the design and should be avoided unless there is a compelling user need or error prevention motivation to utilize a more skeuomorphic design (NRC, 2020).

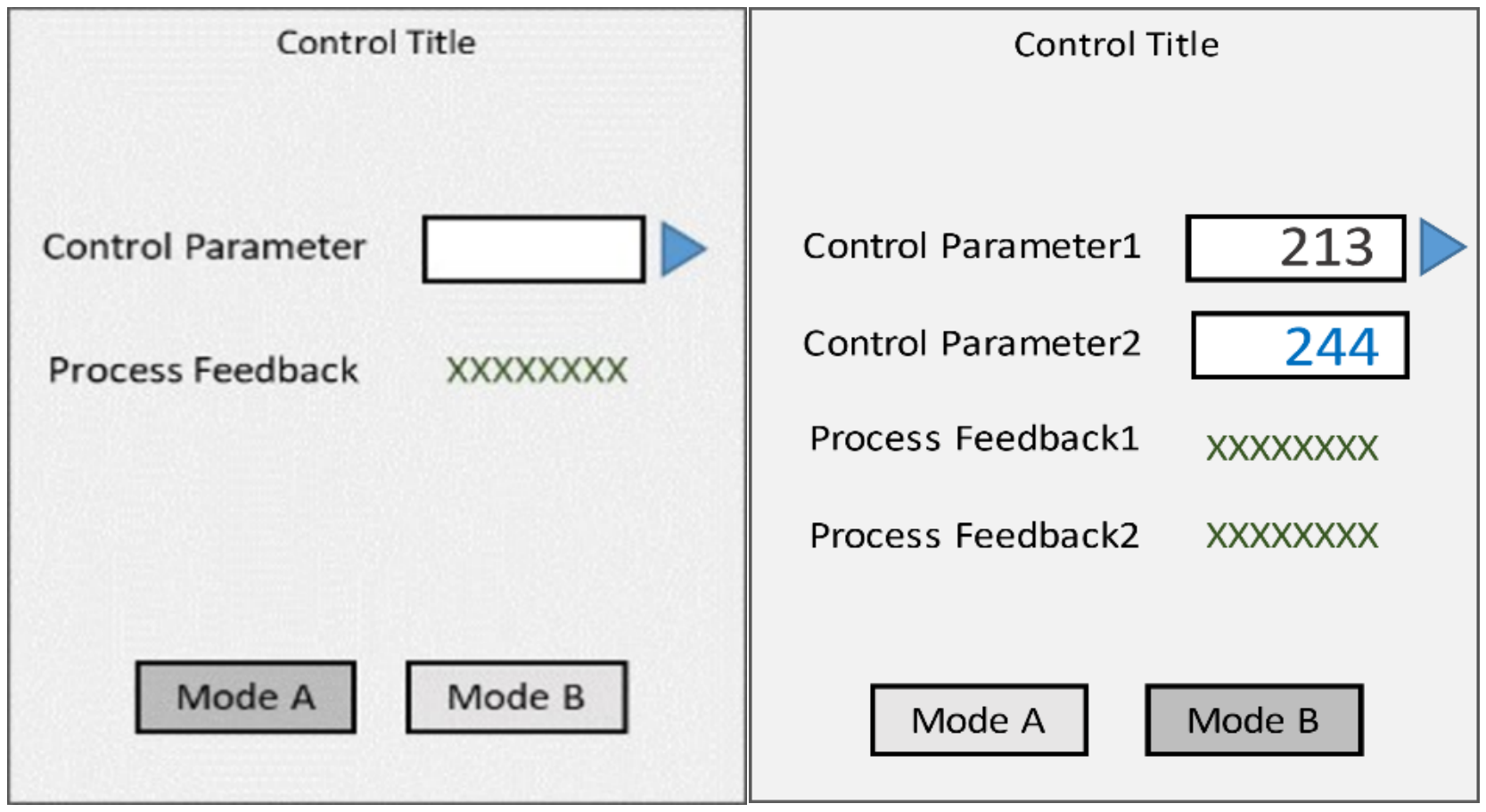

*Figure 12. Example of faceplate or soft control design*

### 3.10.2 Pre-Requisites

When defining and designing controls it is important to have a clear concept for what the control is intending to do and what the context of that action is expected to look like. Ultimately, many control requirements are derived from system, engineering, or technical requirements but how those controls are implemented has a large impact on the success of those actions. The guidance in this section will capture critical components of control design, though the Interaction Design section may have useful insights or requirements as well, since controls are a type of interaction. Particularly as safety impacts rise, but even in more innocuous instances, the design and development of controls must be thoroughly evaluated through user research testing to ensure success is more likely and errors are minimized.

### 3.10.3 Detailed Design Guidance

**HFR-CTRL-01: The system should provide indication of all display elements that include control functionality.**
*Additional Information:* As stated in HFR-IxD-03, it is critical that the design identifies the control elements and communicates this clearly to the user.

**HFR-CTRL-02: All control options for a specific soft controller (i.e., faceplate) should be made accessible by a single click.**
*Additional Information:* Control buttons (generally clicks) or inputs (generally alphanumeric entries) are those that operators use to adjust a component or system status. This includes but is not limited to adjusting setpoints, clearing alarms, navigating menus, and tuning machines. All control designs should have the operational context for controlling, such as current setpoints, values, and predetermined thresholds, viewable while preparing to act.

**HFR-CTRL-03: All frequency performed control actions should be accessible from a soft control faceplate without any additional administrative action.**
*Additional Information:* All available or routinely used functions for control of equipment or components should be continuously visible or a maximum of one click away.

**HFR-CTRL-04: Soft control options should be suitable for characteristics of the task performed.**
Additional Information: See Table below

| Control Type | Example | Suitable Situations |
|---|---|---|
| Discrete Adjustment Controls | Manual Auto | • When control options are limited to a few. |
| Soft Sliders | | • The range of possible values is medium- high.<br>• When changes are incremental and large (gross changes are needed). |
| Arrow Buttons | | • When precise value is needed.<br>• The range of possible values is medium- high.<br>• When changes are incremental and small. |
| Input Field | 78.8 % | • When precise value is needed.<br>• The range of possible values is high.<br>• When changes are variable (rather than incremental). |

*Table 4. Characteristics of various soft control types*

**HFR-CTRL-05: Soft controls should be visually distinguishable from other buttons like navigation buttons.**

**HFR-CTRL-06: Control actions that have severe consequences must be accompanied by a verification step.**
*Additional Information:* As in HFR-IxD-04, communicating the need for confirmation to users is an important feedback and verification step. This should be applied for control actions that carry potential to cause harm to humans, the environment, or system equipment. Severe consequences that could result from an incorrect control action can be avoided by interrupting the action sequence with a verification step and mitigates that risk.

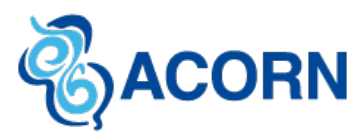

**HFR-CTRL-07: The system should provide confirmations for control actions that are safety important or have potential to disrupt normal operation.**
*Additional Information:* Feedback is most crucial when a user is confirming actions that are safety related or for confirming actions that have the potential to disrupt normal operation (e.g., device shut down).

**HFR-CTRL-08: The system communicates when multiple users are controlling or attempting to control the same equipment simultaneously.**
*Additional Information:* When asked which mistakes are most common, an operator stated that "two operators conducting the same task" (e.g., knobbing a machine) occurs frequently. Typically, it doesn't cause severe issues but sometimes does cause delays in operations.

**HFR-CTRL-09: All control actions should include feedback in the design for operators to understand if a control has been actuated.**
*Additional Information:* Including feedback that is sensory based in some form is critical to enable system-user communication.

Note, as design decisions regarding controls (e.g., buttons) become more defined, a list of acceptable control types and their corresponding design specifications should be added here.

### 3.10.4 Controls in an Accelerator Environment

An accelerator environment is an incredibly complex system that has wide ranging effects, risks, and needs. Controls of components and systems are a critical piece of ensuring the success of the entire operation. Designing controls in these systems should closely follow best practices design guidance and robust user testing to ensure that the controls: work as they should, minimize error, maximize success, and handle off-normal circumstances (e.g., recovery actions).

## 3.11 Alarm Systems

### 3.11.1 General

Alarm systems are a critical component of any industrial control system. Alarm systems serve as automated monitoring systems that can alert operators through various means when a deviation, abnormal state, or threshold may be reached. These systems are critical to industrial systems because the sheer volume of parameters, components, systems, and states are too numerous for individual human operators to monitor and maintain as a whole. In this sense, alarm systems are one of the most foundational aspects of system supports for human performance that is present in any system. It is important to note, however, that alarm systems can also create more problems than they purport to solve if designed incorrectly. Alarm flooding, false positives, unclear alarm messages, poor salience and more can make the overall notification process ineffective or can detract outright from the safe and effective performance of the system. It also bears noting that this section is distinct from the section on Feedback. Alarms are a form of feedback, but due to their critical and unique nature they are identified and described separately. Though many of the points in the Feedback section will also be informative in the design of an alarm system within the accelerator environment.

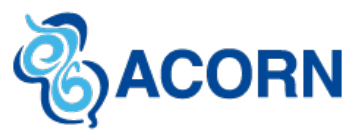

### 3.11.2 Pre-Requisites

Prior to designing an alarm system there is a necessary interplay within the broader operations context and team. System SMEs will provide key component or system parameters that should trigger alarms from a technical basis or safety respect, training coordinators will need to ensure a robust training program around responding to alarms, and then HF/UX professionals will need to evaluate the alarms for usability. Every characteristic of the alarm from sensory feedback aspects to time required to correct issues and clarity of alarm messages must be analyzed and tested with real users. This process can often occur in parallel to the development of other systems. Initially, a list of relevant alarms and their characteristics must be evaluated by HF/UX professionals to begin developing a robust and thorough verification and validation testing framework to ensure that alarms perform effectively and do not pose additional challenges to operations staff.

### 3.11.3 Detailed Design Guidance

**HFR-ALRM-01: Alarms should only be used for off-normal conditions that require timely action by the operator.**

*Additional Information:* The purpose of an alarm system is to direct the user's attention towards plant conditions requiring timely assessment or action. Each alarm should alert, inform, and guide the user. Every alarm presented to the user should be useful and relevant. Every alarm should have a defined response to which the user has adequate time to perform.

**HFR-ALRM-02: The system should provide the user with notifications of any conditions, internal or external, that may impact the accelerator's performance.**

*Additional Information:* A somewhat surprising insight shared by users was how a variety of elements can influence accelerator devices. For example, the weather affects the way some magnets bend the beam. There isn't a direct measurement for this occurrence aside from posting the daily weather report in the control room which means operators must infer on their own why expected beam outputs are slightly off.

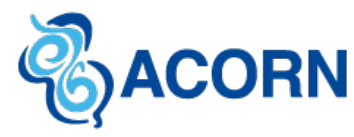

**HFR-ALRM-03: The system should provide an indication that the display is reading data from the control system (i.e., system heartbeat).**
*Additional Information:* The user should know whether data is live or out-of-date. If data is not coming in from the system, this should be made clear to the user.

**HFR-ALRM-04: Live simulation testing with actual operators must be completed prior to deployment of an alarm system to ensure that alarms are effective without becoming an obstacle to safety.**
*Additional Information:* As stated prior, poorly designed or implemented alarm schemas can create more problems than they solve. The way to identify these areas for improvement is with live user tests in upset conditions.

**HFR-ALRM-05: Alarms should provide immediate information regarding data streams that have caused the alarm trigger.**
*Additional Information:* Beyond an identification of the specific system or component that is experiencing off-normal conditions, the alarm system should identify precise data streams that are feeding the indications whenever possible.

**HFR-ALRM-06: Alarm notification messages must be written in plain language first with accompanying technical data such as error codes.**
*Additional Information:* Often alarm systems are developed using 'back end' error identifiers as the error message, however this is ineffective as notifications to the operator. Each alarm instantiation should include the relevant technical error code or data as needed, but the primary feedback to the user must be plain language text that describes the issue succinctly.

**HFR-ALRM-07: Any information regarding time available must be presented in the alarm message in a live, and real-time manner**.
*Additional Information:* Alarms are often describing live temporal states or upsets with practical consequences, e.g., "close this valve in X minutes, or this pump will power down." This creates real time pressure on operators' actions and therefore operators must be made aware of the timelines of any consequences whenever possible.

**HFR-ALRM-08: Any time constraints related to an off-normal condition and their consequences must be tested with live operators in a simulated event framework.**
*Additional Information:* If an alarm has a relevant time constraint, such as described in HFR-ALRM-08, then operators must be evaluated to ensure that the time available is reliably greater than the time required to complete any task. An alarm with a time constraint that is impossible to fulfill is not effective.

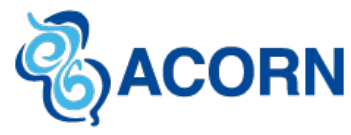

Note, as the design becomes more defined through prototyping and iterative development, additional requirements concerning alarm systems such as alarm hierarchy, prioritization logic, and visual presentation of alarm states should be added here once specific decisions are finalized.

### 3.11.4 Alarm Systems in an Accelerator Environment

In accelerator environments, control room personnel (e.g., operators and crew chiefs) rely on alarm systems as their first indication of system distress. During abnormal events such as beam loss, RF trips, or vacuum failures, a well-functioning alarm design enables operators to triage issues quickly, navigate to the correct subsystem, and take appropriate corrective action, often under time constraints.

Alarms may also escalate to expert users such as system engineers or machine physicists, especially if the root cause lies outside of operational purview of responsibility. Therefore, the alarm design must balance the needs of operator responsiveness with expert diagnostic follow-up ensuring that handovers are smooth and that the system state is reliably communicated between crew members.

## 3.12 Feedback

### 3.12.1 General

Feedback is a core component of effective human-system interaction, particularly in control environments where operator decisions impact complex, real-time systems. Feedback refers to any response the system gives to inform users about the status of their input, the state of the system, or changes within the system. It is essential to ensuring that users remain informed, confident, and in control of their actions.

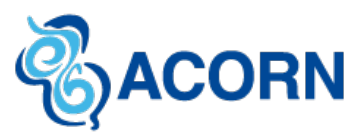

In accelerator control systems, where decisions may have safety, equipment, or mission-critical consequences, feedback must be both timely and contextually appropriate. Feedback can take many forms (e.g., visual, auditory, or haptic) and should be matched to the nature and urgency of the task. Whether a user is issuing a command, navigating a system, or simply monitoring a task, the interface must clearly and consistently communicate what is happening and what is expected next.

Effective feedback helps maintain situational awareness, reduces uncertainty, supports error recovery, and reinforces trust in the system. Just as with navigation, designers should seek to align feedback methods with user expectations and existing mental models while validating decisions through usability testing.

### 3.12.2 Pre-Requisites

While some decisions regarding feedback design are straight forward and will generalize across applications, some are unique and complex and will require deeper understanding. Therefore, before implementing feedback mechanisms, teams must first understand the types of interactions users will have with the system, the criticality of those interactions, and the cognitive demands placed on operators. A recommended method to accomplish this is task analysis and mapping system responses to user actions to determine where feedback is necessary and what type of feedback is most appropriate.

Feedback design should always be informed by user research. Designers should consider factors such as visibility, urgency, interruptiveness, and the user's current cognitive state. For accelerator applications, this will be best accomplished through prototyping and simulation to evaluate how feedback performs in both routine and high-stress scenarios.

### 3.12.3 Detailed Design Guidance

**HFR-FDBK-01: All user actions shall result in immediate and visible system feedback.**
*Additional Information:* Whether clicking a button, toggling a switch, or issuing a command, users should receive immediate visual acknowledgment (e.g., change in button state, 'loading' icon, confirmation message) that the system has received and is processing the input.

**HFR-FDBK-02: System state changes must be reflected in the interface without requiring user polling.**
*Additional Information:* Displays should auto-update to reflect changes in real-time. Users should not have to refresh, navigate away, or perform manual updates to stay informed of system state.

**HFR-FDBK-03: Critical or safety-related feedback must be persistent and prominently displayed.**

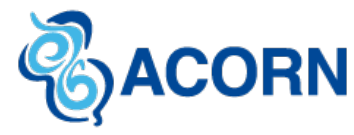

*Additional Information:* Alarms, interlock trips, or overrides must remain visible until explicitly acknowledged. Visual prominence (e.g., color, size, location) should be used to ensure visibility, supported by sound if appropriate.

**HFR-FDBK-04: Feedback should indicate both the result and status of system commands.**
*Additional Information:* When a user takes an action, such as initiating a procedure or applying a setpoint, the system must display whether the action is pending, successful, failed, or incomplete (with reason provided when applicable).

**HFR-FDBK-05: Feedback should align with user mental models and the importance of the action.**
*Additional Information:* Routine actions may require subtle feedback (e.g., animation or color change), while high-impact actions (e.g., beam stop) should require confirmation and deliver prominent response feedback. The level of feedback must match the consequence and user expectations.

**HFR-FDBK-06: Feedback mechanisms must remain consistent across the system.**
*Additional Information:* Similar types of actions should produce the same types of feedback. For example, all "save" actions should trigger a confirmation toast (i.e, a small nonmodal popup that disappears after a few seconds) or status message, and all errors should be styled and positioned consistently (Flaherty, 2024).

Note, as the design continues to progress through prototyping and iterative refinement, additional requirements concerning system feedback such as default response behaviors, visual or auditory confirmation cues, and user-triggered interaction responses should be added here once specific design choices are finalized.

### 3.12.4 Feedback in an Accelerator Environment

In accelerator control systems, feedback design must support operators in both routine monitoring and incident response. Operators must be able to trust that their commands have been received and that any system deviations will be clearly and immediately communicated. Feedback should be closely tied to the accelerator’s mode of operation (e.g., standby, injection, acceleration, fault), and feedback elements should reflect these contexts with appropriate urgency and clarity.

Because of the complexity and long lifecycle of accelerator systems, the feedback design must also accommodate gradual enhancement. Some specific elements such as how setpoint confirmation is displayed or how system delays are visualized may evolve through prototyping and operator testing. As such, feedback guidance in this section will expand in future revisions, particularly for subsystem-specific behaviors and failure modes.

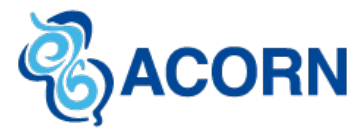

Ultimately, the goal of feedback design in accelerator HSIs is to reinforce the operator's situational awareness, reduce cognitive strain, and ensure that actions lead to predictable and transparent system responses.

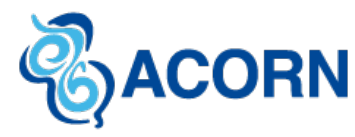

# PART 3: NEXT STEPS & FUTURE REVISIONS

This section defines the governance, maintenance, and evolution process of the accelerator HSI style guide. As the control system environment and user needs continue to evolve, this guide is intended to remain a living document and adapt over time through structured feedback, formal review, and collaboration across teams. This section establishes the foundation for how updates will occur, who will manage them, and how users can engage with the process.

Ensuring the style guide remains relevant and effective requires active participation from developers, designers, operators, and system owners. By embedding mechanisms for version control, continuous feedback, and documentation of change, this section helps protect the long-term usability, coherence, and applicability of the design standards outlined in Parts 1 and 2.

# 4.0 NEXT STEPS AND FUTURE REVISIONS

## 4.1 Governance and Ownership

The HSI style guide is maintained under the stewardship of the UX/Human Factors team, who are responsible for facilitating its revision process, collecting feedback, and ensuring design integrity across control system applications. While the UX/HF team acts as the primary point of contact, collaborative input is expected from interface developers, system engineers, operators, and domain experts.
Key responsibilities of the governance process include reviewing proposed updates or additions to the style guide, maintaining a change log/version history, coordinating usability testing to validate included guidance, and ensuring consistency across documentation, prototypes, and deployed systems.

## 4.2 Feedback & Iteration Process

To ensure the guide remains user-centered and operationally grounded, feedback will be collected continuously through structured and ad hoc channels (e.g., operator feedback during user testing). Additionally, design and code review comments from developers and contributors will also be sought after and implemented. User feedback from prototype testing (e.g., outcomes and lessons learned) as well as user interviews and cognitive walkthroughs will be continuously captured.  If feedback would like to be shared beyond these methods, suggestions for changes, clarifications, or additions may be submitted via a designated change request workflow, managed by the UX/HF team. The process for submission, review, and implementation will be documented separately and made accessible to all contributors.

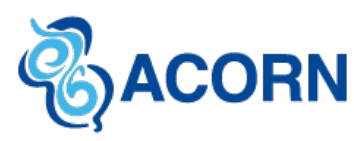

## 4.3 Revision Cadence and Versioning

This style guide will be reviewed and updated on an annual basis at minimum, with interim updates permitted as needed based on project timelines or critical usability findings. Each new version will be published with:

- A version number (e.g., v1.1, v2.0)
- A summary of changes
- A rationale for major updates
- Clear annotations within the document to identify newly added or revised guidance

Sections still under development or awaiting further validation (such as resolution targets, device-specific adaptations, or emerging interaction patterns) will be marked clearly. These areas may be expanded through future revisions informed by user testing and evolving project requirements.

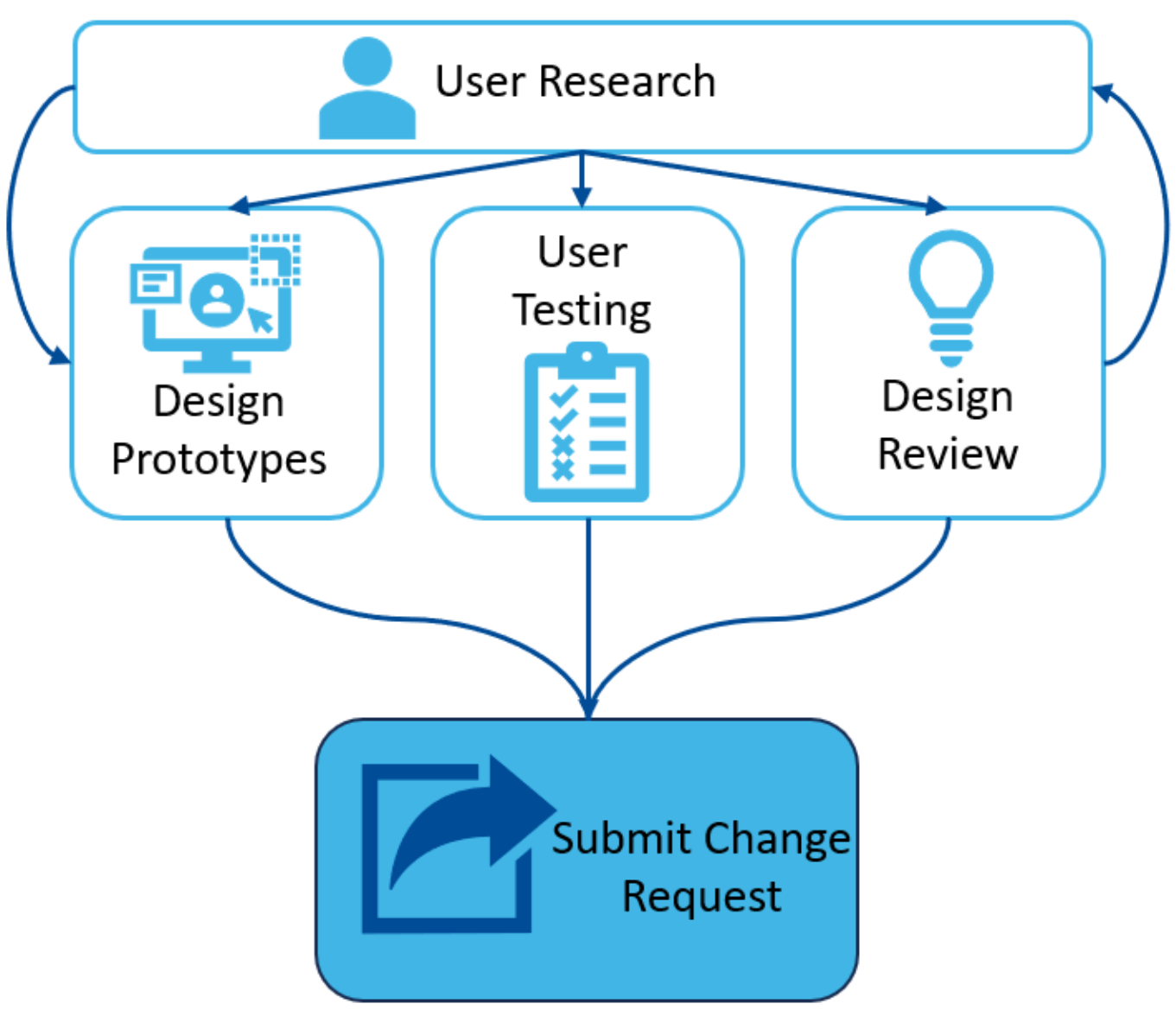

*Figure 13. Revision Process Visualization*

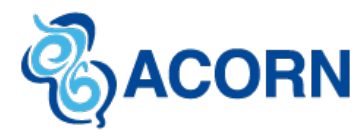

# 5.0 KEY REFERENCES

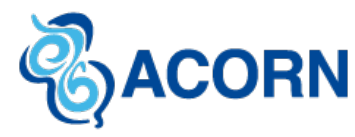

# 6.0 APPENDIX A – KEY INSIGHTS FROM OPERATOR INTERVIEWS

## 6.1 Scope

The key insights described here serve as an input to the technical bases for the requirements described in this document.

## 6.2 Introduction

This document details findings from an initial round of human factors interviews with accelerator operators at Fermilab within a consolidated file instead of raw data files (e.g., individual interview notes). The main findings are organized as key insights and a summary of associated design recommendations. Additionally, this document includes appendix references such as a glossary of accelerator terminology and a task analysis of operations at Fermilab.

### 6.2.1 Interview Protocol

Over the course of multiple months, the INL team met with 15 Fermilab main control room (MCR) operators to complete a semi-structured interview. The interviews were conducted to accomplish two goals: (1) to acquire operator specific knowledge about the accelerator control system and more importantly (2) to elicit operator feedback regarding their overall workflow including pain points.

The interviews took place over a virtual meeting platform. Most operators had the capability to share their screen to demonstrate their responses using ACNET, the primary control system at Fermi, as well as other tools typically used by operators. Each operator was asked to provide some demographic information about their educational background and operations experience at Fermilab.

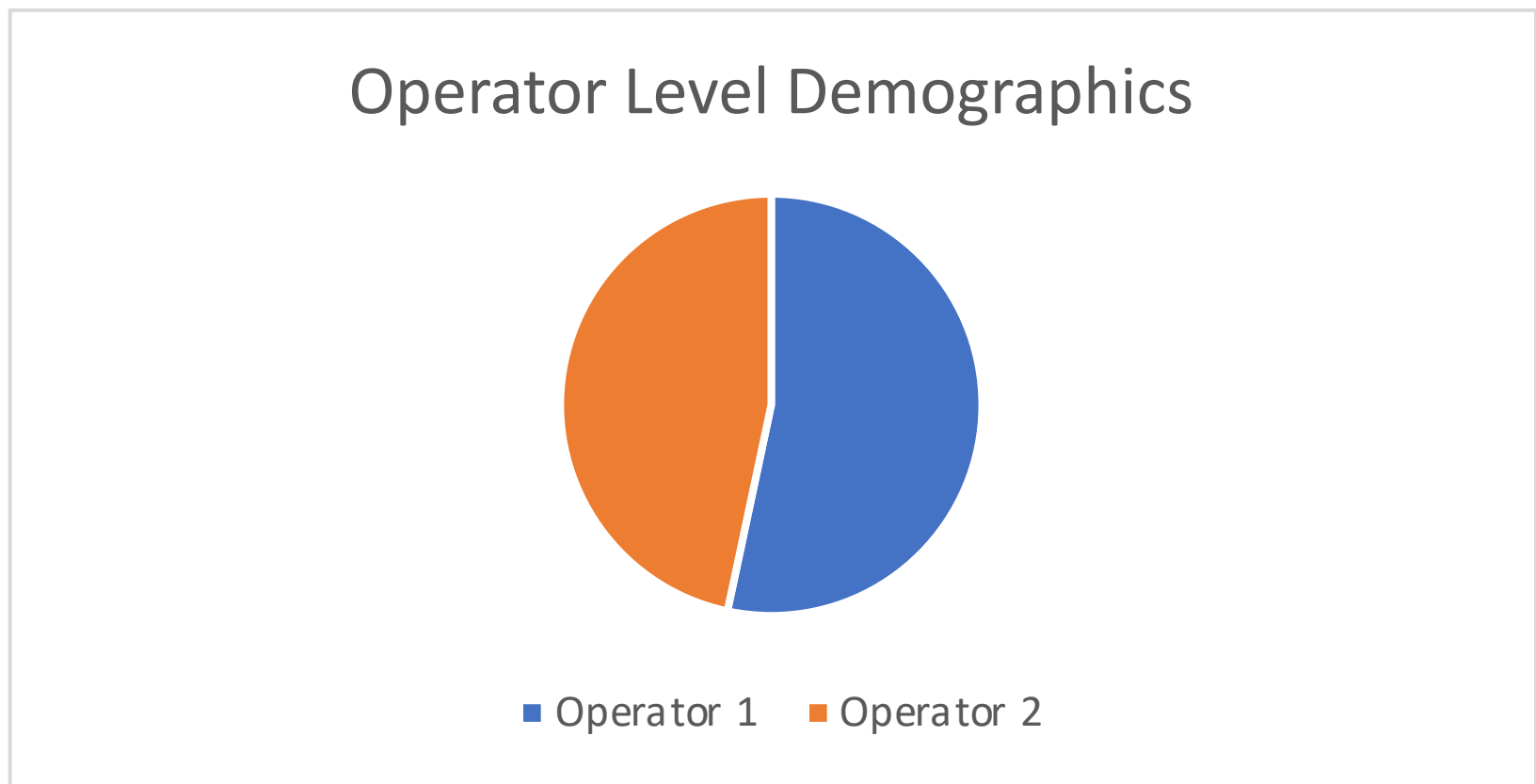

*Figure 1314. Operator level demographics for key insights.*

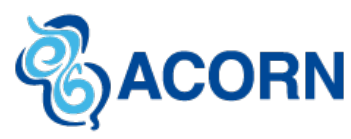

There are two classifications of main control room operators: “operator 1” and “operator 2.” The main difference between these classifications is operator 1s are considered new and require supervision to conduct their work whereas operator 2s are more experienced and can conduct work without supervision. Operator 2s are former operator 1s that have completed the operator training and passed the operator test which is required to advance. Of the 15 operator participants, eight were classified as operator 1s and seven were classified as operator 2s (see figure 13). The average years of operations experience between the participants was 2.32 years. Additionally, each operator had some sort of physics bachelor’s degree, (i.e., engineering physics, nuclear physics, or experimental physics).
Interview questions focused on how operators perform their day-to-day duties and were structured to include four main categories:

- Data and information gathering
- Controls and adjustment making
- Crew dynamics and teamwork
- Available/needed support for operators

The semi-structured nature of the interviews afforded the flexibility for researchers to delve further into specific topics to clarify assumptions about certain operating tasks, gain new perspective on a typical daily task, and further explore the methods employed by any one operator. The flexibility became useful as some topics became stale, hearing the repeated explanations on a topic, and researchers could amend the list of questions to continue pulling new or further enriching information from operators as researchers gained familiarity with the ACNET system.

**Low Hanging Fruit.** Throughout the interviews, operator insights were captured and tallied (see Main Insights section). Some of these insights resulted in identification of design recommendations that would have an immediate and significant impact on control system improvements. These design recommendations are referred to as “low hanging fruit” and are captured in general guidance in [ACORN-doc-700: *Design Philosophy for Accelerator Control Rooms*], and more specific guidance in the style guide. However, it should be noted that the purpose of these documents is to provide comprehensive design guidance for the development of accelerator interfaces. In other words, a succinct summary of “low hanging fruit” design recommendations regarding the first round of operator interviews is not included in either reference. Therefore, an additional purpose of this document is to not only summarize key insights, but also to provide design recommendations based on those insights to address the “low hanging fruit.”

### 6.2.2 Accelerator Control System

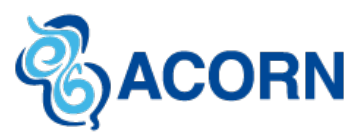

Since one of the primary purposes of these interviews was to provide researchers an opportunity to acquire knowledge of the accelerator control system, this section includes brief descriptions of what the accelerator control system entails, how it is operated, and additional tools that support it.

The high-level purpose of the accelerator control system is somewhat self-explanatory; to control the accelerator. However, a more human/operations centric definition would be to grant front-end access of accelerator equipment (i.e., machines and hardware devices) to accelerator personnel to manipulate said equipment according to stated criteria. The accelerator control system includes thousands of devices (e.g., magnets, sensors, etc.) all of which contain some degree of control functionality. The primary tool utilized by accelerator personnel to control is the Accelerator Control Network (ACNET) which is a connectionless, peer to peer networking protocol used for front ends [3, controls rookie book]. Other tools integrate with ACNET, such as web-based applications, to compile the accelerator control system and provide accelerator personnel the ability to monitor, manipulate, and experiment with accelerator data.

### 6.2.3 User Roles

There are multiple types of roles that interact with the accelerator control system including operators, machine experts, and physicists/engineers. Each of these roles has somewhat unique goals and responsibilities concerning the accelerator control system.

- Operators serve as the first line of defense for detecting and responding to beam line abnormalities. Their primary purpose is to monitor and maintain the beam output as optimally as possible.
- Machine experts serve as equipment specialists for a certain device or a slew of devices. When operators are not able to properly diagnose an event, they defer to machine experts to troubleshoot and restore the beam.
- Physicists/engineers perform experiments and monitor experimental data. They make requests or initiate changes to better meet their experiment requirements.

Each of these roles interact with the accelerator control system for specific purposes, and although each role is important, the initial round of interviews focused solely on main control room operators. This means the documented insights included in this report only reflect operator feedback and not machine experts or physicists/engineers. However, it is the intent of the research team to conduct additional interviews to capture insights that represent all roles as a future effort.

## 6.3 Key Insights

These sections include a list of key insights gathered throughout the operator interviews. The purpose of these insights is to demonstrate knowledge elicitation of operator interaction and workflow of the accelerator control system as well as linking these insights to “low hanging fruit” design recommendations.

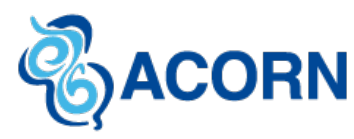

### 6.3.1 Insights without recommendations

| Insight 1. General Operations |
|---|

**1A:** Many operators described their primary role in the control room as being responsible for the following: maintaining beam quality (i.e., keep things running well), troubleshooting when necessary, and identifying/notifying the proper person (i.e., machine expert) to contact when operator diagnosis is unsuccessful. When asked to quantify the frequency of operator diagnosis success rate, one operator estimated that 75-80% of the time, operators can diagnose and solve an issue almost immediately. Sometimes they require more time to diagnose but are still able to handle it themselves, but occasionally they must contact machine experts and hand off the issue for them to solve. A follow up question revealed that there are typically 4-8 experts available at any time to help operators diagnose and solve incidents.

**1B:** Additional insights regarding general operations include how beam requirements are defined. For example, some experiments change beam requirements as often as weekly. The typical protocol for how changes in beam line requirements are handled are as follows: beam line physicists position the beam line where they need it to be according to the requirements of their experiments and then operations is responsible for maintaining the newly defined beam line.

| Insight 2. Tuning |
|---|

**2A:** Tuning is a frequent control behavior exhibited by operators. The purpose of tuning is to manipulate specific accelerator equipment to optimize beam quality (beam quality is sometimes subjective and varies by experiment). When asked about typical tuning behaviors, many operators stated that most expected tuning behaviors occur when an experimenter requests a tune or if an alarm goes off. However, if neither of those things happen, it's up to the discretion of the operator to determine how fine-tuned instruments should be. Another operator stated that each operator makes a lot of judgement calls when it comes to tuning, and they're pickier with their tuning when things are slow (i.e., no events to respond to) during shifts.

| Insight 3. E-log |
|---|

**3A:** E-log is a repository of operator shift records. Multiple operators mentioned how it is common for them to access and scan the e-log for a variety of tasks spanning from tuning to troubleshooting. E-log provides recent event logs as well as descriptions of unique and historical event diagnosis. E-log is an isolated control application (i.e., separate from ACNET) that operators rely heavily on to provide relevant context to their everyday operations.

### 6.3.2 Insights with recommendations

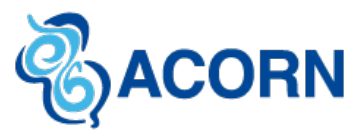

| Insight 4. General Operations |
|---|

**4A:** A somewhat surprising insight that was shared was how a variety of elements can influence accelerator devices. For example, the weather affects the way some magnets bend the beam. There isn't a direct measurement for this occurrence aside from posting the daily weather report in the control room which means operators must infer on their own why expected beam outputs are slightly off.

**Recommendation:** provide a notification to operators when outside temperatures are approaching parameters that are known to affect beam quality instead of relying on an operator's ability to automatically assume the weather is the source of beam deviations.

**4B:** An additional stated insight was console windows are limited to displaying a maximum of five plots. Each operator that mentioned this were asked a follow up question of "why?" and to their knowledge, there was no known reason but suspected it might be a hardware constraint. This constraint has been known to disrupt their workflow in some monitoring tasks, but mostly in diagnosis tasks.

**Recommendation:** if possible, remove the five-plot maximum window console constraint to allow operators more freedom in monitoring and manipulating accelerator data.

| Insight 5. Pain Points |
|---|

**5A:** When asked about pain points of the accelerator control system, multiple operators mentioned how the index pages in ACNET are cumbersome to navigate and cluttered. For example, the programs of index pages are organized alphabetically instead of being prioritized in place of function or frequency of use. Additionally, some programs are obsolete but are still included which causes a lot of unnecessary visual clutter.

**Recommendation:** incorporate a way to prioritize index page programs according to place of function or frequency of use. Additionally, automate a way to eliminate obsolete programs or alternatively, conduct an "obsolete" sweep annually to manually delete obsolete programs.

**5B:** Another consistent pain point stated by operators is the concept of "blowing away plots." Blowing away plots happens because anytime an operator wants to refresh a plot (e.g., view most recent data), the previous data disappears and is replaced by the most recent data. One operator described it as follows, "there are secret buttons, like secret blocks in Mario, that allow [an operator] to refresh the plot but keep the existing data, but if [an operator] doesn't know where that secret button is, they lose that data when they refresh the plot. First, there is no reason to hide necessary control functions, it only increases the overall cumbersomeness of a system. Second, since operators have stated a need to view historical data and most recent data in conjunction, there should be functionality to support that.

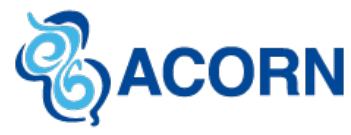

**Recommendation:** make all invisible functionality visible. Provide an ability to automatically view historical data and most recent data simultaneously without erasing one or the other.

## Insight 6. Alarms

**6A:** Alarms play a unique role in accelerator operations. Current alarm functionality includes providing operators a check list of corrective actions as well as maintaining special operating conditions (i.e., bypassing alarms). However, multiple operators stated that the responsibility of remembering which alarms should be in bypass position and which should be active falls on operators. Additionally, there's not a proper resource to validate which alarms should be in which position because that can change often depending on current experiments. If an operator must bypass many alarms to meet experiment requirements, they will typically create a personal alarm list to help them remember and as a source of validation when needed.

A specific example shared was when one operator bypassed an alarm and forgot. Only when they noticed a system irregularity did they remember to switch alarm position. The consequence of this error can vary and result in minor delays or in more extreme situations, can cause beam failure.

**Recommendation:** replace the alarm bypass system with customizable alarms instead. This way, the operator can outsource maintaining special alarm conditions to the system.

## Insight 7. Common Errors

**7A:** Operators are highly trained and extremely skilled at what they do. However, when asked which mistakes are most common, an operator stated that "two operators conducting the same task" (e.g., knobbing a machine) occurs frequently. Typically, it doesn't cause severe issues but sometimes does cause delays in operations.

**Recommendation:** integrate permissions functionality to prohibit multiple users from controlling the same equipment at the same time.

## Insight 8. Restoring from a Save

**8A:** Since beam requirements change so frequently due to experiments, operators rely on a system feature known as "restoring from a save" to avoid manually resetting beam requirements in the event of a system failure that results in erased data. The control system accomplishes this by capturing and storing reference values for accelerator devices only when an operator initiates the "save" functionality. This feature is often utilized when experimenters request a save after they've positioned all accelerator equipment the way they want it. However, operators shouldn't have to remember to save every time new beam requirements are revealed.

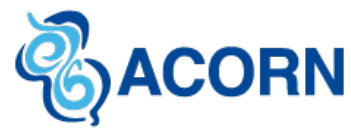

**Recommendation:** Auto-saves should happen automatically each time new beam requirements are set unless an override is manually executed by an operator. Previous save records should be available when desired in case experiments are repeated or previous reference values are needed.

| Insight 9. Flexible Operations |
|---|

**9A:** The troubleshooting and problem solving that occurs within accelerator operations is highly variable. One operator stated that there is a strong need for custom pages (i.e., a flexible program) because diagnosing and solving a unique problem is accomplished differently almost every time. However, an abundance of custom displays leads to visual clutter (see insight 5A) and therefore the need to be disciplined in display quality and review is stated.

**Recommendation:** Flexible operations (i.e., custom pages) should be supported with the caveat that quality control and review is mandatory.

| Insight 10. Requested Functionality |
|---|

**10A:** One operator stated that the scattered nature of the accelerator control system (i.e., isolated applications) prolonged their ability to understand and become proficient in operating. An example of this was mentioned that when operators are tasked with monitoring main injector, the "main injector kickers" are located in a website URL that is completely separate from ACNET. The separation of related information was stated as difficult to overcome and a preference for consolidated information was detailed.

**Recommendation:** evaluate ways to centralize all necessary resources into one location within the control system.

### 6.3.3 Frequency of Insights

Since the interview format was semi-structured, the insights collected across interviews varied. However, a key insight was mentioned in all interviews.

Every single operator described their on-the-job experience as their primary resource of interpreting control system information. In other words, an abundance of tribal knowledge is required to effectively operate the accelerator control system. This causes operators to devote a large portion of their mental workload to learning and remembering unique intricacies of the system whereas they could be devoting that energy to more appropriate tasks such as alarm response. Addressing some of the "low hanging fruit" design recommendations will help reduce operator cognitive burden by eliminating cumbersome functionality, visual clutter, and error prone design.

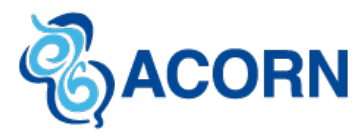

## 6.4 Conclusion and Disclaimer

First and foremost, this document is intended to act as a resource of consolidated findings from the initial round of human factors interviews with accelerator operators at Fermilab. Additionally, this document details a summary of design recommendations to address the current state of the accelerator control system (more specifically ACNET) at the time when interviews were conducted. This document also includes additional resources such as a glossary and an operator task analysis. Although this document is not a comprehensive design guidance, and should not be referenced as one, it is a resource of consolidated findings and documentation of knowledge elicitation.

Incorporating insights from the first round of interviews is crucial to overall project success. However, not only do additional interviews for other accelerator personnel roles need to be conducted, but additional operator interviews need to occur as well. Therefore, a proposed follow-on activity is to develop a gap analysis of operator topics and questions to address with another round of interviews.

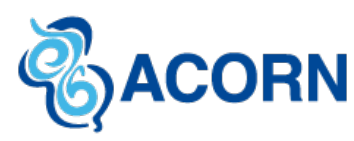

# 7.0 APPENDIX B – TASK ANALYSIS FOR SHIFT OPERATORS

## 7.1 Scope

Another outcome of the interviews was a high-level operator task decomposition. The task decomposition was developed through inferences taken from each interview and two interviews to validate the accuracy of those inferences. Main control room operators have the high-level goal to maintain beam quality within health, safety, and experimental parameters. The task decomposition targets the relative task sequence operators move through to carry out this objective. There is nuance to the sequence of tasks depending on the shift and accelerator state. However, for the bulk of operation the five listed tasks are sequentially executed.

## 7.2 Summary of Performed Tasks

Main control room operators at Fermi National Laboratory are charged with maintaining the quality and efficiency at which the various machines are providing and directing 'beam'. MCR operator's foremost goal is maintaining beam within the constraints of safety codes, to protect the welfare of those working at the lab, and equipment specifications to protect the machines providing and directing the beam. Within those constraints are the various experiment parameters that inform what the beam composition and quality must be. To operate within these parameters MCR operators carry out the following five tasks to perform their duties.

1. Acquire awareness of the current system state and expected operating efficiencies.
2. Monitor the various machines within the system for violations of operating efficiencies.
3. Identify machines at risk of or currently violating expected operating efficiencies.
4. Diagnose what is causing the potential or current violation of expected operating efficiency
5. Act to prevent or restore the machine to operate within expected efficiency range.

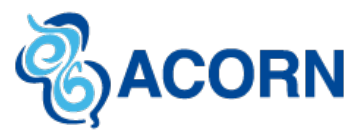

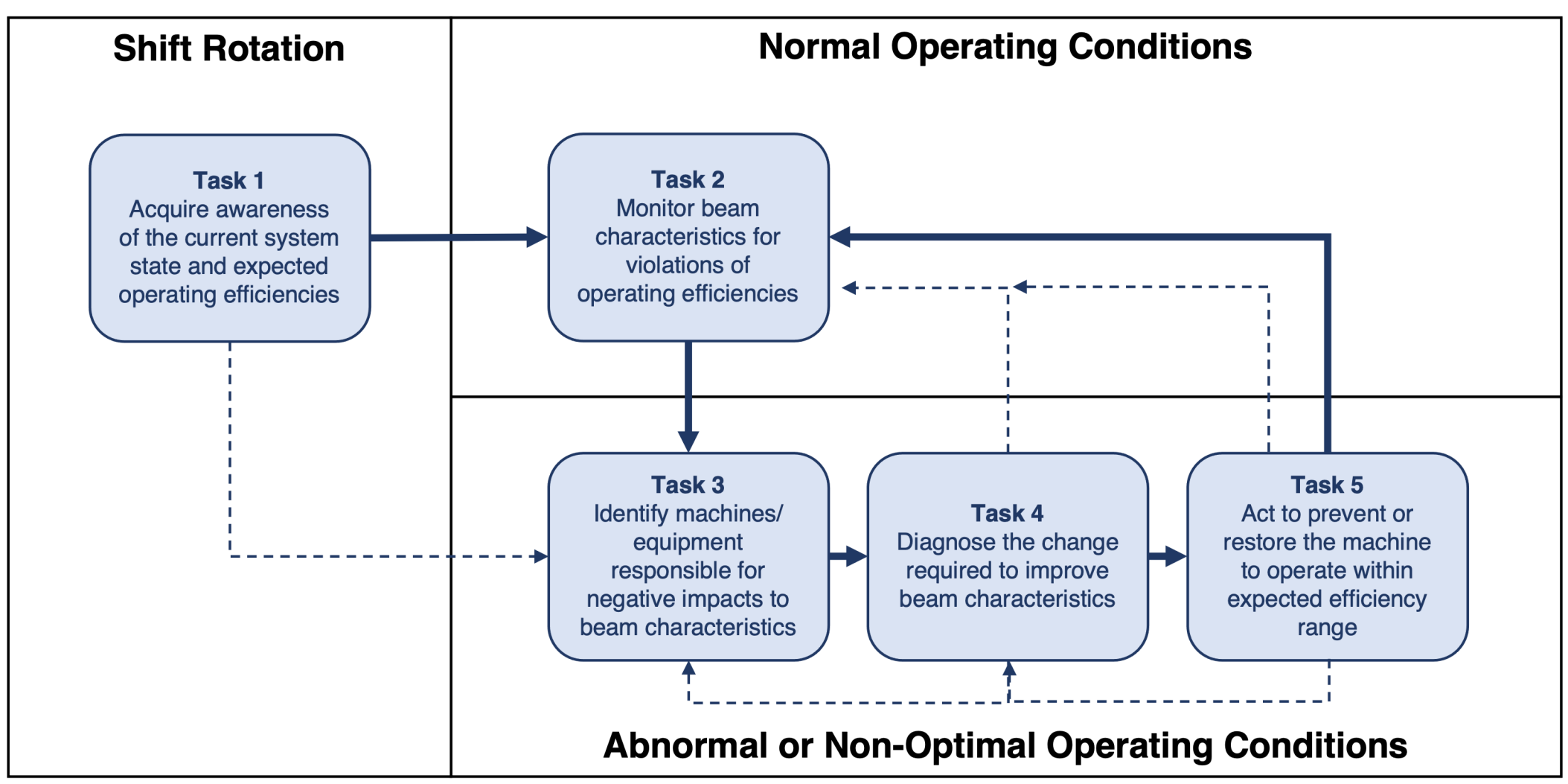

*Figure 1415. High-level tasks performed by shift operators.*

The high-level tasks are cyclical in that once task five is accomplished the operator returns to the second task and continues in that order until a shift is over. At which point the exiting operators support incoming operators with the first task. The tasks at this level are successive in nature in that latter goals are built on the information gathered and actions taken in previous goals. System upgrades that support actions allowing operators to efficiently move from one task to the next will support the upgrade objectives.

During a shift, operators interact with ACNET, crew members, and browser applications to help them complete each task. Therefore, each high-level task can be broken down by the necessary actions or information needed for the operator to move from one high-level task to the next. The information requirements, coordination with crew members or other outside organizations, and specific task sequences operators typically adhere to are described in this task analysis.

## 7.3 Detailed Task Analysis

### 7.3.1 Task 1. Acquire Awareness of the Current System State and Expected Operating Efficiencies

The typical shift rotation means an operator will spend about 16 hours away from operations before returning. Therefore task 1 is regaining awareness of what has happened in the past 16 hours and what is currently happening. This may include but is not limited to machines currently down for maintenance, problematic machines, status of alarm list, and a summary of what has occurred in the previous shift. This orientation process helps operators know where to identify tasking during their shift. It also serves as a filter for operators to perhaps disregard some information in the control room that is otherwise used to identify tasking. One example is disregarding some known alarm statuses that are already being dealt with outside the control room. Operators without this information may attempt to act on a machine that is purposefully offline without knowing.

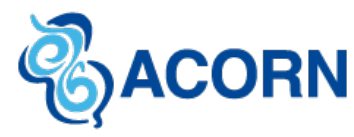

Two formal tasks carried out to inform incoming operators of the system state are 1) the briefing from their control chief occurring outside the control room and 2) a debriefing from operators of the previous shift. The crew chief briefing is a summary report of how the accelerators have been operating and any pertinent or safety-related information operators must know prior to beginning their shift. Then the previous shift hands off the controls to the new shift by presenting a very detailed accounting of what has happened and needs attending. Depending on the state of the system, this could mean directing new operators straight to task 4 or 5 in fixing a machine that is out of alignment.

Once the shift has begun operators use three different resources to acquire and maintain system awareness. First, is a set of windows that provide system wide performance information. Operators typically set up the same five ACNET windows to orient themselves with the system they are operating.
Those windows typically include:

- Main injector losses plot
- Booster losses plot
- Recycler losses plot
- Beam budget liners plot
- Alarm screen

Second are the “comfort displays” that act as large overview displays. The comfort displays are often redundant to the console windows but indicate what the crew chief has identified as important to monitor given the current accelerator statuses. These comfort displays are secondary sources in a nature and their effectiveness suffers due to the physical layout of the MCR that can restrict their visibility from some operator console locations.

Third, is the highly detailed information that accounts for almost every action ever taken in the control room, the electronic logs, or elogs. The elogs are where all operators report and track their actions if a machine has received any kind of attention. Operators needing more detailed information about any part of the system consult the elog. Elog is a browser application and not constrained to the 5 window limit set by ACNET. Operators consult the elogs for detailed descriptions of previous malfunctions, troubleshooting, maintenance, and issue resolutions. The elogs can contain pictures, points-of-contact, and links to other helpful resources. After the briefings and console organization the operator typically has the information required to begin the second high-level task, monitoring beam characteristics for violations of operating efficiencies.

### 7.3.2 Task 2. Monitor Beam Characteristics for Violations of Operating Efficiencies

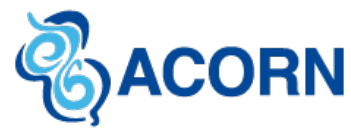

The second high-level goal is monitoring for operating inefficiencies. An inefficiency could be a violation of some beam characteristic that has moved outside the provided experiment, equipment or safety parameters. However, more experienced operators report they also seek out “poor performers” that have not yet violated a parameter threshold but could be made more efficient. Monitoring for inefficiencies requires filtering known statuses and expectations of the system with currently reported system status as presented in ACNET. It is possible some usual indications of violations can be ignored due to bypassed alarms, equipment that is moved offline or other reasons. Hence, acquiring system awareness is important before beginning to monitor beam characteristics. Monitoring tasks occur in no specific order. They include:

- Monitoring personally curated console windows
- Monitor beam profiles
- Monitor power supplies
- Monitor large view displays referred to as “comfort screens”
- Receive assignments to watch machines currently acting less predictably
- Receive calls from Experimenters to tune
- Compare current plot values with beam specifications from experiments

These tasks combine to create the high-level task of monitoring beam characteristics for inefficiencies. Performing these tasks create the opportunities necessary to move to the third high-level task; Identify machines responsible for negative impacts to beam characteristics. Causes for inefficiencies can vary. Often, the tasks in this section are enough to even identify the machines that need tuning; however, some situations require further diagnosis. Those steps are described in the following high-level task.

### 7.3.3 Task 3. Identify Machines and Equipment Responsible for Negative Impacts to Beam Characteristics

The third high-level goal is activated by two situations. 1) The operator may get an indication that a violation has occurred or 2) the operator recognizes a beam characteristic that could be improved. The operator then identifies first the machine where the inefficiency exists. Then identifies the equipment within that machine that is impacting the beam quality.

The information from the second high-level task is the initial input to determining an adjustment is needed. Once a potential adjustment is identified the operator essentially answers the following questions:

- Is the machine linear or cyclical?
- Is there equipment in this machine that has been reported as acting unusual?
- Is someone working on this piece of equipment already causing fluctuations in its efficiency?
- Could the beam quality here be affected by factors upstream of the identified inefficiency?

The first three questions can be answered by accessing information in ACNET such as plots, parameter pages, and downtime logs or browser applications such as the elog, schematics, and training tools. However, at this stage the operator must close out windows that provide more holistic views of the system in favor of more detailed information regarding the inefficiency due to the five window ACNET limit. The fourth question is answered using operator knowledge gained through experience and training. The information provided by ACNET provides some historical data leading up to the inefficiency in question. It also provides at times provides the range of expected values for context. The operator is responsible for understanding what other factors may be impacting the inefficiency in question.

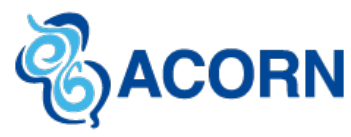

The loss charts operators monitor indicate the machine in which an efficiency exists. Once identified, some diagnostic work is required to identify the right fix for the inefficiency. In general, linear machines are simpler to diagnose because equipment adjustments have only downstream effects. Therefore, beginning at the piece of equipment where the inefficiency is indicated is considered good practice.  Cyclical machines can be more challenging as changes to one piece of equipment can affect the beam on either side of the equipment being tuned.

### 7.3.4 Task 4. Diagnose the Change Required to Improve Beam Characteristics

Once the machine is identified the operator completes the fourth high-level task by comparing the values on the parameter page related to the plot window associated with the equipment responsible for lack in beam quality. Often, the operator will make small adjustments to equipment settings and evaluate the result. If the plot values are moving in a favorable direction, the operator continues to the fifth high-level goal continuing to adjust until the desired plot values are reached. However, if adverse responses occur the operator must continue to diagnose the issue.
The diagnostic strategies vary depending on where the inefficiency or violation is occurring within the system.

- Machine schematics
- Parameter page
- Efficiency plots/loss monitors
- Alignment references (example is bullseye plot)
- Use written instructions for correcting “beam orbit”
- Expected loss/preventable loss (knowledge in the head

### 7.3.5 Task 5. Act to Prevent or Restore the Machine to Operate with Expected Efficiency Range

The fifth high-level goal is the operator acting on the system to restore the beam characteristic to a desired setting. Within this task is a series of acting, monitoring then acting again to ensure that actions taken are having the desired effect. Operators monitor using the equipment plot window and act on a parameter page by either increasing a decreasing a parameter in increments to achieve the desired result. Loss plots and other “first indications” are reviewed to ensure their actions solved the noted issue.

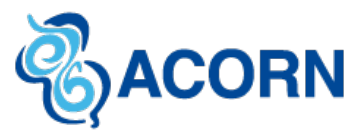

- Knob plots and parameter page
- Adjusting switchyard setting (directing beam to machines) to increase/decrease beam quantity
- Check experiment request (in elogs)
- Loss plots

### 7.3.6 Conclusion

The detailed task analysis presented in this section outlines the structured and cognitively demanding process that accelerator operators follow to maintain system performance and beam quality. From the moment a shift begins, operators must rapidly build situational awareness through briefings, toolsets, and interface interactions. This awareness is essential for monitoring beam characteristics, identifying inefficiencies, and initiating corrective action.
Each high-level task (i.e., acquiring system awareness, monitoring performance, identifying responsible machines, diagnosing problems, and executing system changes) builds upon the last. The effectiveness of each task relies not only on the accuracy of the available tools (e.g., ACNET windows, comfort displays, and e-logs) but also on the operator's ability to interpret and act on complex, dynamic information.

Designing HSIs that support each stage of this workflow can significantly enhance operator effectiveness and system reliability, particularly in minimizing cognitive overload and maximizing clarity. As such, understanding the interconnected nature of these tasks is critical for both system designers and operational support teams.

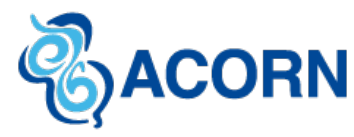

# 8.0 APPENDIX C – SUPPORTING INTERACTION BETWEEN EXTERNAL ROLES AND MCR OPERATIONS AND CONTROL SYSTEM APPLICATIONS (SECONDARY ROLES TASK ANALYSIS)

## 8.1 Introduction

### 8.1.1 Transitioning control system platforms

The control applications for operating FERMIs primary accelerator systems are being migrated to a new control interface system. As part of the migration, all current applications that exist under the Accelerator Controls Network (ACNET) are under review to determine their value to operations and either migrate them to the new system, consolidate duplicate or similar applications, or abandon. Accelerator operations should not be negatively disrupted by the transition to the new control system. The current system houses hundreds of applications developed over the years and it is not feasible to transition every application to the new system. Therefore, an effort is underway to identify the applications important to quality, safe, and successful accelerator operations to enable a smooth transition between control systems.
The transition also provides the opportunity to improve the applications. Previously limited functionality or information presentation can be improved to enhance operator performance. Better information aggregation and abstraction may increase the interpretability and diagnostic capability of main control room (MCR) operators. Applying current usability principles with clear understanding of what each application is used for will make information more accessible to those who have less frequent interactions with the control system but nonetheless rely on it. Deepening our understanding of what roles external to MCR operations rely on what applications and why (functions) help smooth the transition to a new control system designed to meet operational requirements and needs.

External roles contribute to planning, diagnosis, set-up, and safety of accelerator operations. Despite their less frequent and limited interaction with ACNET, the contributions of these external roles remain significant. As we move towards modernizing control interactions, it is important to consider how these roles are supported by the new control interface system. Application usage metrics such as access frequency used and usage time were gathered to generate a ranked list of applications. These metrics, while excellent to measure general application use, do not capture applications used infrequently or by more specialized roles critical to broader accelerator operations and safety. Roles that exist externally to the MCR operators. Such metrics also do not capture the functional value of the applications, though the content of the application provides some inferential information to that end. To capture applications critical to meeting experiment demands efficiently while ensuring safety for humans, the environment, and equipment, researchers conducted interviews with experts connected to accelerator operations but whose responsibilities lie outside of the MCR. These roles may not be directly involved in MCR operations but either support or are supported by MCR operators and applications that currently exist on ACNET.

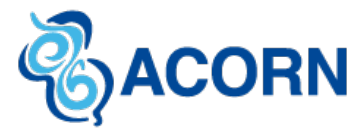

### 8.1.2 Objectives of this effort

Insights derived from semi-structured interviews with main control room operators (appendix B) revealed reliance and multiple communication channels with outside experts and roles external to MCR operations. Staff interacting with ACNET and MCR operators were interviewed to capture the supporting roles goals, tasks, and expertise shared that enable MCR operators to maintain safe and efficient operations.

The two objectives of this effort were to:

1. Identify the ACNET applications that enabled external roles relied on to perform their duties to accelerator operations and function those applications served.
2. Identify the support external roles currently provide to MCR operations that enable safe and efficient operations.

Capturing the interactions between the MCR operators and external roles involved with maintaining beam quality has two benefits to developing new control system applications. New and migrated applications will include features that support external roles and thus support continued safety and efficiency in operations. Second, analyzing how external roles support MCR operators generates insights for designing applications that can provide the same support operators seek from specialists and expand the operational purview of MCR operators.

## 8.2 Method

### 8.2.1 Selecting participants

The external roles we needed to interview had to be tied to MCR operations in either supporting their operations, receiving support from them or otherwise relying on current control system applications to perform their work. To identify external roles for interviews, we started with known workflows. Our initial step involved considering the core operational functions and common tasks necessary to operate the accelerator complex (Table 6). In doing this, a foundational list of the essential processes and activities was created from which to base external role selection.

| Core Operational Functions | Definition |
|---|---|
| Machine Tuning | Adjusting accelerator parameters to optimize beam stability, intensity, and overall performance. |
| Monitoring Beam Efficiency | Tracking how effectively the beam produces the desired output and identifying potential losses or irregularities. |
| Monitoring Utility System Performance | Observing cooling, power, and other support systems to ensure reliable operation. |

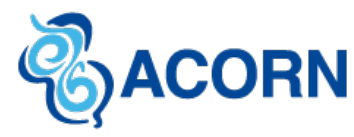

| Responding to Alarms | Investigating and addressing alerts triggered by abnormal system conditions. |
|---|---|
| Monitoring Personnel Safety Systems | Ensuring radiation, interlock, and access controls function correctly to protect staff. |
| Machine Commissioning | Bringing new or upgraded accelerator systems into operation through structured testing and validation. |
| Beam Studies | Conducting experiments on the beam to characterize its properties and test new configurations. |
| Low-level Hardware Communication Troubleshooting | Diagnosing and resolving issues in the direct communication between devices and control systems. |
| System Configuration | Setting up and maintaining parameters, controls, and software needed for reliable accelerator operations |
| Script Development | Writing automation scripts to streamline repetitive tasks and operational procedures |
| Application Development | Designing and maintaining user-facing software tools to support operations, monitoring, and analysis |

*Table 5. Core operational functions identified and their description*

Next, we considered the individuals related to these core operational functions. We collaborated with subject matter experts to identify what roles interact with ACNET to solve operational problems and mapped the relevant personnel, or types of personnel, to these tasks. The roles identified were not based entirely on job title explicitly but could be defined by their interaction with ACNET and the MCR. We then observed commonalities among these individuals. There was significant variation even among individuals with the same title, highlighting the need for a more nuanced approach.

To address this, we identified functional groupings. Rather than imposing a title-based structure, we grouped related activities, and the types of skills or operational knowledge required. This approach led to the creation of a "roles" list, focusing on functional roles relevant to the actual use of the control system.

A total of 21 roles were selected for further investigation. An individual currently holding each role was interviewed about their interactions with ACNET and MCR operations. Their role description is provided within the context of how they support MCR operations or interact with ACNET. The description of the role therefore does not include any other responsibilities outside of the purposes of this effort.

1. MCR Crew Chief
2. Ops Specialist
3. Directorate Manager
4. Controls Department Manager
5. Instrumentation Department Manager

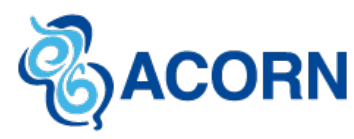

6. Machine Coordinator
7. Run Coordinator
8. Accelerator Physicist
9. Controls Software Engineer
10. Controls Hardware Engineer
11. Mechanical Support Engineer
12. PSED Electrical Engineer
13. PSED Software Engineer
14. Instrumentation Software Engineer
15. Instrumentation Hardware Engineer
16. Fluids System Mechanical Support
17. Controls Operations Technician
18. Mechanical Support Technician
19. PSED Support Technician
20. ES&H Officer (radiation safety)
21. Interlock Specialist

### 8.2.2 Interview method

The interviews were conducted in a semi-structured format. The structured portion of the interview focused on 10 functions (Table 6) determined by subject matter experts (SMEs) as a comprehensive set of functions for maintaining beam performance and safety of the accelerator systems from an operational standpoint. Every interviewee was asked to describe how their role supported one or more of the functions above and to identify, if any, the applications involved in their contribution to the function.

A general questions portion allowed interviewers to help facilitate conversation and gain any additional understanding of the roles and their responsibilities related to the above functions or other interactions that may occur with MCR operators. Part of the interview sought to capture the interviewees' own characterization of how the role they were interviewed for supported MCR operations

### 8.2.3 Data Processing

The team structured the interview data into spreadsheets for further analysis using the following process.

Step 1: Mapped roles to high-level functions

The team mapped external roles to the high-level functions. Mapping was binary, based on a YES or NO system. For clarity, any roles involved downstream in the functions were marked as YES. The team internally reviewed the accuracy of the mappings and identified any potential discrepancies. These discrepancies were discussed and resolved before moving to Step 2.

Step 2: Determined goals and categorized tasks

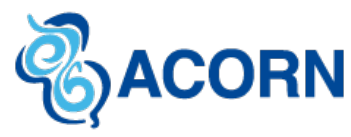

Next, the team reviewed the answers given for each role by function labeled a task type. This categorization helped frame key decisions, actions, and the necessary information for Step 4. The team agreed on the task types to use based on the content found in the interview notes. The task types were:

- Supervise
- Monitor
- Interpret
- Diagnoses
- Decision Support
- Carry Out Action

After categorization, each team member independently reviewed the assignments and flagged any discrepancies. These discrepancies were then discussed and resolved before proceeding to Step 3.

Step 3: Categorized ACNET and MCR Interaction

The team categorized if actions taken by the external role did or did not require ACNET applications and what communication with MCR operators typically occurs. This helped to clarify the nature of communication and Acnet application use. After characterizing external role interaction and communication with ACNET and MCR operations, team members independently reviewed them and flagged any discrepancies. These were discussed and resolved before moving on to Step 4.

Step 4: Identified How Each Task was Performed

In the final step, the team mapped the task type to the function and the current ACNET applications used to perform that task (the HOW). This included detailing the key decisions and actions required and identifying the applications used to support these decisions. The team also documented the key information used from the applications, specifying the parameters referenced. Finally, they defined the key characteristics of the information needed (e.g., trends, setpoints).

### 8.2.4 Application Analysis by Role

| Role | List of applications identified by role.<br><br>(Zach, Michael's, and Casey's Review) | Machine Tuning | Monitoring Beam Efficiency | Monitoring Utility System Performance | Responding to Alarms | Monitoring Personnel Safety Systems | Machine Startup/Commissioning | Beam Studies | Low-Level Hardware troubleshooting | System Configuration | Script Development | Application Development |
|---|---|---|---|---|---|---|---|---|---|---|---|---|
| MCR Crew Chief | *pa4241 or sa1130, pa1438, pa1008, pa4087?, pa1438?* | 0 | High-level (big picture) monitoring (using pa1438 [Beam budget monitor (BBM) PA]?) | Passively monitor system performance through alarms (using pa4087? [Alarm List/Control]) | Delegates alarm response to operators (using pa4087? [Alarm List/Control]) | Passively monitor system performance through alarms (using pa4087? [Alarm List/Control]) | ??? | ??? | Consulted for complex diagnoses (using ???) | High-level (big picture) monitoring (using pa1008 [Dump device database information program]/ D80) | Personal use: ACL automation and diagnosing (using ???) | Personal use: ACL automation and diagnosing (using ???) |

*Figure 15 16. Example of table created to document the applications that support Roles external to the control room that contribute to accelerator performance*

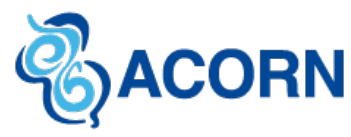

The example table in Figure 15 above demonstrates how roles and the applications were organized by function. The full table provides an interaction map of the applications each role uses to perform the function.

## 8.3 Analysis of Interaction with MCR by Function

The following section will detail the 'when' and 'why' a role outside the control room may get involved with the function. Defining when and why outside roles engage with control room functions informs us how to build a better control system. The better control system should increase the operator's situation awareness of the system without increasing task complexity or operator workload. This analysis identifies the type of support operators rely on, what support operators provide outside roles, and information that, if provided in the new control system, continues support for specialized and planning roles.

Data collected from the interviews were synthesized into the following structure:

**Operator purview** frames the currently performed tasks and context for the MCR operators' role in performing the function being analyzed.

**Expertise support** captures the roles that assist MCR operators with their tasks should they need assistance and is reasonably considered within a typical operators capability to do so. Information here provides insight to potential difficulties with the current control system and where improved application design may support operator ability to perform their functions within the MCR.

**Specialized external support** refers to those roles not considered as an MCR operator. These roles offer specialized knowledge of systems that can support MCR operations when it is not reasonably expected of an operator or the available expertise support roles to have. These roles may provide support for issues with hardware, ACNET interfacing software, or specialty systems. Also captured is how MCR operators support specialized external roles.

**Planning support** is a section to capture the chain of communication from MCR operators to the roles concerned with big picture accelerator operations such as the run and machine coordinators. Though other roles may also be involved in helping coordinate communication.

## 8.4 Results

### 8.4.1 Structure of results

First is the role descriptions and the applications they use. To determine how ACNET is currently used to support roles external to MCR operations, the applications mentioned in interviews were tabulated manually and double-checked using keyword search programs developed by one of the research team members. The key words were pulled from a list of applications already compiled by another ACORN team for their accounting of ACNET applications. The applications categorized as 'core' and 'critical' and 'additional'. Core applications are those indispensable for everyday operations in the main control room. There are 13 in total. Critical applications are those that are indispensable for mission-critical operations across the accelerator complex (i.e., inside and outside of the control room). There are 58 in total. Additional applications are those discovered in interviews that did not overlap with the critical and core applications. The additional applications are those that were not identified by use frequency and time-in-application metrics.

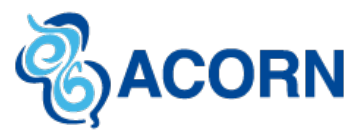

### 8.4.2 Role Descriptions and the applications they use

The following are descriptions of each role, and the applications they use when acting in that role's capacity. None of the descriptions below describe the full scope of responsibilities of any of the interviewees. Each role is a slice of responsibilities and described only as it related to the purposes of this report. Each role is briefly described followed by a table of the applications associated with that role.

**MCR Crew Chief:** The MCR crew chief oversees the MCR operators. Crew chiefs were often MCR operators themselves. They ensure the beam stays within budget. They are responsible for delegating tasks and coordinating the operations crew. They assist with tasks crew operators may find difficult. They ensure the run plan dictated by the run coordinator is followed. Additionally, they manage the timeline application that distributes the beam to different experiments.

**Operations Specialist:** An operations specialist's role is open-ended, providing crucial support to the operators in the control room. He acts as an intermediary between the operators and other departments, specifically focusing on controls. Part of his responsibility is to assist operators through their training program, ensuring they become familiarized with the control system. If operators encounter issues or have questions related to controls, the operations specialist is the primary resource for support and is responsible for delivering training to new trainees, covering the essential control-related information required as a baseline for their roles.

**Directorate Manager:** The directorate manager role involves overseeing accelerator technology and finding a way to quantify the long-term performance of the accelerator complex and its components, which combine to determine the total output. This information is crucial for setting goals for the directorate, such as up-time hours, peak power, or average power. When proposals for work, upgrades, or studies are made, these performance metrics can be referenced to assess their impact on achieving the established goals, thereby aiding in prioritization.

**Controls Department Manager:** The Controls Department Manager supervises projects and ensures the beam stays online, thereby maximizing uptime. Key responsibilities include managing the controls team to maintain a usable and functional control system and interacting with projects, notably the design of PIP-II and ACORN. This role bridges the needs of the existing complex with those of new projects. Additionally, the role requires investigating issues and performing diagnostics to identify problems, which are then passed on to the appropriate personnel for resolution. Ensuring continuous beam delivery is a critical aspect of this position.

**Instrumentation Department Manager** They are responsible for the design of systems within the scope of instrumentation and readings. Their duties include designing, installing, commissioning, and providing support to ensure everything runs smoothly. They work with low voltage electronics and coordinate with support groups to assist machine groups.

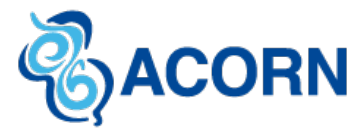

**Machine Coordinator:** This job is a rotational position, typically twice a year, responsible for monitoring machine performance and coordinating with support groups for maintenance and safe access to tunnels and machines. This role requires constant monitoring, performance evaluations, failure tracking, and loss tracking to understand overall machine performance and diagnostics. The goal is to keep the machines in healthy operating condition, minimize dose rates, and minimize beam losses. This coordination is achieved entirely through verbal communication and meetings. Feedback on coordination performance and machine needs is provided in follow-up meetings. Requests and needs are submitted, and the plan is then set forward by the Run Coordinator (Runco) and communicated via PDF or PowerPoint presentations.

**Run Coordinator:** The Run Coordinator manages all work upgrades and develops plans for maintenance activities during planned shutdowns. This role serves as the interface for the director, overseeing failures, and scheduling preventative maintenance to ensure everyone is aligned. They take a global program perspective to avoid creating biased schedules. Their primary goal is to ensure the complex runs as efficiently as possible by balancing the needs of all machines and meeting everyone's expectations with minimal disruption. In the event of failures, they utilize accelerator downtime to make additional repairs and conduct maintenance. There are two run coordinators available, primary and backup, who work in three-week shifts due to the 24-hour on-call nature of the job. They interface with the Accelerator Division, coordinate runs to meet experimental needs, ensure safety, and make decisions that prioritize the overall efficiency of the complex. This role requires keeping track of the big picture, balancing machine requirements, and communicating with management to prioritize actions effectively.

**Accelerator Physicist:** A physicist supporting linear accelerator operations focuses on improving booster performance and exploring global physics topics. They manage detailed scheduling with operations to avoid beam disruptions and are keen on automating processes due to the impracticality of daily manual tuning with thousands of adjustable parameters. The physicist analyzes this data through plots to determine the best operational settings, while real-time monitoring ensures immediate oversight. Essential monitoring systems and alarms track beam performance, radiation levels, and trip events, maintaining safety standards during higher-than-usual losses. By evaluating booster efficiencies, they ensure the booster operates efficiently and safely, continually enhancing its performance.

**Controls Software Engineer:** Their job is to ensure that high-quality applications are available. This includes developing applications for analysis, design, and software development, as well as maintaining, rebuilding outdated applications, and debugging. For example, working on the Beam Budget Monitor to ensure it meets current standards and functionality requirements.

**Controls Hardware Engineer:** The primary responsibility of a Hardware Engineer is designing parts for the accelerator systems, focusing exclusively on hardware rather than software that runs on ACNET. It involves writing a significant amount of embedded software to operate the custom designed devices. The work involves using various vendor tool suites to write the logic codes for Field Programmable Gate Arrays (FPGAs), which contain tens to hundreds of thousands of logic gates.

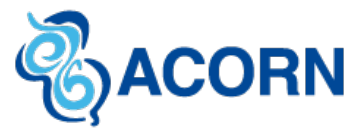

**Mechanical Support Engineer**: The mechanical support engineer interviewed worked on vacuum systems and support structures for the accelerators, while others handle the water systems. Engineers in this group involved with currently operating accelerators monitor vacuum systems and magnet installation. Primary responsibilities include supporting the vacuum systems for the accelerators, ensuring they remain operational through installation, monitoring their performance overtime, and troubleshooting issues if they seem to be failing.

**PSED Electrical Engineer:** Their job is to ensure that power supplies are functioning and regulating within acceptable limits so the beam can be managed effectively. This can include constructing power supplies, which are coordinated by the Tclock. Changes to the Tclock can affect power supply pulse accuracy, so they must diagnose why a power supply is not triggered when expected. This role involves extensive monitoring of power supplies and using plotting packages for performance monitoring.

**PSED Software Engineer:** This role is responsible for developing the embedded systems (front ends) used to feed data to the control system (ACNET). The data made available from the devices to the control application network is used for the console applications developed and deployed by MCR operations.

**Instrumentation Software Engineer:** Their job involves building front ends (i.e., sser-facing interfaces that allow operators and engineers to interact with underlying systems, manage devices, and visualize data) and managing instrumentation from a data production perspective. They facilitate the transfer of large amounts of data and specialize in data acquisition. Most of their interaction with the control room occurs during start-up. If a deeper diagnosis is needed when something breaks, the control room will contact them. If they are away from the control room, operators may need to carry out actions, such as power cycling a device, to help them access the necessary data.

The bulk of their interaction with the control room may occur during start-up. Otherwise, if something breaks and requires deeper diagnosis, the control room will contact an ISE. If they are away from the control room, operators may have to carry out actions for the ISE to access data needs, such as power cycling a device

**Instrumentation Hardware Engineer:** Their job involves data acquisition, ensuring that analog signals are fully digitized and sent to the server. Once the data is on the server, they use LabVIEW to access it and verify that the resolution of the measurements meets the requirements of physics experiments. They also test and commission systems for integration to ensure data flows seamlessly from the source to ACNET. Additionally, they help determine what data is made available to those viewing it in ACNET.

**Fluids System Mechanical Support** They are responsible for designing cooling systems for targets and horns. As an Operations Engineer, they monitor the systems they have worked on or oversee. As a Systems Engineer, they handle new designs or upgrades to existing systems. They monitor trends and real-time data. If an operator or technician does not understand signals or parameters, they reach out to them for assistance.

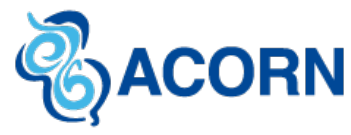

**Controls Operations Technician:** As a System Administrator providing infrastructural support, their official title is Computer Services Specialist. Their responsibilities include troubleshooting malfunctions and monitoring systems to ensure smooth operation. They rely heavily on team members and work primarily with the IT division, AD controls, and central services. Interactions during troubleshooting are often conducted over email and Slack.

**Mechanical Support Technician:** This role involves maintaining and supporting vacuum and fluid systems related to both the main accelerator and testing accelerators. This includes turning pumps on and off, monitoring the systems to ensure they operate smoothly, and checking their health to preemptively address potential issues. By closely monitoring these systems, mechanical support technicians help ensure continuous and efficient operations.

**PSED Support Technician:** The PSED Support Technician works on magnets and kickers, focusing primarily on high voltage systems and low voltage control cards. They handle power supplies for different types of magnets in the accelerator, which directly affect beam performance. High power systems, such as kickers, are used to inject beams into the accelerator. While all these systems are similar in setup and functionality, they serve different purposes within the accelerator. The responsibilities of this role include monitoring, troubleshooting, installation, and overall maintenance of power supplies and kickers for the magnets.

**ES&H Officer (radiation safety):** The ES&H officer is involved in rad worker training, planning, job reviews, ALARA (As Low as Reasonably Achievable) planning, and managing rad work packages, all of which come through me to ensure the ongoing operation of the labs. Regarding operations, this role works with RCT group leaders during shutdown and startup phases of the accelerators to ensure radiological technicians cover all jobs, oversee ALARA planning, and manage run permits and conditions to ensure beam is only sent to approved areas. Those who collaborate with the ES&H officer include the run coordinator, shutdown coordinator, operations team, and the division safety director.

**Interlock Specialist:** Their job involves maintaining safety sensors within service buildings to ensure personnel safety. They primarily monitor and maintain approximately 120 oxygen sensors within these buildings. Sensor readings are viewed within ACNET, exported from the Data Logger. Calibrating and testing sensors require collaboration with the MCR (MCR) to observe sensor readings. She works with the Radiation Safety Interlock System (RSIS), which has permit authority. If high radiation levels are detected in a service building, entry is locked out for everyone. Conversely, if personnel are occupying a service building, the operation of equipment from the MCR is locked out.

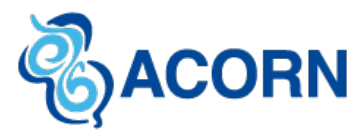

## 8.5 Role interaction by function

### 8.5.1 Machine Tuning

Machine tuning is a daily activity for MCR operations. Primarily it involves making minor adjustments to machine components to maintain efficient beam production and safe beam parameters. However, the tuning MCR ops performs serves additional purposes as well. It can serve as a trouble shooting function when adjusting the temperature and flow conditions of accelerator cooling system (Fluids System Mechanical Support) requiring communication between the tuner in the MCR and the field support adjusting cooling system parameters to achieve smooth operations.

#### 8.5.1.1 Operator Purview:

Operators tune machines as needed to keep the accelerator operating smoothly. The complexity of machine tuning can vary depending on what needs tuning and how that may impact downstream beam performance. Operators have a general understanding of when and how to tune a machine and rely on the information in ACNET to make these determinations.

#### 8.5.1.2 Expertise support (enhancing operator capability):

Operations Specialists are available to assist MCR operators when tuning a machine proves too complex for the operator. Though, they reported only helping in a supervisory role allowing the operators to perform the tuning while operation specialists monitor and provide decision support.

#### 8.5.1.3 Specialized Outside Support (problem occurs outside Ops purview, requiring specialized intervention):

Fluids system mechanical engineers are a resource for tuning and support fluid systems (i.e. water, pressurized gases, compressed air systems) when MCR operators are presented with tuning challenges that may be a result of faults with the fluid systems. Fluid systems engineer can help determine if the data is reliable or if the instrumentation is failing.

The instrumentation software engineer supports auto-tune functionality by developing data supply and automated tuning applications. These applications reduce the real-time operations tasks under the operator purview. If an auto-tune application begins to fail, instrumentation software engineers are the specialized support available to fix it.

PSED Software engineers develop embedded systems within the power supply equipment. They support operations by determining the data and data resolution that is sent to ACNET from the embedded systems. This serves to ensure MCR operations can efficiently tune this equipment. These actions take place outside of Acnet but involve communicating with the control room to determine their data needs.

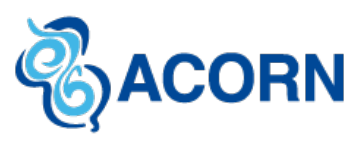

Accelerator Physicist are responsible for tuning large sets of magnets for experiments. They have developed scripts to automate tuning but may also perform manual tuning when necessary. Accelerator Physicists must coordinate with the control room when performing bulk tuning or notice if some parameters have deviated from nominal.

#### 8.5.1.4 Planning Support:

Machine Coordinator: The machine coordinator stays abreast of machine health and status to perform coordinating duties with other stakeholders and planners effectively. In addition to daily monitoring and data log reviews, the Machine Coordinator communicates with MCR operators about tuning trends indicative of declining machine health. Following up with specialists (i.e. technicians and engineers) on the machines and equipment proving difficult to keep properly tuned can further indicate the maintenance urgency to inform efficient maintenance schedules.

### 8.5.2 Monitoring Beam Efficiency

Monitoring beam efficiency is a core responsibility for MCR operations and is performed continuously throughout each shift. This activity involves observing key beam parameters such as intensity, transmission, and loss trends across multiple machines to ensure that the beam is performing within expected operational and experimental specifications. Operators use both personal console windows and shared overview displays to evaluate current performance against known efficiency targets.

In addition to identifying clear threshold violations, experienced operators also look for patterns or gradual degradation in beam performance that may not immediately trigger alarms. This proactive monitoring allows operators to detect emerging inefficiencies early and coordinate with system experts for corrective action. Monitoring beam efficiency also plays a supporting role in experimental operations, where operators adjust parameters based on experimenter requests to optimize beam delivery while maintaining system stability and minimizing losses.

#### *8.5.2.1* Operator Purview*:*

Operators are responsible for identifying areas where beam efficiency could be improved and acting on equipment to tune the system for enhancement. Additionally, the function requires identifying situations where degrading beam efficiency cannot be satisfactorily improved or continues to degrade despite tuning actions. In such cases, the issue is escalated to other roles to inspect the equipment under their purview that may be negatively impacting beam efficiency.

#### 8.5.2.2 Expertise support (enhancing operator capability)

Roles like Crew Chief and Operations Specialist provide additional system knowledge and capability to augment operator actions. They may assist in identifying areas to improve beam efficiency, help delegate operators to attend to issues indicated by the alarm system, or establish the appropriate time to consult outside specialists to inspect the performance of the equipment within their purview.

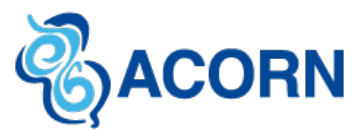

#### 8.5.2.3 Specialized Outside Support (problem occurs outside Ops purview, requiring specialized intervention):

Roles including Instrumentation Hardware Engineer, Fluid System Mechanical Engineer, and Mechanical Support Technicians support the equipment and hardware that makes maintaining beam efficiency possible.

The bulk of these roles are involved with ensuring equipment and hardware is performing as expected before startup, after installation, and immediately following a repair. This may involve real-time communication with operators to help test certain systems to validate how they are functioning appropriately.

The specialized knowledge of these roles is required to maintain accelerator equipment and hardware that impacts beam efficiency, but they rely on control room operators to help by communicating ongoing issues with their related systems, provide a level of troubleshooting support using their knowledge of ACNET.

Roles such as Instrumentation Software Engineer ensure equipment is transmitting the right parameters to the control room to effectively monitor beam efficiency.

It was noted that some operators can see when there is insufficient information and could, if they had access or time, make minor adjustments to instrumentation software (front ends) to correct the issue. However, expertise is required for configuring new software or making complex changes and adjustments to the parameters and equations that support control room operators.

#### 8.5.2.4 Planning Support:

Roles such as Machine Coordinator and Run Coordinator handle planning accelerator operations. Following identification of ongoing or chronic issues to monitoring beam efficiency, either identified by control room operators or by roles in charge of equipment and hardware these planning roles must be informed of the urgency and extent of repairs that may be required. Machine Coordinator is responsible for cataloging and scheduling machine maintenance. The Run Coordinator maintains awareness of available machines to optimize beam delivery to meet the needs of as many experiments as possible.

Generally informed by reports or re-occurring status meetings, these two roles gather information from all roles to determine how to most efficiently schedule the replacement or maintenance of various systems. Further decisions include whether the repair or replacement can be performed while the accelerator is running or if work must be completed while the accelerator is down.

These two roles rely on information from operators to understand what is running and what requires attention. These two roles complement each other, both requiring full awareness of accelerator performance to determine improvements and coordinate beam studies. The

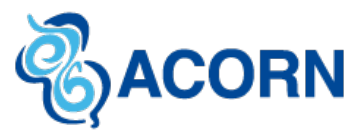

Machine coordinator stays abreast of how all the machines are performing to coordinate how to best schedule accelerator maintenance. The Run Coordinator makes sure all the machines are being used most efficiently to meet experimental needs.

### 8.5.3 Monitoring Utility System Performance

Monitoring utility system performance is a routine but critical aspect of MCR operations. This task involves continuously observing the operational status of support systems that are essential to accelerator function, such as cooling water systems, electrical distribution, compressed air, and vacuum systems. While these systems are not directly involved in beam production, any degradation in their performance can have significant downstream effects on beam stability, machine uptime, and overall facility safety.

Operators monitor telemetry data and alarms associated with utility subsystems through ACNET or dedicated browser-based applications. These tools provide real-time readings of parameters like flow rate, pressure, temperature, and power draw. When irregularities are observed such as a gradual rise in water temperature or a pressure drop in a vacuum line operators may initiate coordination with field personnel or support teams to address the issue before it affects accelerator operations.

In some cases, utility system monitoring overlaps with troubleshooting efforts. For instance, identifying a correlation between magnet instability and cooling flow may prompt a deeper diagnostic check. Thus, maintaining situational awareness of utility system performance is not only preventive but also supports responsive and efficient fault resolution.

#### 8.5.3.1 Operator Purview:

*MCR operators are* tasked with monitoring the performance of various equipment within the Utility System. While they are not responsible for continuous monitoring or diagnosing issues, their role is crucial in supporting specialized roles. They do this by alerting these specialists to potential issues with components of the utility system, enabling further investigation and tracking.

#### 8.5.3.2 Expertise support (enhancing operator capability):

The MCR crew chief and Operations Specialist do not directly attend to Utility System Performance; instead, they delegate this responsibility to the MCR operators. They may, however, intervene if the MCR operators require assistance in managing a situation. Further investigation could help identify the most common barriers that necessitate intervention from the MCR crew chief or Operations Specialist. By labeling these barriers, it would support improved application design and enhance overall performance.

#### 8.5.3.3 Specialized Outside Support (problem occurs outside Ops purview, requiring specialized intervention):

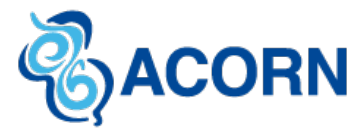

MCR operators provide support to utility system specialists as they are continually monitoring utility system performance, positioning MCR operators as an early warning resource for specialists involved with monitoring the utility system and scheduling and performing maintenance. They feed information to ES&H safety officers for radiation monitoring and may alert Mechanical Support Technicians to suspicious trends, enabling these technicians to take over monitoring. Most specialists monitor utility systems from outside the control room, with specialists primarily using Acnet.

Specialists provide support to MCR operators who find it difficult to interpret system status, they first consult technicians; if the technicians cannot interpret the data, engineers are then consulted. Specialized support roles include Mechanical Support Engineers, Mechanical Support Technicians, Fluids Systems Engineers, and ES&H officers.

#### 8.5.3.4 Planning Support:

Run and Machine Coordinators rely on communication with MCR operators and specialists to stay informed about equipment health status, which is essential for properly scheduling maintenance and beam delivery. Insights gained from MCR operators and specialists support the development of efficient and safe operating schedules that meet the needs of various stakeholders. Machine Coordinators also collect information from mechanical support and fluid systems technicians and engineers to identify machines requiring maintenance. Run Coordinators then use these maintenance schedules to create Beam Delivery plans, ensuring that beam is not sent to areas where people are working or where equipment is down for maintenance. These plans are communicated back to MCR operators to follow.

### 8.5.4 Responding to Alarms

Responding to alarms is a fundamental responsibility of MCR operations, requiring continuous attention and immediate judgment. Alarms serve as the system's primary mechanism for indicating deviations from expected performance or potential faults. Operators must quickly assess the severity and context of an alarm, determine whether immediate action is required, and begin appropriate troubleshooting or coordination steps.

The response process typically begins with identifying the source of the alarm via ACNET or browser-based monitoring tools. Operators review the associated plots, parameter pages, and recent elog entries to gather context. Not all alarms indicate urgent action; experienced operators filter known, benign alarms such as those related to equipment already offline from those requiring immediate diagnosis and resolution.

In many cases, responding to alarms also involves real-time coordination with crew chiefs, support staff, or field technicians to validate conditions, initiate equipment resets, or escalate unresolved issues. When relevant, alarm responses are documented in the elog to maintain traceability and support shift handoffs.

Alarms can signal anything from beam instability to cooling system disruptions or electrical faults. Because of this wide range, operators must maintain a high level of system awareness and draw on their knowledge of current machine conditions to respond effectively. Prompt and accurate alarm response is essential to protecting equipment, ensuring beam quality, and maintaining safe operations across the accelerator complex.

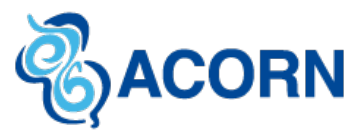

#### 8.5.4.1 Operator Purview:

MCR Operators monitor and respond to alarms to keep the accelerator's health and performance optimized. Alarms help operators attend to system status changes. Alarms at face value are indicators of a status the operator must return to nominal. However, characteristics of an alarm, such as frequency, timing, or the context in which their alarm can have other implications for equipment health that requires further expertise or outside specialized attention from outside the MCR. Operators help route these issues to the proper specialized roles (i.e. technicians) so they may interpret the data available, diagnose, and take appropriate action to resolve the issue.

Expertise support (enhancing operator capability*):*

MCR crew chiefs help delegate operators to attend alarms. Their experience in the control room gives them the knowledge needed to provide alarm response support when needed.
Machine coordinators also support alarm diagnostic and decision support to MCR operators. Their knowledge of machines needing or undergoing maintenance can provide insight into handling alarms. Symbiotically, using machine coordinators as diagnostic and decision support resources has the added impact of informing them of potential issues with machines that may require maintenance.

#### 8.5.4.2 Specialized Outside Support (problem occurs outside Ops purview, requiring specialized intervention):

Providing support to the control room, the interlocks groups help maintain human and system safety-related sensors connected to alarms (i.e. radiation and oxygen sensors). This group works with MCR ops to test sensors and ensure alarms are working as expected. The interlocks group has permissions authority over MCR ops and can override access to some systems.
MCR ops provide support to outside roles by routing alarms that may require localized response or specialized expertise to interpret and respond to.

#### 8.5.4.3 Planning Support:

Machine coordinators benefit by communicating with MCR ops and maintaining awareness of alarms, particularly re-occurring or problematic alarms. Some alarm patterns are indicative of machines that will need placement on the machine coordinator's maintenance schedule.

### 8.5.5 Monitoring Safety Systems

Monitoring safety systems generally involves several key activities. It includes responding to alarms and overseeing safety systems such as safety interlocks and oxygen sensors to ensure that they are functioning within acceptable limits and regulations. This also entails diagnosing any issues and implementing corrective actions as necessary. Additionally, it involves monitoring interlocks connected to ESS and other systems to protect personnel from hazardous conditions, ensuring that sensors and interlocks are reliable and function correctly, and planning maintenance opportunities based on sensor data. Furthermore, it requires coordinating with other roles to schedule maintenance and ensure machine health, supporting operators in their monitoring tasks and interpreting safety data, and using various tools and applications such as ACNET, data loggers, and ESS interlocks to effectively monitor safety systems.

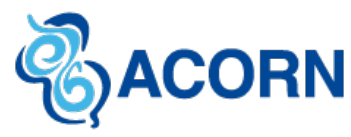

##### 8.5.5.1 Operator Purview:

Operators maintain awareness of safety systems to ensure in-use service areas remain safe for any occupants performing maintenance. They also coordinate with Interlock specialists to identify reliability issues with safety sensors (i.e. radiation and oxygen monitors). Occupied service areas and required maintenance must be scheduled through the machine and run coordinators to properly route beam safely.

##### 8.5.5.2 Expertise support (enhancing operator capability):

The engineer monitors the ESS to confirm that the power supplies are functioning and regulating within acceptable limits. Maintaining proper equipment function allows operators to diagnose issues within their purview without having to first determine that equipment is acting as expected.

The Ops Specialist and MCR Crew Chief plays a supportive role in the operations by assisting in the maintenance of the beam. This role includes periodically monitoring the systems at a high level to provide support to operators in their monitoring tasks. The Ops Specialist ensures that the operators have the necessary support to effectively maintain and monitor the beam operations.

##### 8.5.5.3 Specialized Outside Support (problem occurs outside Ops purview, requiring specialized intervention):

The PSED Electrical Engineer is responsible for ensuring that power supplies operate within the specified parameters. This involves diagnosing any issues that may arise with the power supplies and determining whether the problem lies with the power supply itself or the control system (i.e. a tripped power supply). The PSED Technician monitors ESS interlocks for safety measures and carries out actions related to interlocks and control cards to ensure the safe operation of systems.

The ES&H Officer (rad safety) is responsible for monitoring interlocks connected to the ESS, with a particular focus on radiological impacts. Ensuring the safety systems that protect personnel from hazardous conditions are performing correctly is a key part of this role, as is monitoring service areas for hazardous conditions and maintaining sensor reliability.

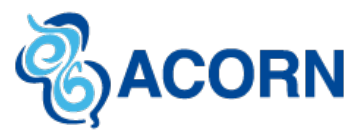

#### 8.5.5.4 Planning Support:

The machine coordinator supports the interlock group and other specialists involved with safety systems to help them plan and coordinate maintenance with minimal impact to safety and beam production. The machine coordinator relies on Acnet, MCR operators, and specialists to maintain awareness of machine health.

The run coordinator uses safety system information to plan beam studies to inform planned shutdowns and maintenance activities. Any machines that are not in operation impact what experiments can be run. The run coordinator uses what is available to balance the needs of safety system maintenance and experiment needs.

### 8.5.6 Machine Commissioning

This function differs from the others as machine commissioning is more focused on those that contribute to the assembly and installation of a machine. MCR operators are the supporting function to help with testing, ensure data accuracy and reliability, and monitor values as components are confirmed as operational. This function is more *specialist centric* rather than *MCR operator* centric.

#### 8.5.6.1 Operator Purview:

The central players in this function are the specialists not the MCR operators. Acnet and the MCR operators support when needed to ramp up power systems and configure the beam line. They carry out actions as needed by the external roles. The MCR operators are coordinated by the operations specialist during this function and communicate with external roles through them. MCR operations support commissioning by sending beam to the equipment being commissioned, helping monitor performance and data from the tests and follow the plan set by the run coordinator.

#### 8.5.6.2 Expertise support (enhancing operator capability):

The nature of function seems to swap the typical paradigm. MCR operators provide their expertise in operating the system such that the machine specialists, machine and run coordinators, and safety officers can calibrate their equipment, test it, and confirm the machine being commissioned is ready to support operations. Therefore, MCR is the expertise support for operating the accelerator to enable the performance and tasks carried out by all external roles involved with machine commissioning.

#### 8.5.6.3 Specialized Outside Support (problem occurs outside Ops purview, requiring specialized intervention):

Machine commissioning requires coordination between multiple specialized support groups. It is the function that includes the roles with direct responsibilities reported by the interviewees. We focused this effort on determining how external roles interact with Acnet and how they currently provide support to MCR operators. However, this function differs from all others in that the MCR provides the support during commissioning.

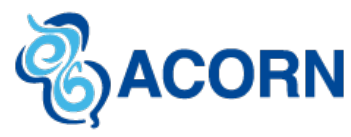

#### 8.5.6.4 Planning Support:

The machine coordinator is responsible for planning how the beam will be directed to the new machine for testing in a safe and stable manner. He must coordinate all up-stream machines, ensure safety systems have been calibrated and pass testing before the new machine can receive a beam.

Run coordinator is the interface between all groups to maintain a mutual awareness of how the event is progressing. The run coordinator may have been who decided to commission the machine to add capabilities or to support better alignment between accelerator capabilities and experiment requirements.

Operations specialists will coordinate MCR actions with external roles to help safely and successfully commission a new machine.

### 8.5.7 Beam Studies

Beam studies are scheduled operational activities that allow accelerator physicists and MCR operators to test, characterize, or optimize beam behavior under non-standard or experimental conditions. Unlike routine operations, beam studies often involve temporarily modifying machine settings, adjusting timing, or bypassing interlocks to investigate specific aspects of system performance, experiment configurations, or beam transport dynamics.

During beam studies, MCR operators work closely with the requesting physicists or systems experts to implement study-specific configurations while maintaining awareness of system safety and equipment limits. Operators may adjust timing, enable alternate beam paths, or change operational modes as directed, all while monitoring for unintended system responses.

Beam studies often occur during dedicated blocks of machine time, separate from standard operational periods, to minimize risk to scheduled experiments and ensure sufficient time for analysis and recovery.

In addition to executing specific changes, MCR operators are responsible for logging each adjustment, documenting system behavior, and supporting troubleshooting efforts when unexpected results occur. These activities are typically captured in the e-log and may include detailed parameter changes, plot captures, and supporting commentary. Beam studies serve as both a tool for scientific exploration and a means of improving overall machine efficiency, flexibility, and reliability.

#### 8.5.7.1 Operator Purview:

MCR Operators focus on maintaining beam health within specific beam study parameters. They monitor and maintain beam quality and delivery to meet the demands of beam experiments, adhering to the run schedule determined by the Run Coordinator.

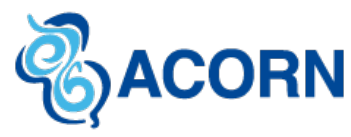

#### 8.5.7.2 Expertise support (enhancing operator capability):

The Ops Specialists is to assist operators in meeting the parameters of beam studies. Their role is primarily secondhand support and supervisory, focusing on monitoring and interpreting data. Ops Specialists use ACNET applications both offline and online to monitor beam efficiency and ensure that the criteria for beam studies are met. This mode of collaboration within ACNET allows them to effectively support operators in maintaining the desired performance levels for beam studies.

#### 8.5.7.3 Specialized Outside Support (problem occurs outside Ops purview, requiring specialized intervention):

Specialized support ensures all equipment and machines are calibrated and tuned such that MCR operations can effectively monitor and operate the accelerator to meet the run schedule determined by the run coordinator and beam studies being performed.
Instrumentation software engineers can connect data from hardware to acnet when needed by MCR operations to run an experiment.
Instrumentation hardware engineers help diagnose unexpected beam performance otherwise undiagnosable from the MCR and its operators.

#### 8.5.7.4 Planning Support:

The run coordinator is responsible for setting the schedule and timing of machines run time optimizing to support as many beam studies as possible. They will monitor performance through acnet to ensure that enough resources are available to successfully carry out scheduled experiments.
The machine coordinator is the communication channel from those designing the experiments to broader accelerator operations. He is charged with working with the run coordinator to ensure all machines required for the beam studies is working properly and the beam studies fit within the operations schedule. Experiments proposed while the beam is running may be considered if it does not disrupt upstream machinery and the outcomes of other experiments.

### 8.5.8 Hardware Troubleshooting

Description: Hardware issues can impact the data transmission from the affected hardware to the MCR control system. Diagnosis and resolving actions cannot take place from the control room and must be performed on the hardware itself. However, issues with hardware can manifest recognizable symptoms that are detectable from ACNET.

#### 8.5.8.1 Operator Purview:

Operators are not responsible for hardware troubleshooting but can sometimes detect a potential issue. Typically, operators encounter situations they realize are beyond the scope of the control room. In these scenarios, regarding hardware troubleshooting, operators “hand off” events that go beyond a reasonable diagnosis (i.e., relatively quick and/or simple) to experts more apt to diagnose and solve them. Fortunately, there are many experts both inside and outside the control room who are responsible for troubleshooting hardware issues such as MCR crew chief, instrumentation department manager, controls hardware engineer, instrumentation hardware engineer, ops specialist, PSED software engineer, PSED technician. Each of these roles is often referred to as work by operations after events that weren't solvable in a timely manner arose and therefore expert help was solicited.

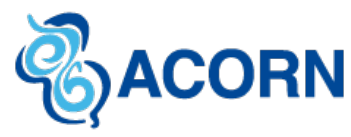

#### 8.5.8.2 Expertise support (enhancing operator capability):

Although hardware troubleshooting aren't direct responsibilities of operators per se, the "hand off" and everything those tasks encompass are. For example, if an operator is tuning the beamline and a machine is unresponsive to tuning, it is usually not the responsibility of the operator to diagnose the exact cause of the machine failure and/or to bring the machine back online. It is, however, the primary responsibility of the operator to communicate the unresponsive machine (or whatever the event is) and to refer the diagnosis to an expert. Across multiple interviews, all additional roles responsible for hardware and troubleshooting expressed the importance of that communication from operators as well as any context that might help the diagnosis. Therefore, any applications that assist in comprehensive communication capabilities between operators and additional roles (e.g., e-log) as well as any collaboration practices between groups should be maintained and potentially evaluated for opportunities for improvement.

#### 8.5.8.3 Specialized Outside Support (problem occurs outside Ops purview, requiring specialized intervention):

When considering applications outside of the operators' purview that are used by additional roles to respond to hardware troubleshooting tasks, both SWIC and CAMAC were mentioned multiple times most notably by the instrumentation and hardware controls engineers. These are physical hardware machines or crates. However, both SWIC and CAMAC have digital interface connections wherein data is being sent from the hardware to software front ends and diagnosis can be performed through both hardware and software perspectives. Therefore, further inspection should occur where opportunities for improvement concerning the interface design and interpretability for hardware such as CAMAC and SWIC are considered in a new and updated control system.

#### 8.5.8.4 Planning Support:

One of the most capable roles concerning planning support for hardware and troubleshooting tasks is ops specialist. Ops specialists are not only previous operators with years of experience as that first line of defense, but they are also an extension of current operators and are able to assess the control system at a higher level compared to the operators who are constantly responding to emergent events. Additionally, when patterns emerge at a high level, ops specialists can spend more time understanding the lower-level hardware, both of which directly strengthen their ability to offer planning support. Although no specific applications were explicitly mentioned by ops specialists relating to planning support, any applications that support their ability to drill down into lower-level hardware data would likely be helpful and is therefore worth consideration for evaluation and improvement.

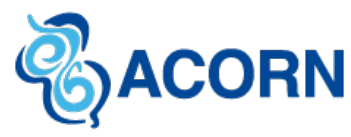

### 8.5.9 System Configuration

System configuration refers to the structured and intentional setup of the accelerator complex hardware, software, control parameters, and relevant operational states to support beamline production. System configuration can also include setup for beamline tuning, experimentation, and maintenance. The purpose of system configuration is to ensure the accelerator functions safely, efficiently, and as expected according to the intended operational scope.

#### 8.5.9.1 Operator Purview:

System configuration is one of the tasks with the highest accelerator personnel responsibility (i.e., nine additional roles are involved with it). It is also a task that operators are directly involved in, and it requires a lot of interdisciplinary coordination between roles and/or groups. Two of the primary tasks that operators are involved in for system configuration are control system configuration and beamline configuration. Examples of sub tasks include setting alarm thresholds, initiating scripts and sequences, and setting up relevant and preferred monitoring screens.

#### 8.5.9.2 Expertise support (enhancing operator capability):

Operators' primary responsibility relating to system configuration is controlling the accelerator through ACNET and front-end computers (e.g., alarm settings, automation sequences, and data logging intervals). As such, it is not only important for operators to have relevant and intuitive control system applications at their disposal, but it is equally important for them to have applications and/or processes that support coordination between the entirety of accelerator personnel, both inside and outside the control system (i.e., all those involved in system configuration).

#### 8.5.9.3 Specialized Outside Support (problem occurs outside Ops purview, requiring specialized intervention):

System configuration is a wide-reaching task that involves many roles including MCR crew chief, controls hardware engineer, fluid systems mechanical support, machine coordinator, instrumentation software engineer, ops specialist, mechanical support engineer, PSED software engineer, and PSED technician. Throughout the interviews for all these roles, only two applications/systems were explicitly mentioned: ACSYS and vacuum systems. However, it should be noted that due to the nature of the system configuration task, the same considerations listed in the operator purview section can and should be applied to all roles relevant to this task (e.g., any application that supports seamless coordination).

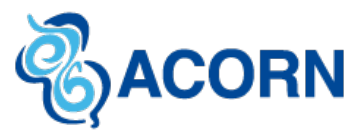

#### 8.5.9.4 Planning Support:

Planning support concerning system configuration relates to each of the additional roles mentioned but is especially important for ops specialists. Since ops specialists work directly with operators, they can observe and document opportunities for improvement across communication and control coordination between operators and additional roles throughout the accelerator complex. Additionally, ops specialists are directly responsible for system configuration and will initiate sequences regularly. Ops specialists have a unique perspective of ancillary support to operators and current hands-on experience, which together can produce the ideal insight for planning support. They can evaluate operations at a higher level and detect potential patterns across emergent events which can then translate to planned maintenance or testing.

### 8.5.10 Script Development

Script development is the creation of automated control programs or operator assist tools that interact with the control system to execute predefined conditions, monitor parameters, or simplify complex tasks. Scripts can help improve efficiency and safety as well as reduce operator workload.

#### 8.5.10.1 Operator Purview:

Script development is often performed by operators, especially when they identify opportunities for efficiency, especially concerning manual, repetitive tasks (e.g., tuning sequence). Some operators use and modify pre-existing scripts whereas other operators develop new scripts. There is a cultural norm in the MCR that the script developer is the script owner and should be responsible for maintaining it. Whom ownership transfers to is unclear when the original owner leaves MCR operations. Occasionally, this results in abandoned scripts that new generations of operators are often unaware of and may recreate a similar script. The ability to track all custom scripts that don’t have an owner or haven’t been used (e.g., 6 months) and automatically decommission them would serve operational efficiency. It would reduce cluttering the list of applications for operators to select leaving only those that are still useful. It would ensure that all scripts are assigned and maintained (i.e., quality assurance) and prevent wasted effort by reducing duplicate work.

#### 8.5.10.2 Expertise support (enhancing operator capability):

Many expert roles are responsible for enhancing operator capability concerning script development including MCR crew chief, ops specialists, and run coordinator. However, regardless of the role type, enhancing operator capability within the framework of script development can be best understood as quality assurance of custom script development and maintenance. In other words, if any of these roles take on the responsibility of developing a custom script, they must also be responsible for assigning the maintenance or maintaining the script themselves. Additionally, script development for external support can reduce task complexity and therefore reduce operator workload. This can be as simple as automating a repetitive manual task or as complex as evaluating workflow improvements. Ops specialists are typically in the best position to identify opportunities for script development. It should be noted that no applications were mentioned throughout the additional role interviews regarding script development or maintenance aside from Phoebus and graphics packages.

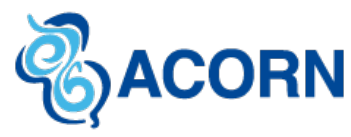

#### 8.5.10.3 Specialized Outside Support (problem occurs outside Ops purview, requiring specialized intervention):

Two roles (i.e., mechanical support technician and accelerator physicist) conduct script development outside of the operator purview. Although both roles/groups conduct script development for their own use (and not operator use), they can sometimes suffer from similar issues of script maintenance. Therefore, a program or application that provides quality assurance of any custom developed scripts across the entire accelerator complex would be beneficial.

#### 8.5.10.4 Planning Support:

Both operations specialists and run coordinators are somewhat involved with script development concerning planning support for operators since both roles can evaluate opportunities for improvement/efficiency that custom scripts might solve. And while both operations specialists and run coordinators engage in script development themselves, they are also known to solicit help from the controls department or even operators to build custom scripts for them. Consequently, a quality assurance application could also help in planning support where a summarized inventory is quickly and easily accessible to determine if a custom script is already available, if an existing script can be modified, or if the development of a new script is assuredly justified.

### 8.5.11 Application Development

Application development is the process of designing, coding, testing, and maintaining custom software applications. Application development can range from a mild effort (i.e., developing a singular new and custom application for a very specific purpose or modifying an existing one) to a major effort (i.e., complete redesign of all applications). Regardless of the development scope however, application development is a task that requires multi-disciplinary coordination to ensure all control system user, most notably operators, have safe, efficient, and user-friendly systems.

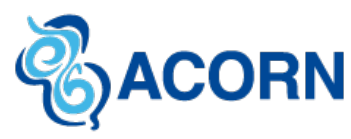

#### 8.5.11.1 Operator Purview:

Operators are heavily involved in both developing and maintaining applications. However, application development is typically reserved for more seasoned operators who have learned the system and have identified ways to supplant or simplify certain tasks throughout operations. In other words, not all operators develop applications, but application development is a key part of the operations group.

#### 8.5.11.2 Expertise support (enhancing operator capability):

Ops specialists and PSED software engineers both provide application development support to enhance operator capabilities. For ops specialists, this typically occurs when they are operators themselves and then they will continue to maintain the applications they developed even when they transition roles into an operations specialist. For PSED software engineers, one of their primary responsibilities is to develop both console and engineering applications to support operators in the control room.

#### 8.5.11.3 Specialized Outside Support (problem occurs outside Ops purview, requiring specialized intervention):

Mechanical support technicians do not typically develop applications; however, they have been known to develop some applications or scripts for different mechanical systems such as vacuum systems.

#### 8.5.11.4 Planning Support:

Ops specialists engage in planning support for operators through cataloging common bottlenecks or error traps within operations and then identifying potential improvements through application development. They are sometimes the developers of new applications and often the maintainers of existing applications.

## 8.6 Global Insights

**Feedback Loop:** The monitoring and interaction in the MCR support external roles in their responsibilities to maintain their systems, thereby enabling operators to manage beam performance more effectively. The natural communication between MCR operators and specialists has created a valuable feedback loop for continuous improvement. Enhancing this feedback loop proactively could further improve system performance and operational efficiency. The informal communication between roles facilitates the execution of accelerator functions and helps all parties fulfill their responsibilities. Capturing and making explicit the knowledge shared during these interactions should be a key design target for applications. Ensuring that this information is communally available will keep everyone mutually aware of the accelerator's status, fostering better coordination and overall system health.

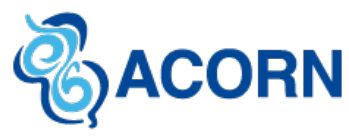

**Interdependent interactions**: The interdependence between external roles and MCR operators supports a broader awareness of accelerator system status, health, and behavior. There is a natural symbiosis of knowledge sharing that impacts all levels of awareness as it relates to operating and maintaining accelerator operations. Supporting and enhancing effective communication, knowledge sharing, and coordination to support the interdependent nature of operations should be considered as functions of the new control system.

MCR operators maintain operations within a narrow operational band, support awareness, and possess deep knowledge in using ACNET for data retrieval. On the other hand, specialists have deep knowledge of their specific equipment for troubleshooting, diagnosis, and action planning, though they may have a narrower understanding of ACNET. MCR operators provide awareness of suspicious conditions to specialist roles, while specialist roles offer expertise in diagnostics and decision support when managing these suspicious conditions. This collaboration ensures that issues are effectively identified and resolved, contributing to the smooth operation of the accelerator system.

**Continuous Improvement:** Application and script development reflect a culture of continuous improvement and present an opportunity for identifying where efficiencies can be gained. MCR operators consistently seek ways to simplify workflows and create tools to support their tasks. Some applications become popular when they effectively address common issues among operators. The MCR crew chief's discussions on efforts to develop scripts that enhance operator workflows serve as a valuable source for pinpointing areas of improvement that the new control system can address.

In summary, ACNET and MCR operations support external roles by providing real-time data, enabling effective communication, assisting with troubleshooting and system configuration, supporting maintenance and commissioning activities, facilitating script and application development, and ensuring emergency response and safety monitoring. This collaborative approach ensures that the accelerator operates smoothly and efficiently.

## 8.7 Current and future work

The migration of control applications for operating Fermi’s primary accelerator systems to a new control interface system involves a thorough review of existing applications under the Accelerator Controls Network (ACNET). This review aims to identify valuable applications for migration, consolidate duplicates, or abandon obsolete ones, ensuring no negative impact on accelerator operations during the transition. The migration presents an opportunity to enhance application functionality and information presentation, improving operator performance and control system accessibility for less frequent users who rely on the control system data. By understanding the roles external to MCR operations and their application needs, the transition can maintain quality, safe, and successful accelerator operations. Application usage metrics and expert interviews have been used to identify critical applications and support provided by external roles, facilitating the development of new control system applications that ensure continued safety and efficiency. This effort also focused on capturing interactions between MCR operators and external roles to design applications that support both parties and expand the operational capabilities of MCR operators.

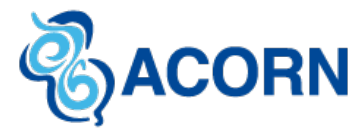

Further work using the data and findings from this effort could map the communication pathways for various functions. Understanding the flow of information in different situations could enable better collaboration and quicker problem resolution. Creating functional workflows, cross-referenced with the applications used, should provide a clear picture of the data and functionality that a new control system could offer. This approach would help maintain the collaborative and interdependent nature of MCR operations and external specialized support.

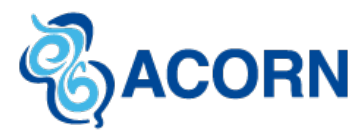

# 9.0 APPENDIX D – INDEX OF REQUIREMENTS

| Section | No. | Requirement |
|---|---|---|
| PART 1: Global Principles for Human Performance | | |
| Design for Primary System Users First and Additional Users Second | HFR-PRI-01 | Operator Workflows Should Drive Initial Design. |
| Design for Primary System Users First and Additional Users Second | HFR-PRI-02 | Core Displays Must Be Optimized for Operational Tasks. |
| Design for Primary System Users First and Additional Users Second | HFR-PRI-03 | Secondary User Needs Must Be Role-Isolated or Configurable. |
| Design for Primary System Users First and Additional Users Second | HFR-PRI-04 | Shared Systems or Displays Must Default to Operator Mode. |
| Support Situation Awareness through Monitoring, Detection, and Selection | HFR-SA-01 | Interfaces must present key system elements in a manner that supports rapid perception and comprehension. |
| Support Situation Awareness through Monitoring, Detection, and Selection | HFR-SA-02 | The interface should support early detection of abnormal or off-nominal conditions. |
| Support Situation Awareness through Monitoring, Detection, and Selection | HFR-SA-03 | Interactive controls must support quick, informed selection, and corrective action. |
| Support Situation Awareness through Monitoring, Detection, and Selection | HFR-SA-04 | Displays must support users' ability to anticipate future system states. |
| Support Situation Awareness through | HFR-SA-05 | The interface must maintain context continuity during transitions and task switching. |

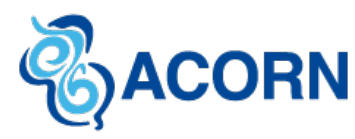

| Section | No. | Requirement |
|---|---|---|
| PART 1: Global Principles for Human Performance | | |
| Monitoring, Detection, and Selection | | |
| Support Situation Awareness through Monitoring, Detection, and Selection | HFR-SA-06 | All aspects of the interface that support SA must be validated through user-centered testing. |
| Embed Usability into Every Phase of Accelerator Interface Design: Consistency | HFR-CONS-01 | Interface elements (e.g., labels, colors, icons, layout zones) must be applied consistently across all pages and subsystems to support user expectations and task fluency |
| Embed Usability into Every Phase of Accelerator Interface Design: Consistency | HFR-CONS-02 | Design systems and visual standards (e.g., widget libraries or pattern sets) must be maintained and referenced in all interface updates to support long-term cohesion |
| Embed Usability into Every Phase of Accelerator Interface Design: Familiarity | HFR-FAM-01 | Common interface patterns (e.g., breadcrumbs, dropdowns) should be used where applicable to match user expectations based on prior experience. |
| Embed Usability into Every Phase of Accelerator Interface Design: Familiarity | HFR-FAM-02 | Terminology, labeling, and iconography must follow known conventions unless specific domain exceptions are required, in which case the UX/HF team should be involved. |
| Embed Usability into Every Phase of Accelerator Interface Design: Simplicity | HFR-SIM-01 | Interfaces must prioritize essential information and controls, removing or hiding secondary details unless explicitly requested by the user (e.g., drill-down). |
| Embed Usability into Every Phase of Accelerator Interface Design: Simplicity | HFR-SIM-02 | Controls for novice tasks must be made immediately visible and clearly labeled. Expert-level functionality may be placed in expandable menus or secondary views. |
| Embed Usability into Every Phase of Accelerator Interface Design: Abstract & Aggregate Data | HFR-AGG-01 | Wherever possible, raw data must be aggregated into summary indicators (e.g., beamline interlock condition as “OPEN” or “CLOSED”). |

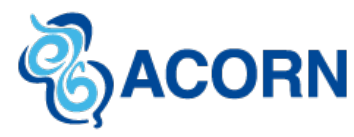

| Section | No. | Requirement |
|---|---|---|
| PART 1: Global Principles for Human Performance | | |
| Embed Usability into Every Phase of Accelerator Interface Design: Abstract & Aggregate Data | HFR-AGG-02 | All aggregated or abstracted indicators must include a method for drilling down into underlying data for expert use or troubleshooting. |
| Embed Usability into Every Phase of Accelerator Interface Design: Transparency | HFR-TRAN-01 | Interfaces must provide immediate, visible feedback when user actions are received, processed, or rejected by the system. |
| Embed Usability into Every Phase of Accelerator Interface Design: Transparency | HFR-TRAN-02 | System controls (e.g., buttons, toggles) must reflect their current state and availability (e.g., enabled, disabled, pending) based on operating mode or user role. |
| Embed Usability into Every Phase of Accelerator Interface Design: Visibility | HFR-VIS-01 | Critical system conditions (e.g., alarms) must be visible on all relevant pages without requiring navigation. |
| Embed Usability into Every Phase of Accelerator Interface Design: Visibility | HFR-VIS-02 | Important information must be presented as distinct (e.g., using layout, color, and formatting) to support rapid recognition, especially during abnormal conditions. |
| Embed Usability into Every Phase of Accelerator Interface Design: Leverage Perceptual Processing | HFR-PER-01 | Related data and controls must be grouped visually and spatially to support fast scanning and reduce visual search effort. |
| Embed Usability into Every Phase of Accelerator Interface Design: Leverage Perceptual Processing | HFR-PER-02 | Pre-attentive features (e.g., high contrast, alignment) may be used to highlight abnormal states but should be sparingly applied to avoid over-alerting or visual noise. |
| Embed Usability into Every Phase of Accelerator Interface Design: Ease of Use | HFR-EASE-01 | All routine tasks should be executable within a minimal number of steps. |

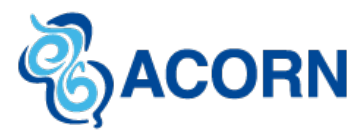

| Section | No. | Requirement |
|---|---|---|
| PART 1: Global Principles for Human Performance | | |
| Embed Usability into Every Phase of Accelerator Interface Design: Ease of Use | HFR-EASE-02 | The interface must be designed to accommodate both novice and expert users. Where appropriate, allow for progressive disclosure of advanced functionality as users become more proficient. |

| Section | No. | Requirement |
|---|---|---|
| PART 2: Style Guide for Accelerator Human-System Interfaces | | |
| Information Architecture | HFR-IA-01 | Understand the role of operator context in developing organizational principles of any interface. |
| Information Architecture | HFR-IA-02 | Operational interfaces should include a "priority" section which enables quick orientation to emergent, novel, or alarm-based information. |
| Display Hierarchy | HFR-DH-01 | Understand the specific forms of awareness the displays are intended to support, supported by user research results, and ensure that all forms of awareness designs are reviewed in user testing. |
| Display Hierarchy | HFR-DH-02 | Ensure that the top level, or overview ("comfort"), information provides general system and accelerator information that supports the broad situation and mutual awareness of equipment and system health. |
| Display Hierarchy | HFR-DH-03 | The design of the top level, or overview ("comfort"), displays should support "at-a-glance" information orientation by operators, supported by user testing results. |
| Display Hierarchy | HFR-DH-04 | The next level, or operator task focused, displays should directly support the operator in performing the tasks and functions assigned to them. |
| Navigation | HFR-NAV-01 | Top or system level navigation should be persistent and visible on all pages. |
| Navigation | HFR-NAV-02 | The systems must provide cues that inform the user of their location in the system and in the overall system informational and navigational hierarchy. |
| Navigation | HFR-NAV-03 | Navigational elements should have a common and persistent design throughout the application. |
| Navigation | HFR-NAV-04 | Designs should group similar navigational elements together. |
| Navigation | HFR-NAV-05**Error! Reference source not found.** | Whenever possible, navigation elements should align with user mental models and common digital design practice. Deviation from these designs or styles should only be done with robust user research. |
| Navigation | HFR-NAV-06 | Accessing system areas that are within one level of a user's current position should only require one interaction to navigate to. |
| Display Formatting and Layout | HFR-DFL-01 | All display pages should contain a header with a unique title at the top of the page. |

| Section | No. | Requirement |
|---|---|---|
| PART 2: Style Guide for Accelerator Human-System Interfaces | | |
| Display Formatting and Layout | HFR-DFL-02 | All display pages should provide a navigation menu at the top left within the header. |
| Display Formatting and Layout | HFR-DFL-03 | All display pages should provide a selectable breadcrumb. |
| Display Formatting and Layout | HFR-DFL-04 | All display pages should provide a primary canvas area that is consistently sized to support the user’s primary task. |
| Display Formatting and Layout | HFR-DFL-05 | The means of saving a display configuration should be explicitly visible to the user. |
| Color | HFR-COL-O1 | Color should be used appropriately to indicate meaning. |
| Color | HFR-COL-O2**Error! Reference source not found.** | Colors should be consistent with accelerator operator convention and expectations. |
| Color | HFR-COL-O3 | A dull screen color scheme should be adopted to reduce display color saturation to increase the value of salient information. |
| Color | HFR-COL-O4 | Saturated colors should be reserved to indicate special meaning. |
| Color | HFR-COL-O5 | Highest priority information (e.g., text or other display elements) must be tested for color blind safety. |
| Color | HFR-COL-O6 | The system should apply the project defined color palette consistently across control system display pages. |
| Typography | HFR-TYP-01 | Typography should be limited to sans serif font options without user research demonstrating an alternative. |
| Typography | HFR-TYP-02 | All alphanumeric text (static and dynamic) should be no less than 9-point font (or 16 minutes of arc) for adequate legibility. |
| Typography | HFR-TYP-03 | All alphanumeric text variations should be consistent throughout all interfaces. |

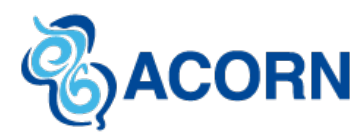

| Section | No. | Requirement |
|---|---|---|
| PART 2: Style Guide for Accelerator Human-System Interfaces | | |
| Typography | HFR-TYP-04 | Labels for UI elements should use a convention that is consistent and intrinsically meaningful to users. |
| Typography | HFR-TYP-05 | All labels to be read should be oriented horizontally on display pages. |
| Typography | HFR-TYP-06 | Labels that contain interaction potential should be visually distinct from information-only labels. |
| Iconography and Symbols | HFR-ICON-01 | Icons and symbols should have clear intentional meaning that is consistent throughout the control system. |
| Iconography and Symbols | HFR-ICON-02 | Icons and symbols should be reserved for common functionality. |
| Iconography and Symbols | HFR-ICON-03 | If color is used as a signifier, an icon or symbol can increase accessibility. |
| Iconography and Symbols | HFR-ICON-04 | Icons and symbols should have contextual content where applicable to communicate intention of use. |
| Information Visualization | HFR-IV-01 | Clearly distinguish contextual information from live plot data |
| Information Visualization | HFR-IV-02 | Present only necessary data on a plot to improve user time to complete task or understand system status. |
| Information Visualization | HFR-IV-03 | Visualization elements should include labels for its title, axes, parameters, colors, shapes, and engineering units. |
| Information Visualization | HFR-IV-04 | Visualization elements should support tooltips and hover pop up functionalities expected by users. |
| Information Visualization | HFR-IV-05 | Visualization elements should include a digital readout of the parameter(s) being represented when precise reading is required of the user. |
| Information Visualization | HFR-IV-06 | Where multiple data points are presented on a single visualization element, each parameter should be coded using color or line type for differentiation. |
| Information Visualization | HFR-IV-07 | When multiple parameters are displayed on a single visualization element, the axes of the |

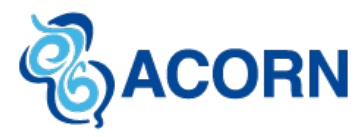

| Section | No. | Requirement |
|---|---|---|
| PART 2: Style Guide for Accelerator Human-System Interfaces | | |
| | | element should be on the same scale, or if differential scales are necessary, the chart should be broken or otherwise provide a visual identifier to users that the scales are different and how the scales should relate. |
| Information Visualization | HFR-IV-08 | All components, line points, and termination points presented on a mimic display should be labeled. |
| Information Visualization | HFR-IV-09 | If a visualization element depicts a physical process, such as a diagram of a system, then any physical processes of flow or movement should be depicted by distinctive arrowheads that demonstrate the flow path. |
| Interaction Design | HFR-IxD-01 | Interactions should be well understood and defined based on the context of the interaction, e.g., what the user is trying to accomplish and how the system will work with the user to achieve that goal. |
| Interaction Design | HFR-IxD-02 | The primary interaction modality should be cursor based, and interactions should consider this in their design. |
| Interaction Design | HFR-IxD-03 | The system should provide indication of all display elements that include interaction functionality. |
| Interaction Design | HFR-IxD-04 | Data entry actions should be accompanied by a verification step. |
| Interaction Design | HFR-IxD-05 | Visual feedback should be provided across all user interactions with the system. |
| Interaction Design | HFR-IxD-06 | Visual feedback should be applied consistently across the control system. |
| Interaction Design | HFR-IxD-07 | System latency should be 0.2 seconds or less for real-time responses. |
| Interaction Design | HFR-IxD-08 | Blinking/flashing should be used only for alerting the user to events that require immediate attention. |
| Interaction Design | HFR-IxD-09 | No more than two blink/flash rates should be used. |

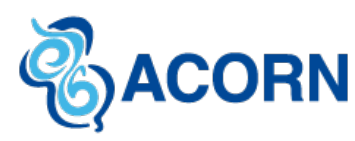

| Section | No. | Requirement |
|---|---|---|
| PART 2: Style Guide for Accelerator Human-System Interfaces | | |
| Controls | HFR-CTRL-01 | The system should provide indication of all display elements that include control functionality. |
| Controls | HFR-CTRL-02 | All control options for a specific soft controller (i.e., faceplate) should be made accessible by a single click. |
| Controls | HFR-CTRL-03 | All frequency performed control actions should be accessible from a soft control faceplate without any additional administrative action. |
| Controls | HFR-CTRL-04 | Soft control options should be suitable for characteristics of the task performed. |
| Controls | HFR-CTRL-05 | Soft controls should be visually distinguishable from other buttons like navigation buttons. |
| Controls | HFR-CTRL-06 | Control actions must be accompanied by a verification step. |
| Controls | HFR-CTRL-07 | The system should provide confirmations for control actions that are safety important or have potential to disrupt normal operation. |
| Controls | HFR-CTRL-08 | The system should prohibit multiple users from controlling the same equipment. |
| Controls | HFR-CTRL-09 | All control actions should include feedback in the design for operators to understand if a control has been actuated. |
| Alarm Systems | HFR-ALRM-01 | Alarms should only be used for off-normal conditions that require timely action by the operator. |
| Alarm Systems | HFR-ALRM-02 | The system should provide the user with notifications of any conditions, internal or external, that may impact the accelerator's performance. |
| Alarm Systems | HFR-ALRM-03 | The system should provide an indication that the display is reading data from the control system (i.e., system heartbeat). |
| Alarm Systems | HFR-ALRM-04 | Live simulation testing with actual operators must be completed prior to deployment of an alarm system to ensure that alarms are effective without becoming an obstacle to safety. |

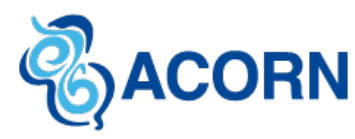

| Section | No. | Requirement |
| --- | --- | --- |
| PART 2: Style Guide for Accelerator Human-System Interfaces | | |
| Alarm Systems | HFR-ALRM-05 | Alarms should provide immediate information regarding data streams that have caused the alarm trigger. |
| Alarm Systems | HFR-ALRM-06 | Alarm notification messages must be written in plain language first with accompanying technical data such as error codes. |
| Alarm Systems | HFR-ALRM-07 | Any information regarding time available must be presented in the alarm message in a live, and real-time manner. |
| Alarm Systems | HFR-ALRM-08 | Any time constraints related to an off-normal condition, and their consequences must be tested with live operators in a simulated event framework. |
| Feedback | HFR-FDBK-01 | All user actions shall result in immediate and visible system feedback. |
| Feedback | HFR-FDBK-02 | System state changes must be reflected in the interface without requiring user polling. |
| Feedback | HFR-FDBK-03 | Critical or safety-related feedback must be persistent and prominently displayed. |
| Feedback | HFR-FDBK-04 | Feedback should indicate both the result and status of system commands. |
| Feedback | HFR-FDBK-05 | Feedback should align with user mental models and the importance of the action. |
| Feedback | HFR-FDBK-06 | Feedback mechanisms must remain consistent across the system. |

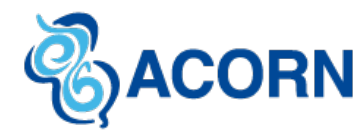

# 10.0 APPENDIX E – COLOR GLOSSARY

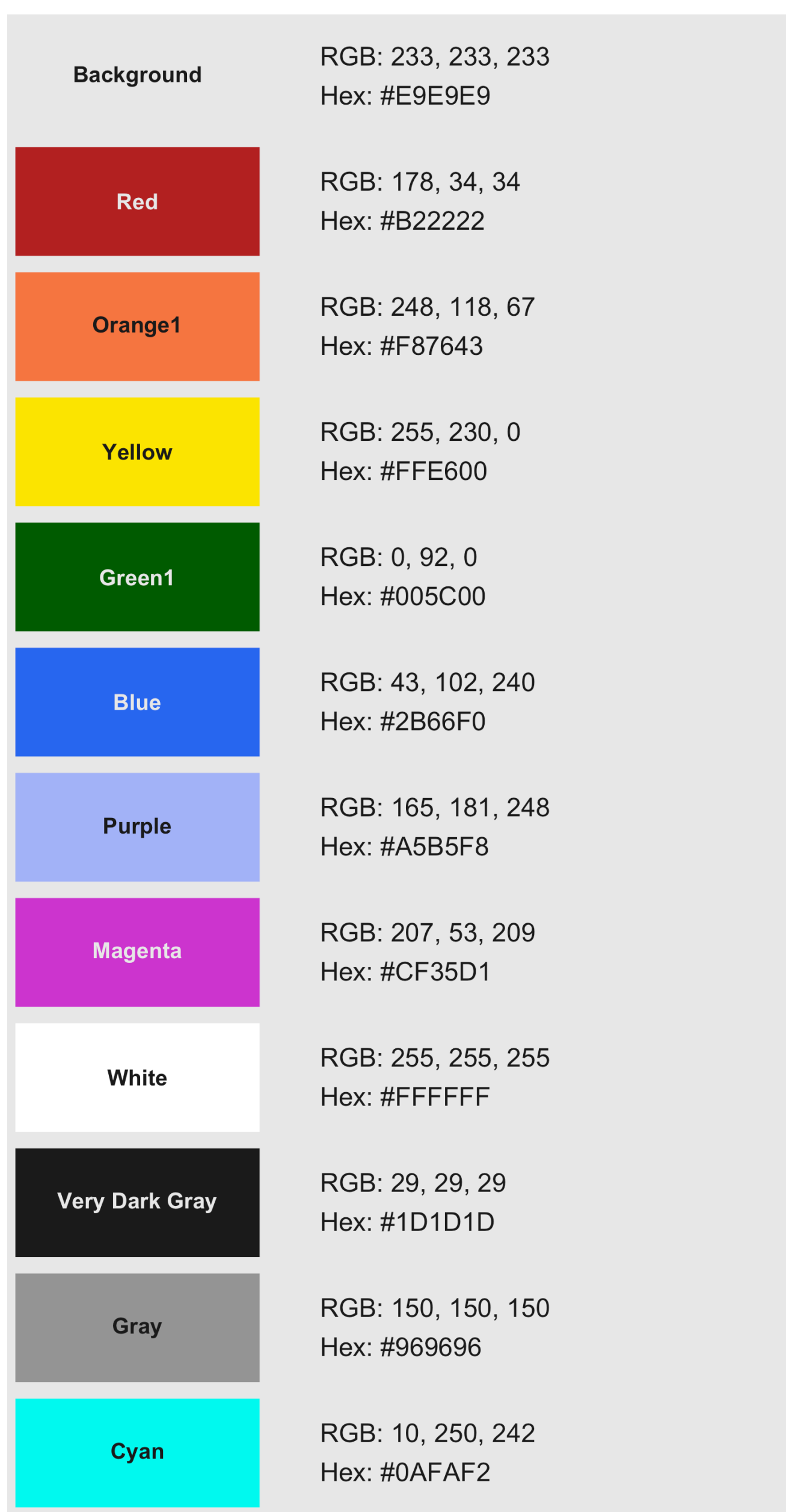

|  |  |
|---|---|
| Background | RGB: 233, 233, 233<br>Hex: #E9E9E9 |
| Red | RGB: 178, 34, 34<br>Hex: #B22222 |
| Orange1 | RGB: 248, 118, 67<br>Hex: #F87643 |
| Yellow | RGB: 255, 230, 0<br>Hex: #FFE600 |
| Green1 | RGB: 0, 92, 0<br>Hex: #005C00 |
| Blue | RGB: 43, 102, 240<br>Hex: #2B66F0 |
| Purple | RGB: 165, 181, 248<br>Hex: #A5B5F8 |
| Magenta | RGB: 207, 53, 209<br>Hex: #CF35D1 |
| White | RGB: 255, 255, 255<br>Hex: #FFFFFF |
| Very Dark Gray | RGB: 29, 29, 29<br>Hex: #1D1D1D |
| Gray | RGB: 150, 150, 150<br>Hex: #969696 |
| Cyan | RGB: 10, 250, 242<br>Hex: #0AFAF2 |

*Figure 1617. ACORN Light Theme Color Palette*

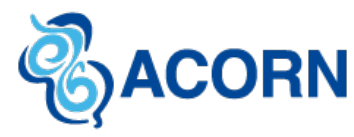

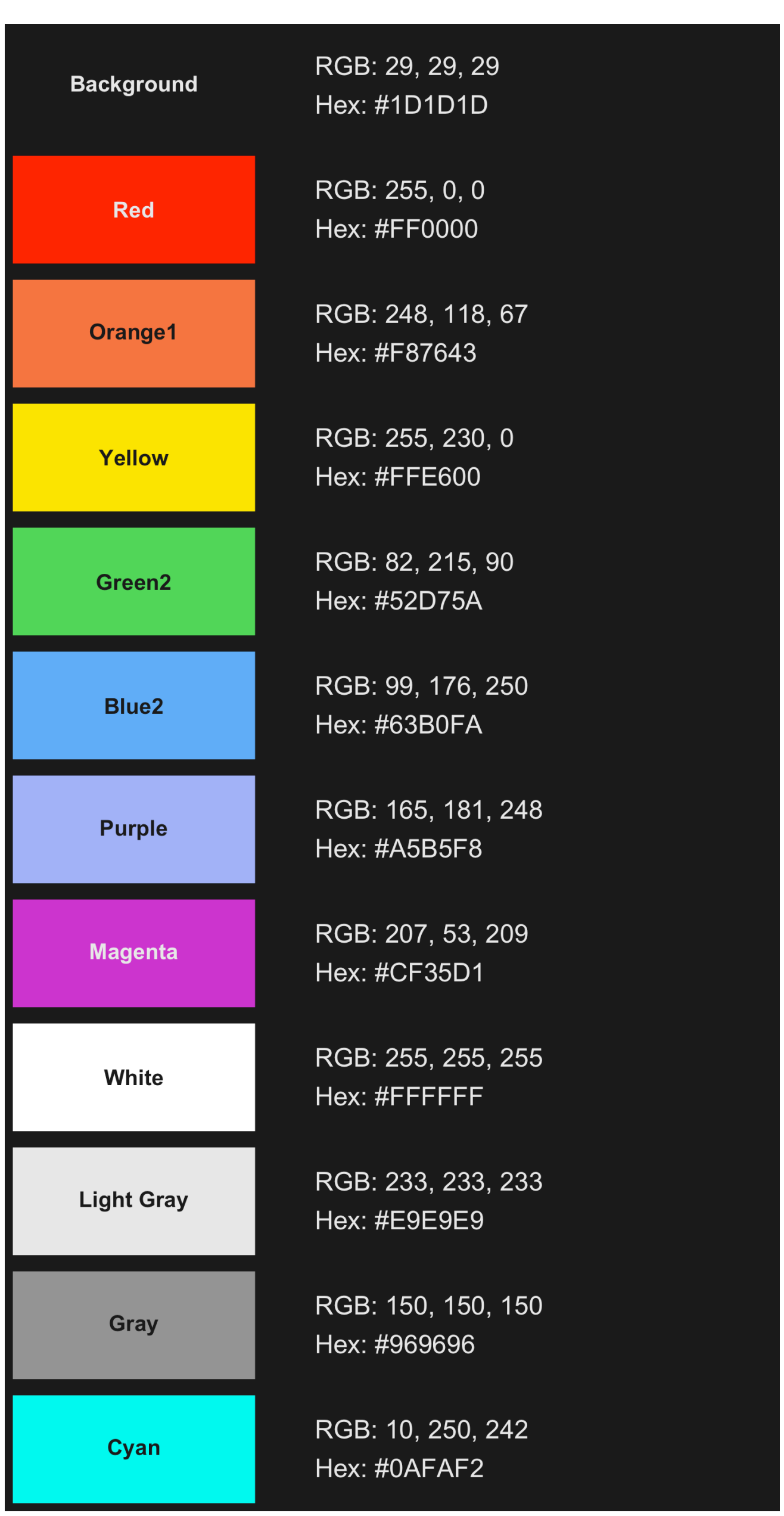

*Figure 1718. ACORN Dark Theme Color Palette*

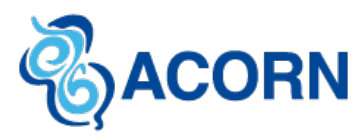
ACORN

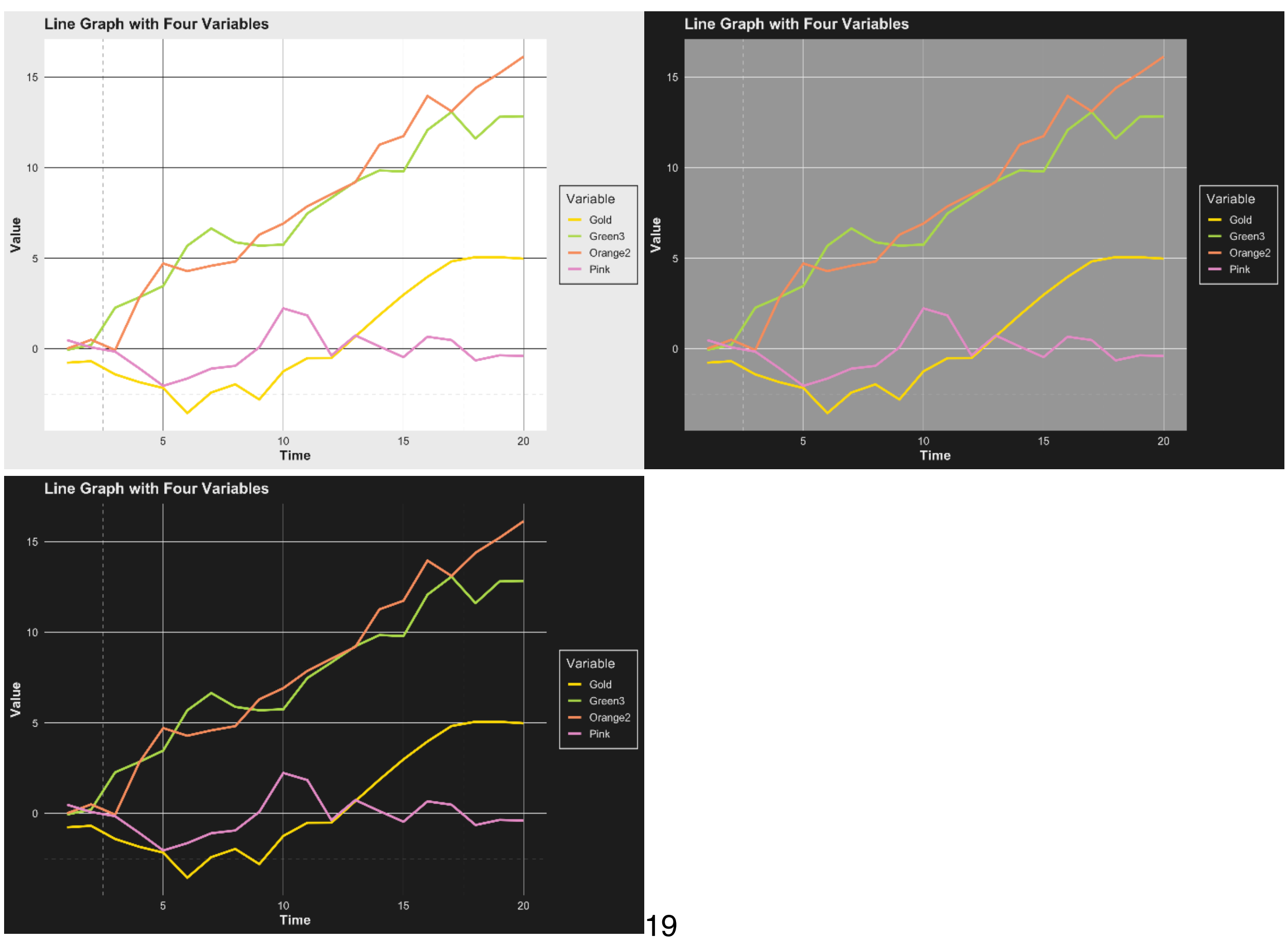
Line Graph with Four Variables
Value
Time
Variable
Gold
Green3
Orange2
Pink
Line Graph with Four Variables
Value
Time
Variable
Gold
Green3
Orange2
Pink
Line Graph with Four Variables
Value
Time
Variable
Gold
Green3
Orange2
Pink

19

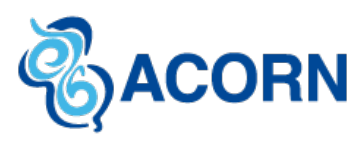

|  |  |  |  |
|---|---|---|---|
| Light Gray | RGB: 233, 233, 233<br>Hex: #E9E9E9 | Light Gray | RGB: 233, 233, 233<br>Hex: #E9E9E9 |
| Gray | RGB: 150, 150, 150<br>Hex: #969696 | Gray | RGB: 150, 150, 150<br>Hex: #969696 |
| Very Dark Gray | RGB: 29, 29, 29<br>Hex: #1D1D1D | Very Dark Gray | RGB: 29, 29, 29<br>Hex: #1D1D1D |
| White | RGB: 255, 255, 255<br>Hex: #FFFFFF | White | RGB: 255, 255, 255<br>Hex: #FFFFFF |
| Green3 | RGB: 170, 216, 74<br>Hex: #AAD84A | Green3 | RGB: 170, 216, 74<br>Hex: #AAD84A |
| Gold | RGB: 251, 216, 0<br>Hex: #FBD800 | Gold | RGB: 251, 216, 0<br>Hex: #FBD800 |
| Orange2 | RGB: 244, 140, 93<br>Hex: #F48C5D | Orange2 | RGB: 244, 140, 93<br>Hex: #F48C5D |
| Pink | RGB: 225, 138, 196<br>Hex: #E18AC4 | Pink | RGB: 225, 138, 196<br>Hex: #E18AC4 |

20

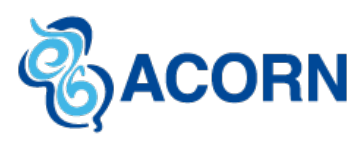

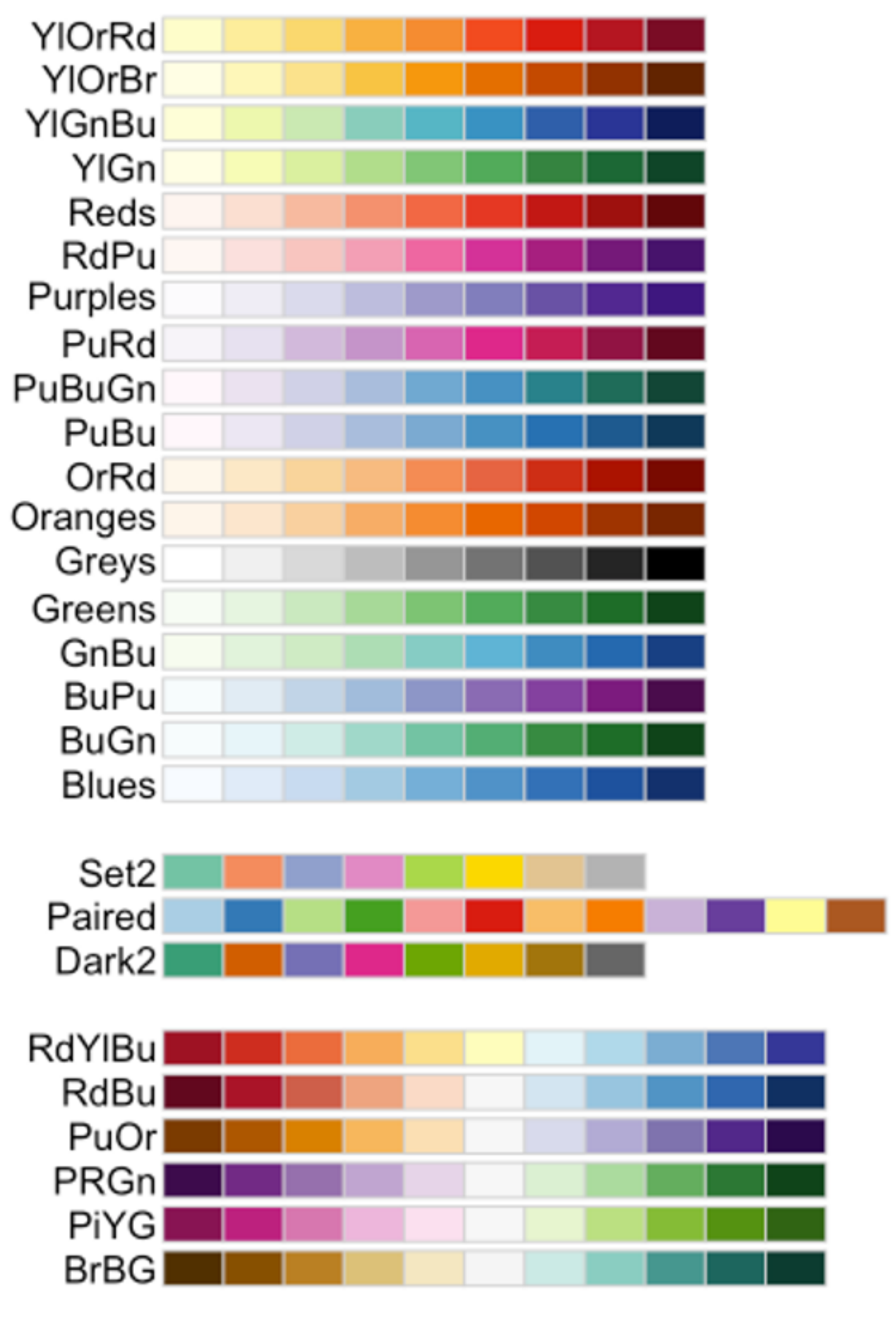

*Figure 2021. Colorblind-friendly ColorBrewer® Palettes for data visualization.*

Note, ColorBrewer can be accessed from the Dart language from:

- https://pub.dev/documentation/colorbrewer/latest/

```
import 'package:colorbrewer/colorbrewer.dart';
```

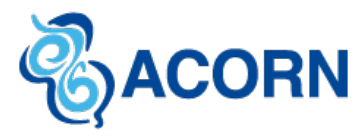

# 11.0 APPENDIX F – DISCUSSION OF TRADE-OFFS

Every interface has different requirements due to targeting different users, with different skill-sets, trying to accomplish different goals. This variability will ultimately lead to decisions that may sacrifice one design element in favor of another. Trade-offs of this kind are a natural part of the design process and must be considered and a direction decided upon. It can be hard to predict where trade-off decision will be required. However, there are some general principles that will likely clash during the design of an accelerator control system such as:

1. Designing for inherent and perceived usability
2. Designing for power user capability and novice user familiarity
3. Meeting user wants while designing for user needs

Inherent and perceived usability can both have implications for the user experience and their ability to perform functions. Perceived usability is the user's impression of the control system, how they perceive the interface visibility and transparency. Their feeling while using the system can also contribute to perceived usability. High perceived usability generally results in a positive experience and great initial impressions of the HSI. Inherent usability is the quantifiable capability of a system to support users in completing their goals. Inherent usability refers to the HSI's actual ability to address the gulf of evaluation and execution. Both types of usability are important aspects of a design. However, when the two conflict in a design decision it is recommended, in the case of developing an HSI for accelerator operators, to opt for inherent usability.

Both power and novice users will be interacting with essentially the same HSI design for accelerator operations. At times, it is difficult to assess which user persona takes priority in the design process. At these times, it is prudent to return to the high-level goals for the upgraded HSI design such as decreasing the time to train new control room operators. As discussed in later sections, novice users will require greater system visibility as they familiarize themselves with the accelerators and the HSI. Therefore, power user functionality can be added such that they require a little more understanding and experience with the system to use.  That way as users progress from novice to power users they will gradually engage with more HSI functionality and learn how to apply it to their tasks.

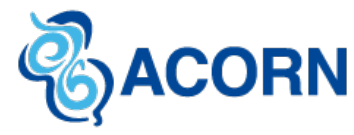

Occasionally, user wants and user needs will clash. It is important to have an HSI that users want and enjoy using but ultimately it must be functional and support their ability to perform in the control room. As these trade-offs are encountered the designers must weigh the implications of the different decisions and how they impact the operator's ability to perform in the control room. If the cost to functionality is low and the benefit of improving the perceived usability is great enough, there may be justification for acquiescing to user wants.

There may be other instances where a trade-off decision is encountered. Each trade-off must be evaluated for the cost to HSI functionality and inherent usability. When these decisions are made, there must be a method for ensuring that consistency across HSIs is maintained.

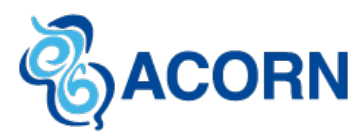

# 12.0 APPENDIX G – REFERENCES

## 12.1 Scope of Document

## 12.2 Support Situation Awareness through Monitoring, Detection, and Selection

## 12.3 Embed Usability into Every Phase of Accelerator Interface Design

## 12.4 Information Architecture

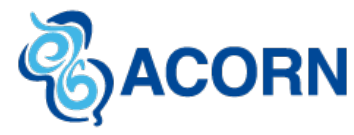

## 12.5 Display Formatting

## 12.6 Color

## 12.7 Iconography & Symbols

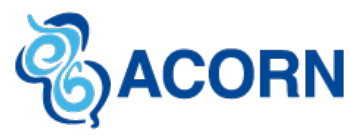

## 12.8 Interaction Design

## 12.9 Alarm Systems

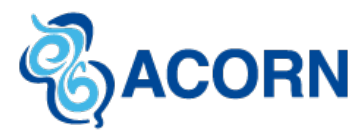

# 13.0 APPENDIX H – RELEVANT USER EXPERIENCE AND HUMAN FACTORS METHODS

The following methods allow the design team to capture user input when implementing the guidance provided in the style guide.

These methods represent a focused selection of human factors techniques commonly used to inform the development of HSIs. It is important to note that these are not the only valid approaches. Many other methods exist within the broader human factors and user research toolbox that can provide similar insights, and their use may vary depending on the project context, timeline, or user population.

The methods included in this section were selected because they are practical, flexible, and well-suited for evaluating how accelerator users perceive, organize, and navigate complex information. These techniques have been successfully applied in other high-reliability environments and are particularly valuable in systems like accelerator controls, where clarity and usability across a range of users and experience levels is critical. Teams are encouraged to work with human factors or UX specialists to tailor their approach, and to treat these methods as a starting point, not a fixed prescription.

- *Card Sorting:* Card sorting is a specific user research task where users are asked to group, arrange, order, or hierarchically sort information. This can include categories, content types, or use cases. There are multiple methodologies that can be used in a card sorting task, work with your human factors or user experience advisor to find the right one for your project (Tankala and Sherwin, 2024).

- *Tree Testing:* Tree testing is a task-based user research test that can evaluate a hierarchy for findability and usability. Users are asked to find specific types or objects of content within an existing hierarchy. This enables the user research team to understand how effective the hierarchy is, how intuitively useful it is, and where some problems may arise for users (Laubheimer, 2023).

- *User Feedback Sessions:* User feedback sessions are informal gatherings of users either after or during use of the system. They can occur after you have users interact with the system for some amount of time, or you can bring users into the room without seeing the system and demonstrate it to them. Those are the most common forms, but there is a nearly infinite number of options. The goal is to capture honest feedback from users on things they like and do not like about the controls and interactions.

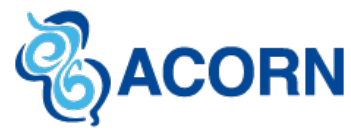

- *Usability Testing (moderated):* Usability testing (moderated), sometimes called 'user testing,' involved a user research team member sitting with the user and observing their usage of the system, often collecting direct feedback simultaneously. This enables the user research team member to see problems as they unfold, as users become lost or confused with what to do. Those user experiences are directly linked to a flaw in the communication between the users and the system.

- *Think-aloud Sessions:* In a think-aloud session, sometimes called 'cognitive walkthrough,' the user is often placed in front of the interface for the first time and instructed to narrate their thoughts as they use the system. This can be done with task direction such as "find this screen" or undirected where the user is told to explore the interface. During the defined think-aloud time, often 10-15 minutes, the user research team member is not to speak or respond to the user. Even if the user asks a question, unless they are in a critical failure state where they cannot continue due to an error or something, the team member can note the question but must remain silent. Some users struggle with this task for a minute or so but then the thoughts will start coming quickly. Think-aloud sessions can provide quality, unfiltered, feedback on the system and its design.

- *Click Testing:* Click testing is often done with an external software program or overlay that tracks the user's clicks on the site. This can show in a very precise way where there are interaction flaws. If users do not follow the "correct" path, click incorrect UI elements, or fail to complete a task click testing can often capture where that problem lies in the interface. Often there will be a qualitative component, debrief, or match with user feedback sessions to then gain the context of the issue. Click testing can only show where the issue is, not what it is. So it is a valuable tool but not in isolation.

These methods can be applied to address specific guidance in this style guide such as seen in the following table. To note, user feedback sessions, usability testing, and think-aloud sessions offer a core set of approaches to address the guidance in this document. User feedback sessions offer a practical way of addressing particular inconsistencies or tradeoffs between guidelines from this document. Think-aloud sessions are particularly useful when first implementing guidelines as it allows for rich information of users' expectations and experiences when interacting with the HSI. Finally, formal usability testing can be applied to verify that the guidance has been effectively implemented.

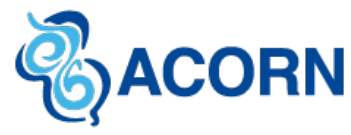

With this, specialty methods like card sorting, tree testing, and click testing offer effective approaches for addressing considerations with information architecture, display hierarchy, and interaction design considerations.

In all cases, it is important to include a member from the UX/Human Factors team when implementing a specific method to ensure such approaches are applied appropriately without inadvertent bias or error resulting in invalid results.

| Method | Use | Support Situation Awareness | Embed Usability in Every Phase | Information Architecture | Display Hierarchy | Navigation | Display Formatting & Layout | Color | Typography | Iconography & Symbols | Information Visualization (Dynamic Display) | Interaction Design | Controls | Alarm Systems | Feedback | | | | | |
|---|---|---|---|---|---|---|---|---|---|---|---|---|---|---|---|---|---|---|---|---|
| Card Sorting | Use to collect feedback of how users categorize information or displays | | | X | X | | | | | | | | | | | | | | | |
| Tree Testing | Use to evaluate the effectiveness of information categorized or display hierarchy | | | X | X | | | | | | | | | | | | | | | |
| User Feedback Sessions | Collect feedback to address potential inconsistencies or tradeoffs in guidance | X | X | X | X | X | X | X | X | X | X | X | X | X | X | | | | | |
| Usability Testing (Moderated) | | | Use to verify guidelines are successfully applied | X | X | X | X | X | X | X | X | | | | X | X | X | X | X | X |
| Think-aloud Sessions | | | Collect verbal feedback to inform the application of specific guidance | X | X | X | X | X | X | X | X | | | | X | X | X | X | X | X |

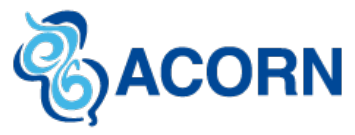

| Method | Use | Support Situation Awareness | Embed Usability in Every Phase | Information Architecture | Display Hierarchy | Navigation | Display Formatting & Layout | Color | Typography | Iconography & Symbols | Information Visualization (Dynamic Display) | Interaction Design | Controls | Alarm Systems | Feedback | | | | | |
|---|---|---|---|---|---|---|---|---|---|---|---|---|---|---|---|---|---|---|---|---|
| Click Testing | | | Used to evaluate how users find information | | | X | X | | | | | | | | | | X | | | |